\documentclass{emulateapj}
\usepackage{graphicx}
\shorttitle{The ALMA Spectroscopic Survey Large Program: The Infrared Excess of $z\geq2$ galaxies}
\shortauthors{Bouwens et al.}

\def\lsim{\mathrel{\rlap{\lower 3pt \hbox{$\sim$}} \raise 2.0pt \hbox{$<$}}}
\def\gsim{\mathrel{\rlap{\lower 3pt \hbox{$\sim$}} \raise 2.0pt \hbox{$>$}}}
\begin{document}
\title{The ALMA Spectroscopic Survey Large Program: The Infrared
  Excess of $z=1.5$--10 UV-selected Galaxies and the Implied
  High-Redshift Star Formation History} \author{Rychard
  Bouwens\altaffilmark{1}, Jorge
  Gonz{\'a}lez-L{\'o}pez\altaffilmark{2}, Manuel
  Aravena\altaffilmark{2}, Roberto Decarli\altaffilmark{3}, Mladen
  Novak\altaffilmark{4}, Mauro Stefanon\altaffilmark{1}, Fabian
  Walter\altaffilmark{4,5}, Leindert Boogaard\altaffilmark{1}, Chris
  Carilli\altaffilmark{5,6}, Ugn\.{e}
  Dudzevi{\v{c}}i{\={u}}t{\.{e}}\altaffilmark{7}, Ian
    Smail\altaffilmark{7}, Emanuele Daddi\altaffilmark{8}, Elisabete
    da Cunha\altaffilmark{9}, Rob Ivison\altaffilmark{10}, Themiya
    Nanayakkara\altaffilmark{1,11}, Paulo Cortes\altaffilmark{12,13},
    Pierre Cox\altaffilmark{14}, Hanae Inami\altaffilmark{15}, Pascal
    Oesch\altaffilmark{16,17}, Gerg{\" o} Popping\altaffilmark{4,10},
    Dominik Riechers\altaffilmark{18,19}, Paul van der
    Werf\altaffilmark{1}, Axel Weiss\altaffilmark{20}, Yoshi
    Fudamoto\altaffilmark{16}, Jeff Wagg\altaffilmark{21}}
\altaffiltext{1}{Leiden Observatory, Leiden University, NL-2300 RA
  Leiden, Netherlands}
\altaffiltext{2}{N{\'u}cleo de Astronom{\'i}a de la Facultad de
  Ingenier{\' i}a y Ciencias, Universidad Diego Portales,
  Av. Ej{\'e}rcito Libertador 441, Santiago, Chile}
\altaffiltext{3}{INAF-Osservatorio di Astrofisica e Scienza dello
  Spazio, via Gobetti 93/3, I-40129, Bologna, Italy}
\altaffiltext{4}{Max-Planck-Institut f{\"u}r Astronomie,
  K{\"o}nigstuhl 17, D-69117 Heidelberg, Germany}
\altaffiltext{5}{National Radio Astronomy Observatory, Pete V. Domenici Array Science Center, P.O. Box O, Socorro, NM 87801, USA}
\altaffiltext{6}{Battcock Centre for Experimental Astrophysics,
  Cavendish Laboratory, Cambridge CB3 0HE, UK}
\altaffiltext{7}{Centre for Extragalactic Astronomy, Department of
  Physics, Durham University, South Road, Durham, DH1 3LE, UK}
\altaffiltext{8}{Laboratoire AIM, CEA/DSM-CNRS-Universite Paris
  Diderot, Irfu/Service d’Astrophysique, CEA Saclay, Orme des
  Merisiers, 91191 Gif-sur-Yvette cedex, France}
\altaffiltext{9}{International Centre for Radio Astronomy Research,
  The University of Western Australia, 35 Stirling Highway, Crawley,
  WA 6009, Australia}
\altaffiltext{10}{European Southern Observatory, Karl Schwarzschild
  Strasse 2, 85748 Garching, Germany}
\altaffiltext{11}{Centre for Astrophysics \& Supercomputing, Swinburne
  University of Technology, PO Box 218, Hawthorn, VIC 3112, Australia}
\altaffiltext{12}{Joint ALMA Observatory - ESO, Av. Alonso de
  C{\'o}rdova, 3104, Santiago, Chile}
\altaffiltext{13}{National Radio Astronomy Observatory, 520 Edgemont
  Rd, Charlottesville, VA, 22903, USA}
\altaffiltext{14}{Institut d’Astrophysique de Paris IAP, 98 bis
  Blvd. Arago, 75014 Paris, France}
\altaffiltext{15}{Hiroshima Astrophysical Science Center, Hiroshima
  University, 1-3-1 Kagamiyama, Higashi-Hiroshima, Hiroshima,
  739-8526, Japan}
\altaffiltext{16}{Department of Astronomy, University of Geneva, 51
  Ch. des Maillettes, 1290 Versoix, Switzerland}
\altaffiltext{17}{International Associate, Cosmic Dawn Center (DAWN) at
  the Niels Bohr Institute, University of Copenhagen and DTU-Space,
  Technical University of Denmark, Copenhagen, Denmark}
\altaffiltext{18}{Department of Astronomy, Cornell University, Space
  Sciences Building, Ithaca, NY 14853, USA}
\altaffiltext{19}{Humboldt Research Fellow}
\altaffiltext{20}{Max-Planck-Institut f{\"u}r Radioastronomie, Auf dem H{\"u}gel 69, 53121 Bonn, Germany}
\altaffiltext{21}{SKA Organization, Lower Withington Macclesfield,
  Cheshire SK11 9DL, UK}

\begin{abstract}
We make use of sensitive (9.3$\mu$Jy$\,$beam$^{-1}$ RMS)
$1.2\,$mm-continuum observations from the ASPECS ALMA large program of
the Hubble Ultra Deep Field (HUDF) to probe dust-enshrouded star
formation from 1362 Lyman-break galaxies spanning the redshift range
$z=1.5$--10 (to $\sim$7-28 M$_{\odot}\,$yr$^{-1}$ at $4\sigma$ over
the entire range).  We find that the fraction of ALMA-detected
galaxies in our $z=1.5$--10 samples increases steeply with stellar
mass, with the detection fraction rising from 0\% at $10^{9.0}$
$M_{\odot}$ to 85$_{-18}^{+9}$\% at $>$10$^{10}$ $M_{\odot}$.
Moreover, stacking all 1253 low-mass ($<$10$^{9.25}$ $M_{\odot}$)
galaxies over the ASPECS footprint, we find a mean continuum flux of
$-$0.1$\pm$0.4$\mu$Jy$\,$beam$^{-1}$, implying a hard upper limit on
the obscured SFR of $<$0.6 $M_{\odot}$$\,$yr$^{-1}$ ($4\sigma$) in a
typical low-mass galaxy.  The correlation between the infrared excess
IRX of $UV$-selected galaxies ($L_{IR}/L_{UV}$) and the $UV$-continuum
slope is also seen in our ASPECS data and shows consistency with a
Calzetti-like relation at $>$$10^{9.5}$ $M_{\odot}$ and a SMC-like
relation at lower masses.  Using stellar-mass and $\beta$ measurements
for $z\sim2$ galaxies over CANDELS, we derive a new empirical relation
between $\beta$ and stellar mass and then use this correlation to show
that our IRX-$\beta$ and IRX-stellar mass relations are consistent
with each other.  We then use these constraints to express the
infrared excess as a bivariate function of $\beta$ and stellar mass.
Finally, we present updated estimates of star-formation rate density
determinations at $z>3$, leveraging current improvements in the
measured infrared excess and recent probes of ultra-luminous far-IR
galaxies at $z>2$.
\end{abstract}

\section{Introduction}

One significant focal point in studies of galaxy formation and
evolution has been a careful quantification of the cosmic star
formation history.  Knowing when most of the stars were formed across
cosmic time is important for understanding the build-up of metals, for
interpreting the stellar populations in both dwarf galaxies and
stellar streams in the halo of our galaxy, and for interpreting cosmic
reionization.  At the present, there is a rough consensus that the
overall cosmic star formation increases from early times to $z\sim3$,
reaching an approximate peak at a redshift of $z\sim2$--3, 2 billion
years after the Big Bang, and then finally decreases at $z<1$ (Madau
\& Dickinson 2014).

Because of the different observational techniques required,
determinations of the cosmic star formation rate (SFR) density have
typically been divided between that fraction of star formation
activity directly observable from rest-$UV$ light and that obscured by
dust which can be inferred from the far-IR emission from galaxies.
Determinations of the unobscured rest-$UV$ SFR density has shown
generally good agreement overall in terms of different results in the
literature (e.g., Madau \& Dickinson 2014; Stark 2016) thanks to the
relatively straightforward procedures for selecting such sources
(e.g., Steidel et al.\ 1996) and substantial sensitive near-IR probes
to 1.6$\mu$m allowing for an efficient probe of such star formation to
$z\sim10$ (e.g., Oesch et al.\ 2018).  Determinations of the obscured
SFR density out to $z\sim 3$ are also mature thanks to the significant
amounts of long wavelength {\it Spitzer} and {\it Herschel}
observations acquired over a wide variety of legacy fields (Reddy et
al.\ 2008; Daddi et al.\ 2009; Magnelli et al.\ 2009, 2011, 2013;
Karim et al.\ 2011; Cucciati et al.\ 2012; {\'A}lvarez-M{\'a}rquez et
al.\ 2016).

In samples of star forming galaxies with both obscured and unobscured
star formation rate estimates, there has been great interest in
determining the ratio of the two quantities, which has traditionally
been expressed in terms of the ratio of the IR luminosity $L_{IR}$ and
$UV$ luminosity $L_{UV}$ of a galaxy.  This quantity is known as the
infrared excess IRX ($IRX = L_{IR}/L_{UV}$), and the correlation of
IRX with the $UV$-continuum slope $\beta$ (or stellar mass)
conveniently allows for an estimate of the $IR$ luminosity or obscured
star formation rate of galaxies where no far-IR observations are
available.

In spite of the significant utility of {\it Herschel} and {\it
  Spitzer}/MIPS for probing obscured star formation out to $z\sim3$,
it has been much more challenging to use these same facilities to
probe such star formation at $z>3$.  The availability of
high-resolution ALMA observations over extragalactic legacy fields has
significantly revolutionized our attempt to probe obscured star
formation in this regime, both in normal star-forming galaxies and
also in more extreme star-forming galaxies which are almost entirely
obscured at rest-$UV$ wavelengths (e.g., Hodge et al.\ 2013; Stach et
al.\ 2019).  Particularly impactful have been the targeted
observations of modest samples of bright star-forming galaxies at
$z\sim5$--8 (Capak et al.\ 2015; Bowler et al.\ 2018; Hashimoto et
al.\ 2018; Harikane et al.\ 2019; B{\'e}thermin et al.\ 2020;
S. Schouws et al.\ 2020, in prep) and deep studies of star-forming
galaxies in the Hubble Ultra Deep Field (Aravena et al.\ 2016; Bouwens
et al.\ 2016; Dunlop et al.\ 2017; McLure et al.\ 2018).

While there are clearly some $z>3$ sources which are well detected in
the far-IR continuum with ALMA (Watson et al.\ 2015; Knudsen et
al.\ 2017; Hashimoto et al.\ 2019), the vast majority of $UV$-selected
$z>3$ sources are not detected individually in the available ALMA
continuum observations, suggesting that only a fraction of the star
formation activity at $z>3$ is obscured by dust.  However, this
interpretation depends significantly on the assumed SED shape of
galaxies in the far-IR, which are needed to infer the total infrared
luminosity from single-band ALMA measurements.  Specifically, a hotter
dust temperature would also make galaxies fainter in the band 6 and 7
(1mm and 870$\mu$m, respectively) observations available for most
$z>4$ galaxies (e.g., Bouwens et al.\ 2016; Barisic et al.\ 2017;
Faisst et al.\ 2017; Bakx et al.\ 2020; but see however Simpson et
al.\ 2017; Casey et al.\ 2018; Dudzevi{\v{c}}i{\={u}}t{\.{e}} et
al.\ 2020).  As a result of this, there are a number of ongoing
efforts to determine how the dust temperature of star-forming galaxies
evolves with cosmic time (Symeonidis et al.\ 2013; Magnelli et
al.\ 2014; Faisst et al.\ 2017; Knudsen et al.\ 2017;
Dudzevi{\v{c}}i{\={u}}t{\.{e}} et al.\ 2020).

Meanwhile, ALMA has been instrumental in identifying modest numbers of
far-$IR$ bright but $UV$ faint galaxies in the $z>3$ universe (e.g.,
Simpson et al.\ 2014; Franco et al.\ 2018; Williams et al.\ 2019;
Yamaguchi et al.\ 2019; Casey et al.\ 2019; Wang et al.\ 2019;
Dudzevi{\v{c}}i{\={u}}t{\.{e}} et al.\ 2020).  The contributed SFR
density of these galaxies to the total SFR density varies from study
to study, but in some cases appears to be comparable to the total SFR
density of Lyman-Break galaxies at $z\sim5$ (Wang et al.\ 2019; Casey
et al.\ 2019; Dudzevi{\v{c}}i{\={u}}t{\.{e}} et al.\ 2020).  Given the
faintness and rarety of these galaxies in the rest-$UV$, they need to
be identified from far-IR detections and their redshifts determined
through constraints on the far-IR SED shape or line scans.

Despite progress with ALMA, current constraints on dust obscuration in
galaxies at $z>3$ is limited, especially for galaxies at low stellar
masses ($<$10$^{9.5}$ $M_{\odot}$).  For these lower mass galaxies,
there has been some debate on whether these galaxies show a steeper
SMC-like extinction curve (see e.g., Reddy et al.\ 2006; Bouwens et
al.\ 2016; Reddy et al.\ 2018) or instead exhibits a shallower
Calzetti-like form (e.g., McLure et al.\ 2018).

Fortunately, new sensitive dust continuum observations have been
acquired over a contiguous 4.2 arcmin$^2$ region with the Hubble Ultra
Deep Field (HUDF) thanks to the 150 hour ALMA Spectroscopic Survey in
the HUDF (ASPECS) large program, obtaining 60 hours of band 3
observations and 90 hours of band 6 observations over the field
(Gonz{\'a}lez-L{\'o}pez et al.\ 2020).  The region chosen for
targeting by ASPECS is that region of the HUDF containing the deepest
near-IR, optical, X-ray, and radio observations available anywhere on
the sky (Beckwith et al.\ 2006; Bouwens et al.\ 2011; Ellis et
al.\ 2013; Illingworth et al.\ 2013; Teplitz et al.\ 2013; Rujopakarn
et al.\ 2016).  These deep, multi-band photometric observations have
made it possible to identify 1362 $UV$-selected star-forming galaxies
at $z\sim1.5$--10 and to systematically quantify their obscured SFRs
as a function of a wide variety of physical properties.  The new 1-mm
continuum ASPECS observations are sufficiently sensitive to probe
dust-obscured SFRs of 7-28 $M_{\odot}$$\,$yr$^{-1}$ at 4$\sigma$ over a
$\sim$5$\times$10$^4$ Mpc$^3$ comoving volume in the distant universe.
The 4.2 arcmin$^2$ targeted with our large program is $\sim$4$\times$
wider than in our ASPECS pilot program (Walter et al.\ 2016; Aravena
et al.\ 2016; Bouwens et al.\ 2016).

The purpose of this paper is to leverage these new observations from
the ASPECS program to probe dust obscured SFR from 1362 star-forming
galaxies at $z=1.5$--10 found over this 4.2 arcmin$^2$ ASPECS
footprint.  The significantly deeper observations not only make it
possible for us to conduct a sensitive search for dust obscured star
formation in individual $z>3$ galaxies, but also allow us to reassess
the dependence of the infrared excess on quantities like the $UV$
slope $\beta$ and stellar mass, while looking at how the dust-obscured
SFRs varies from source to source for a given set of physical
properties.  Thanks to the sensitivity and area of the ASPECS
observations, we can derive particularly tight constraints on the
obscured star formation from galaxies at lower ($<$10$^{9.5}$
$M_{\odot}$) stellar masses.  Probing to such low stellar masses has
been difficult with telescopes like {\it Herschel} (e.g., Pannella et
al.\ 2015) due to challenges with source confusion.

In making use of even more sensitive ALMA observations over wider
areas to revisit our analyses of the infrared excess from our pilot
program (Bouwens et al.\ 2016), we can leverage a number of advances.
For example, new measurements of the dust temperature at $z>3$ from
Pavesi et al.\ (2016), Strandet et al.\ (2016), Knudsen et
al.\ (2017), Schreiber et al.\ (2018), and Hashimoto et al.\ (2019)
plausibly allow us to set better constraints on the dust temperature
evolution to $z\sim5$ and beyond.  In addition, improved constraints
on the obscured SFR density now exist from far-IR bright but UV-faint
galaxies based on a variety of wide-area probes (e.g., Simpson et
al.\ 2014; Franco et al.\ 2018, 2020a; Yamaguchi et al.\ 2019; Wang et
al.\ 2019; Casey et al.\ 2019; Dudzevi{\v{c}}i{\={u}}t{\.{e}} et
al.\ 2020).  Given these improvements and our more sensitive ALMA
observations over the HUDF, a significant aim of the present study
will be to obtain improved constraints on the total SFR density of the
universe.

Here we provide an outline for our paper.  \S2 provides a brief
summary of the ALMA observations we utilize in our analysis,
$z=1.5$--10 galaxy samples, derived stellar masses and $UV$-continuum
slopes, and fiducial scenario for dust temperature evolution.  \S3
presents the small sample of $z=1.5$--10 galaxies where we find
dust-continuum detections in our ASPECS observations as well as our
stack results on the infrared excess.  In \S4, we look at the
implications of our results for dust obscured star formation rate and
cosmic SFR density at $z\gtrsim2$.  \S5 provides a summary of the new
results obtained from our ASPECS large program.

We refer to the {\it HST} F225W, F275W, F336W, F435W, F606W, F775W,
F814W, F850LP, F105W, F125W, F140W, and F160W bands as $UV_{225}$,
$UV_{275}$, $U_{336}$, $B_{435}$, $V_{606}$, $i_{775}$, $I_{814}$,
$z_{850}$, $Y_{105}$, $J_{125}$, $JH_{140}$, and $H_{160}$,
respectively, for simplicity.  For consistency with previous work, we
find it convenient to quote results in terms of the luminosity
$L_{z=3}^{*}$ Steidel et al.\ (1999) derived at $z\sim3$, i.e.,
$M_{1700,AB}=-21.07$.  Throughout the paper we assume a standard
``concordance'' cosmology with $H_0=70$ km s$^{-1}$ Mpc$^{-1}$,
$\Omega_{\rm m}=0.3$ and $\Omega_{\Lambda}=0.7$, which are in
agreement with recent cosmological constraints (Planck Collaboration
et al.\ 2016).  Stellar masses and obscured SFRs are quoted assuming a
Chabrier (2003) IMF.  Magnitudes are in the AB system (Oke \& Gunn
1983).

\section{Observations and Sample}

\subsection{ASPECS Band 6, {\it HST}, and {\it Spitzer} Data}

The principal data used are the band-6 ALMA observations from the
2016.1.00324.L program over the HUDF.  Those observations were
obtained through a full frequency scan in band 6 (212 $-$ 272 GHz)
with ALMA in its most compact configuration.  The observations are
distributed over 85 pointings separated by 11$''$ and cover an
approximate area of $\sim$4.2 arcmin$^2$ to near uniform depth.  Our
construction of a continuum mosaic from ALMA data is described in
Gonz{\'a}lez-L{\'o}pez et al.\ (2020).  The peak sensitivity in our 1.2
mm continuum observations is 9.3$\mu$Jy ($1\sigma$) per synthesized
beam (1.53$''$ $\times$ 1.08$''$: Gonz{\'a}lez-L{\'o}pez et
al.\ 2020).

For {\it HST} optical ACS/WFC and near-infrared WFC3/IR observations,
we make use of the XDF reductions (Illingworth et al.\ 2013), which
incorporated all ACS+WFC3/IR data available over the HUDF in 2013.
The XDF reductions are $\sim$0.1-0.2 mag deeper than original Beckwith
et al.\ (2006) reductions at optical wavelengths and also provide
coverage in the F814W band.  The WFC3/IR reductions made available as
part of the XDF release include all data from the original HUDF09
(Bouwens et al.\ 2011), CANDELS (Grogin et al.\ 2011; Koekemoer et
al.\ 2011), and the HUDF12 (Ellis et al.\ 2013) programs.  Subsequent
to the XDF release, only 17 additional orbits of {\it HST} imaging
data have been obtained with {\it HST} over the XDF region (5 of which
are in the F105W band and 12 in the F435W band).  Given that this is
$<$4\% the integration time already included in the XDF release, we
elected to use the XDF release due to the effort putting into using
super sky flats to optimize the sensitivity.\footnote{We do
  nevertheless note the existence of a new Hubble Legacy Field data
  release (Illingworth et al.\ 2016; Whitaker et al.\ 2019), which
  does include 5 additional orbits of F105W observations from the FIGS
  (Pirzkal et al.\ 2017) and CLEAR (Estrada-Carpenter et al.\ 2019)
  programs over the XDF region.}

For the $0.2$-0.4$\mu$m WFC3/UVIS data over the ASPECS field, we made
use of the v2 release of the UVUDF epoch 3 data (Teplitz et al.\ 2013;
Rafelski et al.\ 2015) which included imaging data in the F225W,
F275W, and F336W bands.  The {\it Spitzer}/IRAC observations we
utilize are from the $\sim$200-hour stacks of the IRAC observations
over the HUDF from the GREATS program (M. Stefanon et al.\ 2020: PI:
Labb{\'e}).

\begin{figure*}
\epsscale{0.8} \plotone{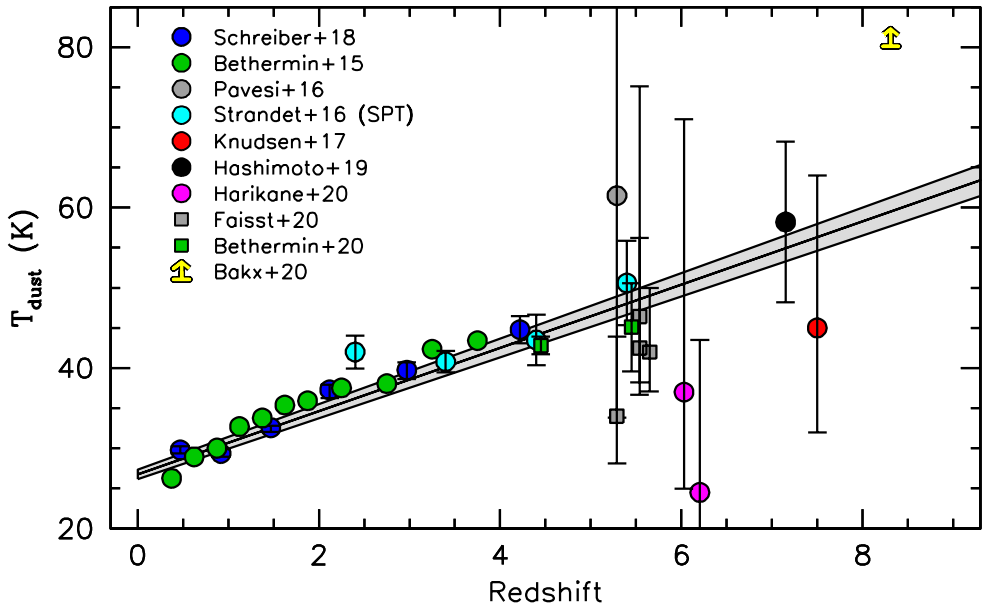}
\caption{Dust temperature estimated for galaxies of various stellar
  masses versus redshift.  Included are temperature measurements from
  Schreiber et al.\ (2018: \textit{blue circles}) for sources with
  stellar masses from $10^{10.0}$ to $10^{11.0}$ $M_{\odot}$,
  B{\'e}thermin et al.\ (2015: \textit{green circles}), Pavesi et
  al.\ (2016) for a $z\sim5.25$ source (\textit{gray circle}),
  Strandet et al.\ (2016: \textit{cyan circles}) for SPT selected
  sources, Knudsen et al.\ (2017: \textit{red circle}) for the lensed
  $z\sim7.5$ galaxy behind Abell 1689 (Bradley et al.\ 2008; Watson et
  al.\ 2015), Hashimoto et al.\ (2019: \textit{black circle}) for a
  bright $z\sim7.13$ source, Harikane et al.\ (2020: \textit{magenta
    circles}) for two bright $z\sim6.1$ galaxies, Faisst et
  al.\ (2020: \textit{gray squares}) for four $z\sim5.5$ galaxies,
  B{\'e}thermin et al.\ (2020: \textit{green squares}) stacking
  $z=4$-5 and $z=5$-6 galaxies, and Bakx et al.\ (2020: \textit{yellow
    lower limit}) for the Tamura et al.\ (2019) $z=8.31$ galaxy.  The
  shaded gray line shows the best-fit linear relationship we derive
  between dust temperature and redshift.\label{fig:tvsz}}
\end{figure*}

\subsection{Flux Measurements}

Photometry for sources in our samples is performed in the same way as
in the Bouwens et al.\ (2016) analysis from the ASPECS pilot program.
{\it HST} fluxes are derived using our own modified version of the
SExtractor (Bertin \& Arnouts 1996) software.  Source detection is
performed on the square-root of $\chi^2$ image (Szalay et al.\ 1999:
similar to a coadded image) constructed from the $V_{606}$, $i_{775}$,
$Y_{105}$, $J_{125}$, $JH_{140}$, and $H_{160}$ images.  After
PSF-correcting fluxes to match the $H_{160}$-band image, color
measurements are made in Kron-style (1980) scalable apertures with a
Kron factor of 1.6. ``Total magnitude'' fluxes are derived by (1)
correcting up the fluxes in smaller scalable apertures to account for
the additional flux seen in a larger-scalable aperture (Kron factor of
2.5) seen on the square root of $\chi^2$ image and (2) correcting for
the flux outside these larger scalable apertures and on the wings of
the PSF using tabulations of the encircled energy, appropriate for
point sources (Dressel 2012).

As in our earlier analysis and many other analyses (e.g., Shapley et
al.\ 2005; Labb{\'e} et al.\ 2006, 2010, 2015; Grazian et al.\ 2006;
Laidler et al.\ 2007; Merlin et al.\ 2015), {\it Spitzer}/IRAC
photometry was performed using the {\it HST} observations as a
template to model the fluxes of sources in the {\it Spitzer}/IRAC
observations and thus perform photometry below the nominal confusion
limit.  In performing photometry, a simultaneous fit of the flux of a
source of interest and its neighbors is performed, the flux from
neighboring sources is subtracted, and then aperture photometry on the
source of interest is performed.  Photometry is performed in
1.8$''$-diameter circular apertures for the {\it Spitzer}/IRAC
3.6$\mu$m and 4.5$\mu$m bands and 2.0$''$-diameter circular apertures
for the 5.8$\mu$m and 8.0$\mu$m bands.  The observed fluxes are
corrected to total based on the inferred growth curve for sources
after PSF-correction to the {\it Spitzer}/IRAC PSF.

A similar procedure is used to derive fluxes for sources from the deep
ground-based $K$-band observations available from the VLT/HAWK-I HUGS
(Fontana et al.\ 2014), VLT/ISAAC, and PANIC observations over the
HUDF ($5\sigma$ depths of 26.5 mag).

\subsection{Fiducial SED Template and Dust Temperature Evolution}

The purpose of this subsection is to summarize our approach in
modeling the far-IR SED of faint, $UV$-selected $z=1.5$--10 galaxies.
Having accurate constraints on the overall form of the far-IR SED for
these galaxies is potentially important for interpreting far-IR continuum
observations of the distant universe to quantify the dust-obscured
SFRs.  The goal of this subsection will be to use a variety of
published observations from the literature to motivate the approach we
will utilize throughout the balance of this manuscript.

As is common practice (e.g., Casey 2012), we will adopt a modified
blackbody (MBB) form to model the far-IR spectral energy distributions
of galaxies (e.g., Casey 2012), with a dust emissivity power-law
spectral index of $\beta_d = 1.6$, which is towards the center of the
range of values 1.5 to 2.0 frequently found in the observations (Eales
et al.\ 1989; Klaas et al.\ 1997).  MBB SEDs have the advantage of
being relatively simple in form, but are known to show less flux at
mid-IR wavelengths than galaxies with a prominent mid-IR power-law
component.  Fortunately, the impact of such differences on the
conversion factors from the 1.2mm flux densities we observe and the
total IR luminosity is relatively modest (i.e., factors of
$\lesssim$1.5: see e.g. Casey et al.\ 2018), especially relative to
other issues like the dust temperature.

Characterizing the evolution of the dust temperature as a function of
redshift is challenging due to both selection bias and the significant
dependence the dust temperature can show on other quantities like the
bolometric luminosity, specific star formation, and the wavelength
where dust becomes opaque (e.g., Magnelli et al.\ 2014; Liang et
al.\ 2019; Ma et al.\ 2019) which are arguably larger and more
significant than the impact of redshift on the dust temperature.

\begin{deluxetable*}{cccccccccc}
\tablewidth{0cm}
\tablecolumns{10}
\tabletypesize{\footnotesize}
\tablecaption{$4\sigma$ Sensitivity Limits for our Probe of Obscured Star Formation from Individual $z\gtrsim1.5$ Galaxies and the Dependence on SED\label{tab:limsfr}}
\tablehead{\colhead{Far-Infrared} & \multicolumn{9}{c}{4$\sigma$ Sensitivity Limits ($10^{10}$ $L_{\odot}$)}\\
\colhead{SED Model} & \colhead{$z$$\sim$2} & \colhead{$z$$\sim$3} & \colhead{$z$$\sim$4} & \colhead{$z$$\sim$5} & \colhead{$z$$\sim$6} & \colhead{$z$$\sim$7} & \colhead{$z$$\sim$8} & \colhead{$z$$\sim$9} & \colhead{$z$$\sim$10}}
\startdata
 Fiducial Evolving \tablenotemark{a,b} & 6.8 & 9.0 & 11.2 & 13.6 & 16.1 & 18.7 & 21.7 & 24.9 & 28.4 \\
35K greybody\tablenotemark{b} & 7.1 & 6.3 & 5.5 & 5.1 & 4.8 & 4.7 & 4.7 & 4.9 & 5.2 \\
50K greybody\tablenotemark{b} & 30.8 & 25.2 & 20.9 & 17.8 & 15.7 & 14.4 & 13.6 & 13.3 & 13.4 \\\\
  & \multicolumn{9}{c}{4$\sigma$ Limit for Probes of the Obscured SFR (M$_{\odot}$$\,$yr$^{-1}$)\tablenotemark{c}}\\
 SED Model & $z$$\sim$2 & $z$$\sim$3 & $z$$\sim$4 & $z$$\sim$5 & $z$$\sim$6 & $z$$\sim$7 & $z$$\sim$8 & $z$$\sim$9 & $z$$\sim$10 \\
\tableline
Fiducial Evolving \tablenotemark{a,b} & 6.8 & 9.0 & 11.2 & 13.6 & 16.1 & 18.7 & 21.7 & 24.9 & 28.4 \\
35K greybody\tablenotemark{b} & 7.1 & 6.3 & 5.5 & 5.1 & 4.8 & 4.7 & 4.7 & 4.9 & 5.2 \\
50K greybody\tablenotemark{b} & 30.8 & 25.2 & 20.9 & 17.8 & 15.7 & 14.4 & 13.6 & 13.3 & 13.4 \\\\
  & \multicolumn{9}{c}{Dust Temperatures for Fiducial Evolving SED Model ($\deg$ K)}\\
  & 34.6 & 38.5 & 42.5 & 46.4 & 50.4 & 54.3 & 58.2 & 62.2 & 66.1 
\enddata
\tablenotetext{a}{Using Eq.~\ref{eq:tvsz}}
\tablenotetext{b}{Standard modified blackbody form (e.g., Casey 2012)
  with a dust emissivity power-law spectral index of $\beta_d = 1.6$
  (Eales et al.\ 1989; Klaas et al.\ 1997).}
\tablenotetext{c}{The Kennicutt (1998) conversion factor from IR luminosity to SFR is adopted.}
\end{deluxetable*}

\begin{figure*}
\epsscale{1.15} \plotone{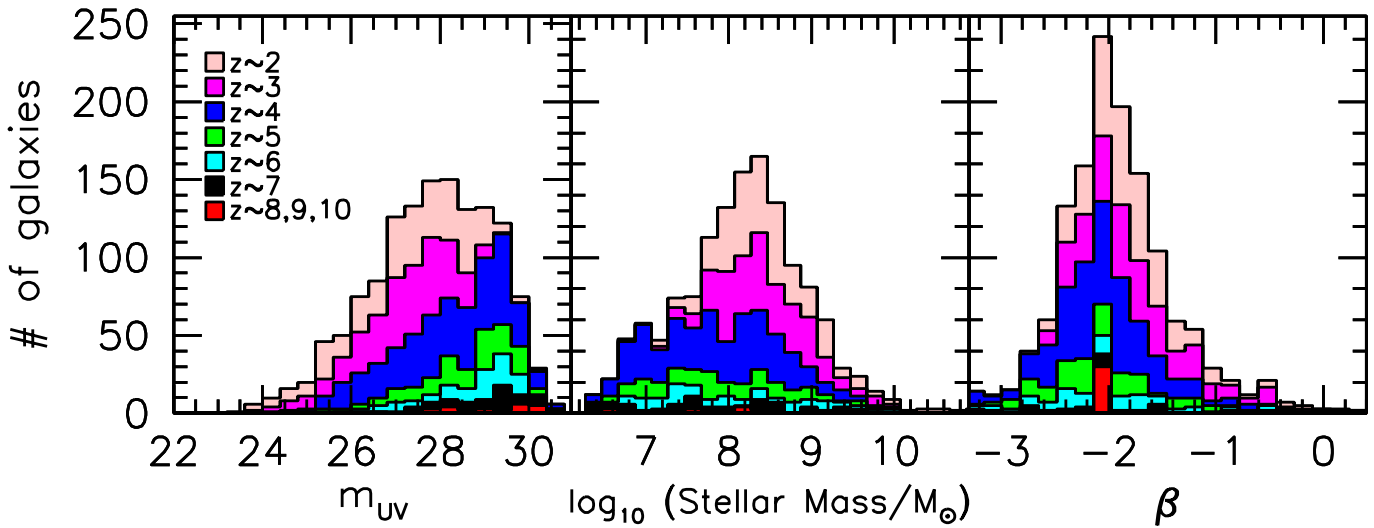}
\caption{Cumulative histograms showing the composition of the HUDF
  samples examined with our deep ASPECS 1.2-mm continuum observations
  as a function of apparent magnitude (measured at wavelengths probing
  the $UV$ continuum), stellar mass, and $UV$-continuum slope $\beta$
  (\textit{left, central, and right panels, respectively}).  Shown are
  our $z\sim2$, $z\sim3$, $z\sim4$, $z\sim5$, $z\sim6$, $z\sim7$, and
  $z\sim8$--10 samples (\textit{pink, magenta, blue, green, cyan,
    black, and red shaded histograms, respectively}).  The
  $UV$-continuum slopes $\beta$ of $z=8$--10 sources are all taken to
  be $-2.2$ consistent with the results of Bouwens et
  al.\ (2014a).\label{fig:sample}}
\end{figure*}

\begin{deluxetable}{cccc}
\tablewidth{0cm}
\tablecolumns{4}
\tabletypesize{\footnotesize}
\tablecaption{Number of $UV$-selected $z\sim2$, $z\sim3$, $z\sim4$, $z\sim5$, $z\sim6$, $z\sim7$, $z\sim8$, $z\sim9$, and $z\sim10$ Galaxies Located within our 4.2 arcmin$^2$ ASPECS footprint\label{tab:sample}}
\tablehead{\colhead{} & \colhead{} & \colhead{\# of} & \colhead{}\\
\colhead{Redshift} & \colhead{Selection Criterion} & \colhead{Sources} & \colhead{Ref\tablenotemark{a}}}
\startdata
$z\sim2$ & $UV_{275}$-dropout or &  & \\
         & $1.5<z_{phot}<2.5$ & 447 & R15/This Work\\
$z\sim3$ & $U_{336}$-dropout or &  & \\
         & $2.5<z_{phot}<3.5$ & 203 & R15/This Work\\
$z\sim4$ & $B_{435}$-dropout or &  & \\
         & $3.5<z_{phot}<4.5$ & 395 & B15/This Work\\
$z\sim5$ & $V_{606}$-dropout & 139 & B15\\
$z\sim6$ & $i_{775}$-dropout & 94 & B15\\
$z\sim7$ & $z_{850}$-dropout or \\
         & $6.5<z_{phot}<7.5$ & 54 & B15/This Work\\
$z\sim8$ & $Y_{105}$-dropout & 24 & B15\\
$z\sim9$ & $Y_{105}$-dropout & 4 & This Work\\
$zi\sim10$ & $J_{125}$-dropout & 2 & This Work\\
\multicolumn{2}{c}{Total} & 1362
\enddata
\tablenotetext{a}{References: B15 = Bouwens et al.\ (2015), R15 = Rafelski et al.\ (2015)}
\end{deluxetable}

Nevertheless, there have been multiple studies looking at the
evolution of dust temperature in galaxies with redshift for fixed
values of the bolometric luminosity (e.g., B{\'e}thermin et al.\ 2015;
Schreiber et al.\ 2018).  One particularly comprehensive recent study
on this front has been by Schreiber et al.\ (2018), who consider the
apparent evolution in dust temperatures from $z\sim4$ to $z\sim0$
using stacks of the available {\it Herschel} observations.

In Figure~\ref{fig:tvsz}, we present the same observations that
Schreiber et al.\ (2018) consider, and then add to their constraints
earlier results from B{\'e}thermin et al.\ (2015).  Finally, we also
include the dust temperature measurements obtained by Pavesi et
al.\ (2016) on a $z\sim5.25$ galaxy, by Knudsen et al.\ (2017) on a
$z\sim7.5$ galaxy, by Hashimoto et al.\ (2019) on a $z=7.15$ galaxy,
by Harikane et al.\ (2020) on two $z\sim6.1$ galaxies, by Bakx et
al.\ (2020) on the Tamura et al.\ (2019) $z=8.31$ galaxy, by Faisst et
al.\ (2020) on four $z\sim5.5$ galaxies, and by B{\'e}thermin et
al.\ (2020) on stacks of $z=4$-5 and $z=5$-6 galaxies, as well as the
median dust temperatures measured by Strandet et al.\ (2016) on their
sample of bright South Pole Telescope (SPT) sources.  Each of these
temperature measurements is reported to be corrected for the impact of
CMB radiation (da Cunha et al.\ 2013).

To make the present dust temperature measurements in
Figure~\ref{fig:tvsz} as consistent as possible, all measurements have
been converted to their equivalent values using an emissivity index
$\beta_d$ of 1.6 and using the light-weighted dust temperatures
(converting the Schreiber et al.\ 2018 temperatures from the
mass-weighted temperatures to light-weighted temperatures using their
Eq. 6).  Pursuing a joint fit to all dust temperature measurements in
Figure~\ref{fig:tvsz}, we derive the following relationship between
dust temperature and redshift:
\begin{equation}
T_d [K] = (34.6 \pm 0.3) + (3.94 \pm 0.26)(z-2)
\label{eq:tvsz}
\end{equation}
The best-fit evolution we derive for the dust temperature is higher
than what Schreiber et al.\ (2018) derive ($T_d [K] = (32.9 \pm 2.4) +
(4.60 \pm 0.35) (z-2)$) due to our use of light-weighted dust
temperatures where the dust temperatures are higher.  Our best-fit
relation for the temperature evolution does, however, evolve slightly
less steeply with redshift, largely as a result of our inclusion of
constraints from SPT sources, the four Faisst et al.\ (2020)
$z\sim5.5$ galaxies, and the new B{\'e}thermin et al.\ (2020) stack
constraints for $z=4$-6 galaxies.  This best-fit evolution is also not
especially dissimilar from the trends found in theoretical models such
as those by Narayanan et al.\ (2018), Liang et al.\ (2019), and Ma et
al.\ (2019).  In the Narayanan et al.\ (2018) results, the dust
temperature increases from 40-50 K in galaxies at $z\sim2$--3 galaxies
to 55-70 K at $z\sim6$--7.  In Liang et al.\ (2019) and Ma et
al.\ (2019), the evolution in dust temperature expected on the basis
of the evolution of the \textsc{MASSIVEFIRE} sample is
$(1+z)^{0.36\pm0.06}$ (their Table 2), similar to that implied by
Eq.~\ref{eq:tvsz} above.

Despite the clear evolution in temperature found here and earlier by
B{\'e}thermin et al.\ (2015) and Schreiber et al.\ (2018), other
recent studies find no less evolution in dust temperature with
redshift.  For example, Ivison et al.\ (2016) infer only $\sim$50\% as
much evolution in the dust temperature as we find, while other
studies, e.g., Dudzevi{\v{c}}i{\={u}}t{\.{e}} et al.\ (2020), find no
significant evolution in the dust temperature of galaxies with
redshift when a purely luminosity-limited sample is studied (see also
Strandet et al.\ 2016).  Dudzevi{\v{c}}i{\={u}}t{\.{e}}
et al.\ (2020) have argued that the apparent temperature evolution
that studies such as Schreiber et al.\ (2018) have found is likely a
consequence of luminosity variations in that study.  Given this, we
also consider there being less evolution of the dust temperature of
galaxies with cosmic time than in our fiducial models.

Assuming that the effective dust temperature of obscured SF in
$z\sim1.5$--10 galaxies follows the same evolution as given by
Eq.~\ref{eq:tvsz}, we can derive the limiting dust-obscured star
formation rate we would be able to detect as a function of redshift
from our program.  Adopting a modified blackbody form for the SED
shape described at the beginning of this section and accounting for
the impact of the CMB (e.g., da Cunha et al.\ 2013: \S3.1.1), we
estimate that we should be able to detect at $4\sigma$ any
star-forming galaxy at $z>2$ with an IR luminosity (8-1000$\mu$m
rest-frame) in excess of 6.8$\times10^{10}$ $L_{\odot}$ at $z\sim2$,
9.0$\times10^{10}$ $L_{\odot}$ at $z\sim3$, and
$\sim$11.2-28.4$\times10^{10}$ $L_{\odot}$ at $z\sim4$--10.  We
verified that use of potentially more realistic far-IR SED templates
than a modified blackbody form, following e.g.
{\'A}lvarez-M{\'a}rquez et al.(2016) with a mid-IR power-law, yields
similar 1.2mm to IR luminosity conversion factors (see also Appendix A
of Fudamoto et al.\ 2020a).

Adopting the Kennicutt (1998) conversion between IR luminosity and the
star formation rate (SFR), these limits translate to $4\sigma$ limits
on the obscured SFRs of 6.8 $M_{\odot}$$\,$yr$^{-1}$, 9.0
$M_{\odot}$$\,$yr$^{-1}$, and 11.2-28.4 $M_{\odot}$$\,$yr$^{-1}$, respectively,
at these redshifts.  If we instead allow for much less evolution in
the dust temperature, such that the typical dust temperature at
$z\sim4$--8 is 35 K, the $4\sigma$ limits from ASPECS translates to
limits on the obscured SFRs of 4-5 $M_{\odot}$$\,$yr$^{-1}$.

In Table~\ref{tab:limsfr}, we provide these limiting luminosities and
SFRs in tabular form, while providing for context these limits for
modified blackbody SEDs if the dust temperature is fixed at 35K or
50K.

\subsection{Selections of $z=1.5$--10 Galaxies}

In constructing samples of $z=1.5$--10 galaxies for examination with
the ASPECS ALMA data, we utilize both Lyman-break selection criteria
as well as a photometric redshift selection to ensure our samples are
as comprehensive as possible.

For consistency with earlier results from our pilot study (Bouwens et
al.\ 2016), we have adopted essentially identical color-color and
photometric redshift selection criteria to those applied in Bouwens et
al.\ (2016).  $z=1.5$--3.5 sources are identified using the same
Lyman-break color criteria we had earlier used in Bouwens et
al.\ (2016) and identified by running the EAZY photometric redshift
code (Brammer et al.\ 2008) on our own {\it HST} WFC3/UVIS, ACS, and
WFC3/IR photometric catalogs.  Our $z\sim2$ and $z\sim3$ color
criteria are as follows:
\begin{eqnarray*}
z\sim2: (UV_{275}-U_{336}>1)\wedge (U_{336}-B_{435}<1)\wedge \\
(V_{606}-Y_{105}<0.7)\wedge (S/N(UV_{225})<1.5)
\end{eqnarray*}
\begin{eqnarray*}
z\sim3: (U_{336}-B_{435}>1)\wedge(B_{435}-V_{606}<1.2)\wedge \\
(i_{775}-Y_{105}<0.7)\wedge(\chi_{UV_{225},UV_{275}}^2<2)
\end{eqnarray*}
where $\wedge$, $\vee$, and S/N represent the logical \textbf{AND},
\textbf{OR} symbols, $\chi^2$ is the $\chi^2$ parameter defined in
Bouwens et al. (2011), and signal-to-noise in our smaller scalable
apertures, respectively.  We also made use of the photometric catalog
of Rafelski et al.\ (2015) and included those sources in our samples,
if not present in the other selections.

Our $z=4$--8 samples are drawn from the Bouwens et al.\ (2015) samples
and include all $z=3.5$--8.5 galaxies located over the 4.2 arcmin$^2$
ASPECS region.  The Bouwens et al.\ (2015) samples were based on the
deep optical ACS and WFC3/IR observations within the HUDF.  $z=4$--8
samples were constructed by applying Lyman-break-like color criteria
to the XDF reduction (Illingworth et al.\ 2013) of the Hubble Ultra
Deep Field.  Those criteria are the following for our $z\sim4$, 5, 6,
7, and 8 selections:
\begin{eqnarray*}
z\sim4: (B_{435}-V_{606}>1)\wedge (i_{775}-J_{125}<1) \wedge \\
(B_{435}-V_{606} > 1.6(i_{775}-J_{125})+1)
\end{eqnarray*}
\begin{eqnarray*}
z\sim5: (V_{606}-i_{775}>1.2)\wedge (z_{850}-H_{160}<1.3) \wedge \\
(V_{606}-i_{775} > 0.8(z_{850}-H_{160})+1.2)
\end{eqnarray*}
\begin{eqnarray*}
z\sim6: (i_{775}-z_{850}>1.0)\wedge (Y_{105}-H_{160}<1.0) \wedge \\
(i_{775}-z_{850} > 0.777(Y_{105}-H_{160})+1.0) 
\end{eqnarray*}
\begin{eqnarray*}
z\sim7: (z_{850}-Y_{105}>0.7)\wedge (J_{125}-H_{160}<0.45) \wedge \\
(z_{850}-Y_{105} > 0.8(J_{125}-H_{160})+0.7) 
\end{eqnarray*}
\begin{eqnarray*}
z\sim8: (Y_{105}-J_{125}>0.45)\wedge (J_{125}-H_{160}<0.5) \wedge \\
(Y_{105}-J_{125} > 0.75(J_{125}-H_{160})+0.525)
\end{eqnarray*}
The six galaxies in our $z=9$--10 samples are identified by applying
the following $Y_{105}$/$J_{125}$-dropout Lyman-break color criteria
to the available {\it HST} data:
\begin{eqnarray*}
z\sim9: ((Y_{105}-H_{160})+2(J_{125}-JH_{140})>1.5)\wedge\\
((Y_{105}-H_{160})+2(J_{125}-JH_{140})>\\
~~~~~~~1.5+1.4(JH_{140}-H_{160}))\wedge\\
(JH_{140}-H_{160}<0.5)\wedge(J_{125}-H_{160}<1.2)
\end{eqnarray*}
\begin{eqnarray*}
z\sim10: (J_{125}-H_{160}>1.2)\wedge ((H_{160}-[3.6]<1.4) \vee \\
(\textrm{S/N}([3.6]) < 2))
\end{eqnarray*}
Selected sources are required to be undetected ($<$2$\sigma$) in all
{\it HST}s passbands blueward of the break both individually and in a
stack.  Potential stars are excluded from our selection using the
measured SExtractor (Bertin \& Arnouts 1996) stellarity criterion.

We adopted a $UVJ$-like criterion (Williams et al.\ 2009) which allow
us to exclude passive galaxies from our $z\gtrsim1.5$ selection of
star-forming galaxies.  Specifically, we adopt the prescription given
in Pannella et al.\ (2015):
\begin{eqnarray*}
(U-V < 1.3) \wedge (V-J > 1.6) \wedge \\
(U-V < 0.88(V-J)+0.59)
\end{eqnarray*}
which is very similar to the prescription given in Williams et
al.\ (2009).  Application of this criteria to our $z\sim1.5$--10
selection results in the exclusion of just one source from our
selection.

The $z\sim2$, 3, 4, 5, 6, 7, 8, 9, and 10 selections we consider over
the ASPECS footprint include 447, 203, 395, 139, 94, 54, 24, 4, and 2
distant sources, respectively (Table~\ref{tab:sample}).  The expected
contamination levels in these color-selected samples by lower-redshift
galaxies (or stars) is estimated to be on the order of 3-8\% (e.g.,
Bouwens et al.\ 2015).  Sources in our selection have apparent
magnitude in the $UV$-continuum extending from 23.5 mag to 30.5 mag
(Figure~\ref{fig:sample}: \textit{left panel}).

\subsection{$UV$-continuum slopes $\beta$ and Stellar Masses for Individual Sources over ASPECS}

Based on an abundance of previous work, it is well known that the
infrared excess is correlated with the measured $UV$-continuum slope
of galaxies (e.g., Meurer et al.\ 1999) and also the stellar mass
(e.g., Whitaker et al.\ 2017).

For each of the sources over ASPECS, we derive $UV$-continuum slope
$\beta$ fitting the {\it HST} photometry in various bands probing the
$UV$-continuum to a power-law $f_{1600}(\lambda/1600\AA)^{\beta}$ to
derive a mean flux at $\sim$1600$\AA$ and also a spectral slope
$\beta$.  Flux measurements in band passes that could be impacted by
IGM absorption or rest-frame optical $\gtrsim$3500\AA$\,$light are
excluded.  The inclusion of photometric constraints on the
$UV$-continuum even to $\sim$3000$\AA$ is expected to have little
impact on the derived $\beta$ given the general power-law-like shape
of the $UV$ continuum (e.g., see Appendix A in Wilkins et al.\ 2016).
Due to the limited wavelength leverage available to derive
$UV$-continuum sources for sources at $z=8$--10, we take the
$UV$-continuum slope $\beta$ to be uniformly $-2.2$ consistent with
the results of Bouwens et al.\ (2014a).

\begin{figure*}
\epsscale{1.15}
\plotone{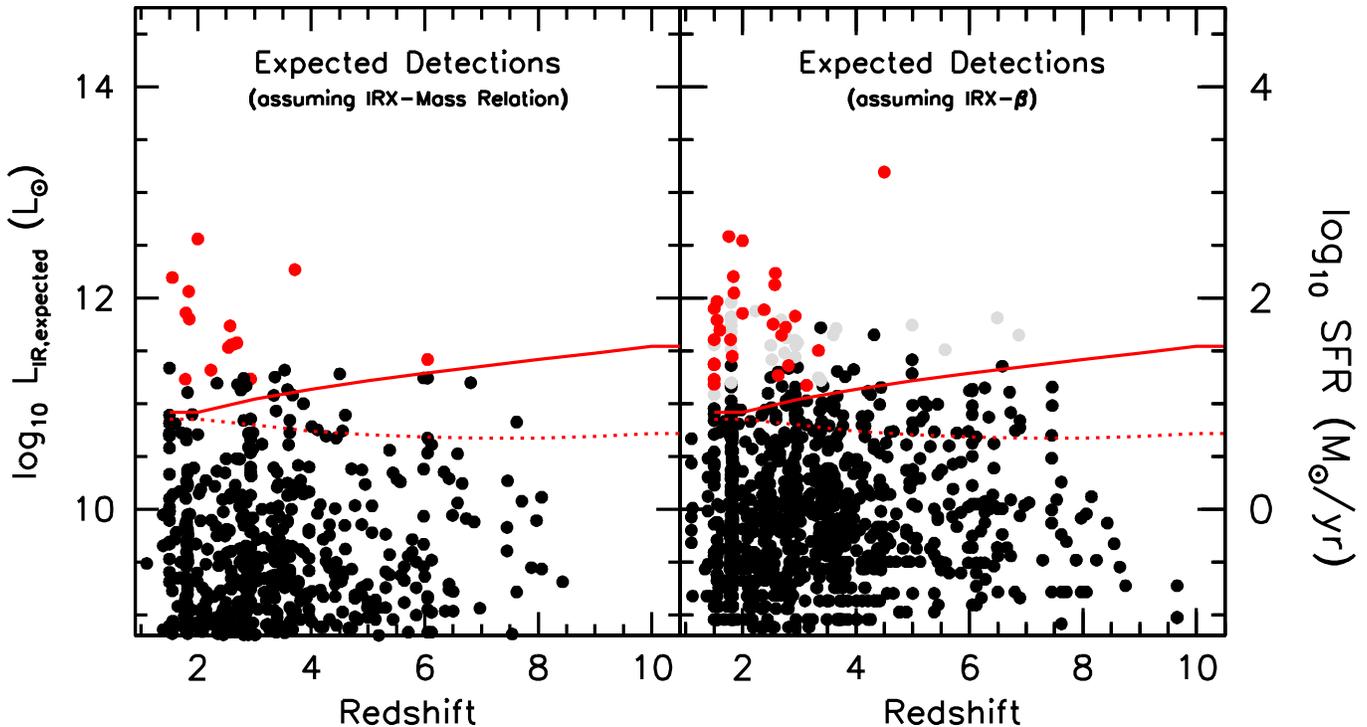}
\caption{Expected IR luminosities (per $L_{\odot}$) versus photometric
  redshift of $z=1.5$--10 galaxies (\textit{circles}) within the 4.2
  arcmin$^2$ ASPECS footprint.  Expected IR luminosities are based on
  (1) the consensus IRX-stellar mass relationship from Bouwens et
  al.\ (2016: \textit{left panel}) and (2) the consensus low-redshift
  IRX-$\beta$ relationship (\textit{right panel}: see Appendix B).
  The equivalent dust-obscured SFR using the Kennicutt (1998)
  conversion factor is shown on the right vertical axis.  The solid
  and dotted red lines indicate the $4\sigma$ limiting luminosities to
  which ASPECS can probe as a function of redshift in the deepest
  regions of our ALMA mosaic adopting the fiducial dust temperature
  evolution given in Figure~\ref{fig:tvsz} and adopting a fixed dust
  temperature of 35 K, respectively.  The solid red circles correspond
  to sources where $4\sigma$ detections are expected, while the black
  circles indicate sources where a $4\sigma$ detection is not expected
  (adopting the fiducial dust temperature evolution we assume).
  Sources predicted to show $>$$4\sigma$ detection using the
  IRX-$\beta$ relationship, but with stellar masses less than
  $10^{9.5}$ $M_{\odot}$ are shown in gray.  Black sources can appear
  above the red lines if these sources fall in regions of ASPECS where
  the sensitivities are lower than the maximum.\label{fig:probe}}
\end{figure*}

As in other work (e.g., Sawicki \& Yee 1998; Brinchmann \& Ellis 2000;
Papovich et al.\ 2001; Labb{\'e} et al.\ 2006; Gonz{\'a}lez et
al.\ 2014), we estimate stellar masses for individual sources in our
samples by modeling the observed photometry using stellar population
libraries and considering variable (or fixed) star formation
histories, metallicities, and dust content.

For $z\sim1.5$--10 sources in our catalogs, we make use of the
publicly-available code \textsc{FAST} (Kriek et al.\ 2009) to perform
this fitting.  We assume a Chabrier (2003) IMF, a metallicity of 0.2
$Z_{\odot}$, a stellar population age from 10 Myr to the age of the
universe, and allow the dust extinction in the rest-frame V to range
from zero to 2 mag, which we acknowledge may be inadequate for some
especially dust rich galaxies (e.g., Simpson et al.\ 2017).  We assume
an $e^{-t/\tau}$ star formation history and allow the $\tau$ parameter
to have any value from 1 Gyr to 100 Gyr.  Our fixing the fiducial
metallicity to 0.2 $Z_{\odot}$ is motivated by studies of the
metallicity of individual $z\sim2$--4 galaxies (Pettini et al.\ 2000)
or as predicted from cosmological hydrodynamical simulations (Finlator
et al.\ 2011; Wise et al.\ 2012).  While the current choice of
parameters can have a sizeable impact on inferred quantities like the
age of a stellar population (changing by $>$0.3-0.5 dex), these
choices typically do not have a major impact ($\gtrsim$0.2 dex) on the
inferred stellar masses.

In deriving the stellar masses for individual sources, use is made of
flux measurements from 11 {\it HST} bands ($UV_{225}$, $UV_{275}$,
$U_{336}$, $B_{435}$, $V_{606}$, $i_{775}$, $z_{850}$, $Y_{105}$,
$J_{125}$, $JH_{140}$, $H_{160}$), 1 band in the near-IR from the
ground ($K_s$), and 4 {\it Spitzer}/IRAC bands (3.6$\mu$m, 4.5$\mu$m,
5.8$\mu$m, and 8.0$\mu$m).  The {\it HST} photometry we use for
estimating stellar masses is derived applying the same procedure as
used for selecting our $z\sim1.5$--3.5 LBG samples (see \S2.2).

\begin{deluxetable*}{cccccccccc}
\tablewidth{0cm}
\tablecolumns{10}
\tabletypesize{\footnotesize}
\tablecaption{$z\gtrsim 1.5$ $UV$-selected galaxies showing 4$\sigma$ detections in our deep ALMA continuum observations\label{tab:detect}}
\tablehead{
\colhead{} & \colhead{} & \colhead{} & \colhead{} &           \colhead{} & \colhead{$\log_{10}$} & \colhead{} & \colhead{Measured} & \colhead{Inferred}\\
\colhead{} & \colhead{} & \colhead{} & \colhead{$m_{UV,0}$} & \colhead{} & \colhead{$M/$} & \colhead{} & \colhead{$f_{1.2mm}$} & \colhead{$L_{IR}$}\\
\colhead{ID\tablenotemark{a}} & \colhead{R.A.} & \colhead{DEC} & \colhead{[mag]} & \colhead{$z$} & \colhead{$M_{\odot}$} & \colhead{$\beta$} & \colhead{[$\mu$Jy]\tablenotemark{b}} & \colhead{[$10^{10}$$L_{\odot}$]} & \colhead{Ref$^{**}$}}
\startdata
XDFU-2435246390 (C06)  &  03:32:43.52  &  $-$27:46:39.0  &  27.6  &  2.696\tablenotemark{$\dagger$}  &  10.92  &  $-$0.3$\pm$0.4  &  1071$\pm$46  &  259$\pm$11 &  3\\
XDFU-2385446340 (C01)  &  03:32:38.54  &  $-$27:46:34.0  &  24.4  &  2.543\tablenotemark{$\dagger$}  &  9.90  &  $-$1.2$\pm$0.1  &  752$\pm$10  &  226$\pm$3 &  1,2,3\\
XDFU-2397246112 (C05)  &  03:32:39.72  &  $-$27:46:11.2  &  24.9  &  1.551\tablenotemark{$\dagger$}  &  11.10  &  $-$0.4$\pm$0.1  &  461$\pm$14  &  112$\pm$3 &  1,2,3\\
XDFU-2369747272 (C02)  &  03:32:36.97  &  $-$27:47:27.2  &  26.9  &  1.76\tablenotemark{*}  &  10.66  &  1.3$\pm$0.2  &  432$\pm$9  &  104$\pm$2 &  3\\
XDFU-2400547554 (C10)  &  03:32:40.05  &  $-$27:47:55.4  &  23.6  &  1.997\tablenotemark{$\dagger$}  &  10.83  &  $-$0.4$\pm$0.1  &  342$\pm$18  &  83$\pm$4 &  3\\
XDFU-2410746315 (C04)  &  03:32:41.07  &  $-$27:46:31.5  &  27.0  &  2.454\tablenotemark{$\dagger$}  &  9.39  &  $-$0.8$\pm$0.1  &  316$\pm$11  &  95$\pm$3 &  3\\
XDFU-2433446471 (C11)  &  03:32:43.34  &  $-$27:46:47.1  &  28.2  &  2.76\tablenotemark{*}  &  11.00  &  0.5$\pm$0.2  &  289$\pm$21  &  87$\pm$6 &  3\\
XDFU-2350746475 (C07)  &  03:32:35.07  &  $-$27:46:47.5  &  26.6  &  2.58\tablenotemark{$\dagger$}  &  10.89  &  0.5$\pm$0.2  &  233$\pm$11  &  56$\pm$3 &  3\\
XDFU-2416846554 (C14a)  &  03:32:41.68  &  $-$27:46:55.4  &  27.4  &  1.999\tablenotemark{$\dagger$}  &  10.47  &  0.6$\pm$0.3  &  185$\pm$10  &  45$\pm$2 &  \\
XDFB-2380246263 (C08)  &  03:32:38.02  &  $-$27:46:26.3  &  25.4  &  3.711\tablenotemark{$\ddagger$}  &  10.81  &  2.9$\pm$0.1  &  163$\pm$10  &  59$\pm$4 &  1\\
XDFB-2355547038 (C09)  &  03:32:35.55  &  $-$27:47:03.8  &  26.2  &  3.601\tablenotemark{$\dagger$}  &  9.47  &  $-$0.8$\pm$0.1  &  155$\pm$9  &  56$\pm$3 &  \\
XDFU-2387248103 (C24)  &  03:32:38.72  &  $-$27:48:10.3  &  26.0  &  2.68\tablenotemark{*}  &  9.45  &  $-$0.5$\pm$0.1  &  134$\pm$24  &  40$\pm$7 & \\
XDFU-2373546453 (C18)  &  03:32:37.35  &  $-$27:46:45.3  &  23.9  &  1.845\tablenotemark{$\ddagger$}  &  10.49  &  $-$0.7$\pm$0.1  &  107$\pm$10  &  26$\pm$2 &  1,2\\
XDFU4596 (C17)  &  03:32:38.80  &  $-$27:47:14.8  &  24.5  &  1.848\tablenotemark{$\ddagger$}  &  10.46  &  $-$0.6$\pm$0.1  &  97$\pm$9  &  23$\pm$2 &  \\
XDFU-2361746276 (C19)  &  03:32:36.17  &  $-$27:46:27.6  &  25.4  &  2.574\tablenotemark{$\dagger$}  &  10.59  &  $-$0.2$\pm$0.1  &  85$\pm$12  &  20$\pm$3 &  1\\
XDFU9838 (C26)  &  03:32:34.68  &  $-$27:46:44.5  &  25.5  &  1.552\tablenotemark{$\ddagger$}  &  10.31  &  $-$0.2$\pm$0.1  &  65$\pm$15  &  16$\pm$4 & \\
XDFU-2359847256 (C21)  &  03:32:35.98  &  $-$27:47:25.6  &  25.2  &  2.69\tablenotemark{*}  &  10.24  &  $-$1.0$\pm$0.1  &  58$\pm$10  &  18$\pm$3 & \\
XDFU-2370746171\tablenotemark{c} (C31)  &  03:32:37.07  &  $-$27:46:17.1  &  23.7  &  2.227\tablenotemark{$\ddagger$}  &  9.49  &  $-$1.3$\pm$0.1  &  47$\pm$11  &  14$\pm$3 & 2
\enddata
\tablenotetext{$^{**}$}{References previously reporting continuum detections of the identified sources: [1] Aravena et al.\ 2016, [2] Bouwens et al.\ 2016, [3] Dunlop et al.\ 2017}
\tablenotetext{$^*$}{Photometric Redshift}
\tablenotetext{$\dagger$}{Spectroscopic redshift from the detection of a CO line in the ASPECS ALMA data (Boogaard et al.\ 2019).}
\tablenotetext{$\ddagger$}{Spectroscopic redshift available for this source from the MUSE GTO observations over the HUDF (Bacon et al.\ 2017).}
\tablenotetext{$^a$}{The source IDs included inside the parentheses are as in Gonz{\'a}lez-L{\'o}pez et al.\ (2020) and Aravena et al.\ (2020).}
\tablenotetext{$^b$}{Measurements as in Gonz{\'a}lez-L{\'o}pez et al.\ (2020).}
\tablenotetext{$^c$}{This source was previously reported as a tentative $2.3\sigma$ detection in Bouwens et al.\ (2016).}
\end{deluxetable*}

\begin{figure*}
\epsscale{0.98} \plotone{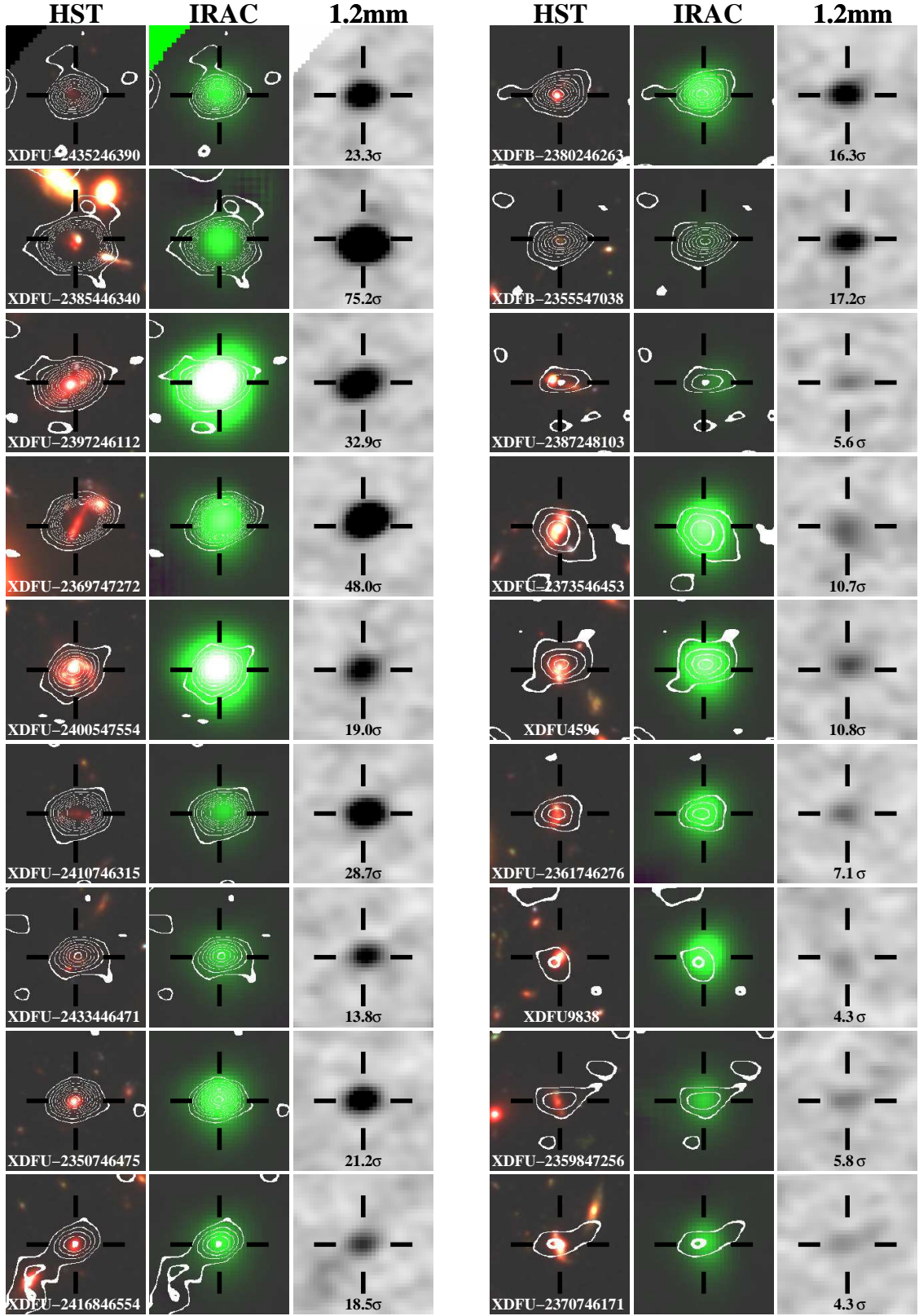}
\caption{{\it HST} composite $B_{435}i_{775}H_{160}$ (\textit{left}),
  IRAC 3.6$\mu$m (\textit{middle}), and 1.2$\,$mm ALMA-continuum
  images (\textit{right}) for 18 $z\sim1.5$--3.7 galaxies that we
  detect at 4$\sigma$ in our 4.2 arcmin$^2$ ASPECS program.  The size
  of the stamps is 7.2''$\times$7.2''.  The position of our
  1.2$\,$mm-continuum detections relative to the position of sources
  in our {\it HST} or {\it Spitzer}/IRAC images are illustrated in the
  left and center stamps with the $2\sigma$, $4\sigma$, $6\sigma$,
  $8\sigma$, $10\sigma$, ..., $20\sigma$ contours (\textit{white
    lines}).  Light from neighboring sources on the IRAC images have
  been removed for clarity.\label{fig:stamps}}
\end{figure*}

\begin{figure*}
\epsscale{1.05}
\plotone{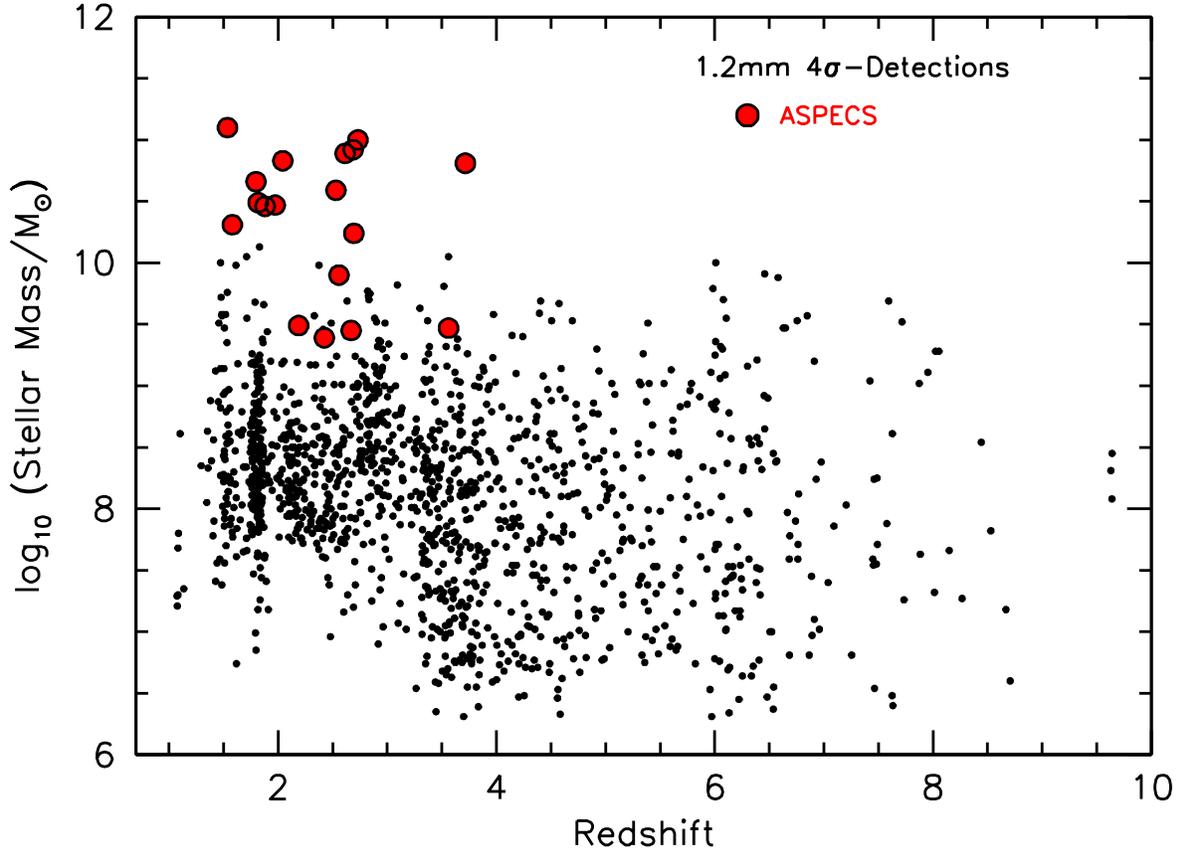}
\caption{Inferred stellar mass versus redshift for galaxies identified
  over the $\sim$4.2 arcmin$^2$ region in the HUDF with the deepest
  WFC3/IR imaging observations from the HUDF09 and HUDF12 programs
  (Bouwens et al.\ 2011; Ellis et al.\ 2013; Illingworth et
  al.\ 2013).  Large filled red circles indicate those sources which
  are detected at 4$\sigma$, while the small black circles indicate
  those sources from the $\sim$4.2 arcmin$^2$ ASPECS footprint that
  are not detected at 1.2$\,$mm in the ASPECS observations.  This
  figure is similar in design to Figure 6 from both Bouwens et
  al.\ (2016) and Dunlop et al.\ (2017) and leads to a similar
  conclusion.  It is clear that stellar mass is a particularly useful
  predictor of IR luminosity over a wide range in
  redshift.\label{fig:massz}}
\end{figure*}

\begin{figure}
\epsscale{1.15}
\plotone{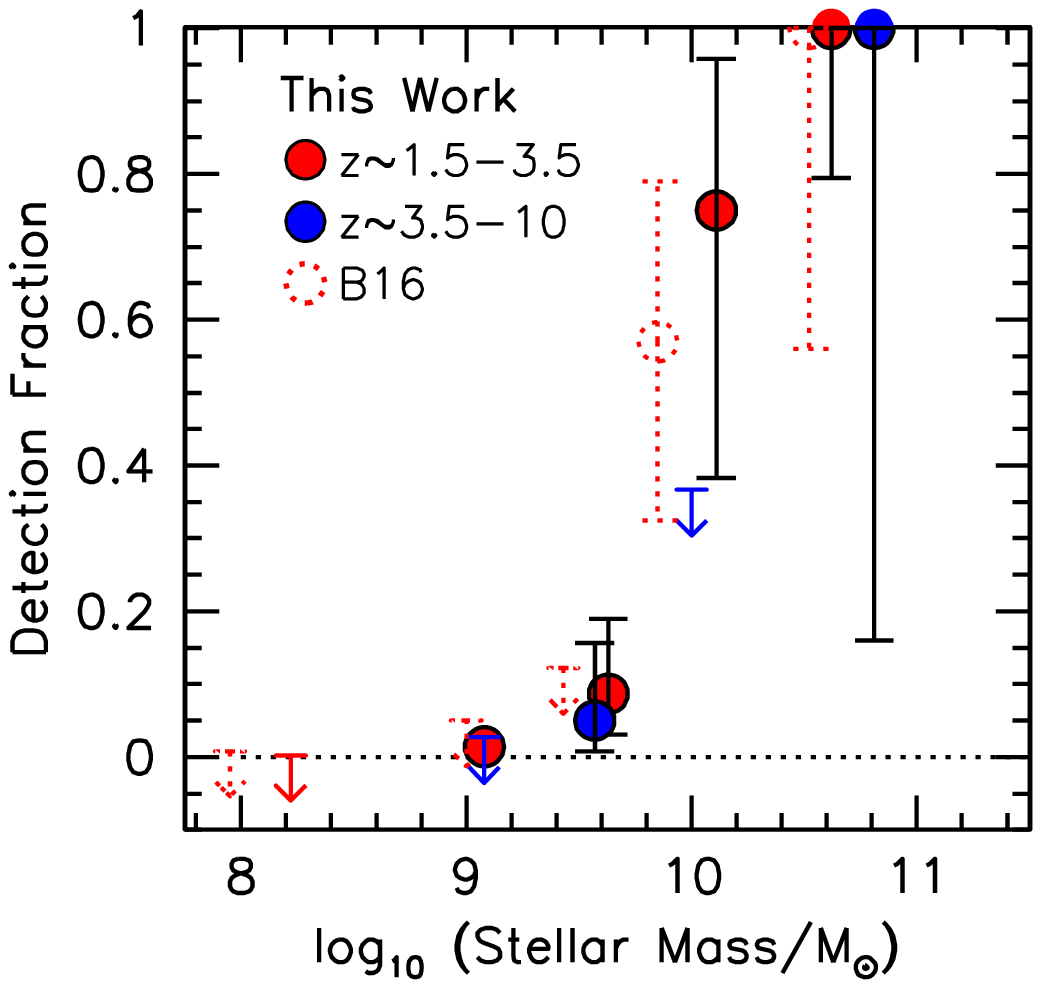}
\caption{Fraction of $z=1.5$--3.5 and $z=3.5$--10 galaxies that are
  detected at $4\sigma$ in our ALMA $1.2\,$mm continuum observations
  versus the inferred stellar mass (\textit{solid red circle} and
  \textit{solid blue circles}, respectively).  Errors and upper limits
  are $1\sigma$.  Only the 939 $z=1.5$--10 galaxies where our
  $1\sigma$ continuum sensitivity is highest
  ($<$20$\mu$Jy$\,$beam$^{-1}$) are included in this determination.
  The dotted open red circles show the results from our ASPECS pilot
  study (Bouwens et al.\ 2016).  Stellar mass appears to be a good
  predictor of dust emission in $z=1.5$--10 galaxies, with 11 of the
  13 $>10^{10}$ $M_{\odot}$ galaxies detected at
  4$\sigma$.\label{fig:detsm}}
\end{figure}

A modest correction is made to the Spitzer/IRAC 3.6$\mu$m and
4.5$\mu$m photometry to account for the impact of nebular emission
lines on the observed IRAC fluxes.  Specifically, the 3.6$\mu$m and
4.5$\mu$m band fluxes of galaxies in the redshift ranges $z=3.8$--5.0
and $z=5.1$--6.6, respectively, are reduced by 0.32 mag and 0.35 mag,
respectively, to remove the contribution of the H$\alpha$+[NII]
emission lines to the broadband fluxes.  A 0.32 mag and 0.35 mag
correction is appropriate for a rest-frame equivalent width of
$\sim$500\AA$\,$ and $\sim$540\AA$\,$, respectively, for the
H$\alpha$+[NII] emission lines, consistent with most determinations of
the H$\alpha$+[NII] emission line EW over the range $z=3.8$-5.4 (Stark
et al.\ 2013; M{\'a}rmol-Queralt{\'o} et al.\ 2016; Faisst et
al.\ 2016; Smit et al.\ 2016; Rasappu et al.\ 2016).  For galaxies in
the redshift ranges, $z=5.4$--7.0 and $z=7.0$--9.1, the measured
fluxes in the 3.6$\mu$m and 4.5$\mu$m bands are reduced by 0.5 mag.  A
0.5 mag correction is appropriate for a rest-frame equivalent width of
$\sim$680\AA$\,$ for the H$\alpha$+[NII] emission lines, consistent
with most determinations of the H$\alpha$+[NII] emission line EW over
the range $z=3.8$-5.4 (Labbe et al.\ 2013; Smit et al.\ 2014, 2015;
Faisst et al.\ 2016; Endsley et al.\ 2020).  The fiducial stellar mass
estimates we derive using \textsc{FAST} are typically $\sim$0.1 dex
lower than using other stellar population codes like \textsc{MAGPHYS}
and \textsc{Prospector} (see Appendix A).

The middle panel of Figure~\ref{fig:sample} illustrates the effective
range in stellar mass probed by our $z=1.5$--10 sample.  Most sources
from our HUDF $z=1.5$--10 sample have stellar masses in the range
$10^{7.5}$ $M_{\odot}$ to $10^{9.5}$ $M_{\odot}$.  The most massive
sources probed by our program extend to $10^{11.5}$ $M_{\odot}$.
Beyond the stellar mass itself, Figure~\ref{fig:sample} also
illustrates the range in $UV$-continuum slope $\beta$ probed by our
samples (see \S3.1 for details on how $\beta$ is derived).  Since the
measured $\beta$ has been demonstrated to be quite effective in
estimating the infrared excess for lower-redshift $UV$-selected
samples (e.g., M99; Reddy et al.\ 2006; Daddi et al.\ 2007), it is
useful for us to probe a broad range in $\beta$.  As can be seen from
Figure~\ref{fig:sample}, our samples probe the range $\beta\sim-1.5$
to $\sim-2.5$ quite effectively.

\section{Results} 

In this section, we quantify the infrared excess (IRX) of star-forming
galaxies in the intermediate to high-redshift universe $z>1.5$.  As in
previous work (e.g., Meurer et al.\ 1999; {\'A}lvarez-M{\'a}rquez et
al.\ 2016; Whitaker et al.\ 2017) we define the infrared excess (IRX)
to be
\begin{equation}
IRX = \frac{L_{IR}}{L_{UV}}
\end{equation}
where $L_{IR}$ is the infrared luminosity of galaxies (including all
rest-frame emission from 8$\mu$m to 1000$\mu$m) and $L_{UV}$ is the
$UV$ luminosity of galaxies, which we take to be $\nu f_{\nu}$.  $\nu$
is evaluated at $c / \lambda_{1600\AA}$ in computing the $UV$
luminosities $L_{UV}$ of sources.

\subsection{Expected Number of Continuum Detections from $z\sim1.5$--10 Galaxies within ASPECS}

Thanks to the limited evolution seen in the IRX vs. stellar mass and
IRX vs. $\beta$ results over the entire redshift range $z\sim3$ to
$z\sim0$ (Reddy et al.\ 2006; Whitaker et al.\ 2017; Fudamoto et
al.\ 2020a), we might expect these relations to be at least
approximately valid to even higher redshifts.

Before looking in detail at which sources show continuum detections
and what their properties are, let us briefly calculate how many
sources we would expect to detect based on published IRX vs. stellar
mass and IRX vs. $UV$-continuum slope $\beta$ relations.  Given the
limited evolution in these relations, we expect the predicted results
to be reasonably accurate in estimating the overall numbers from our
program.  For our baseline IRX - stellar mass $M$ relation, we take
the relation derived in our pilot program (Appendix A from Bouwens et
al.\ 2016):
\begin{equation}
\log_{10} IRX_{M,0} = \log_{10} M - 9.17
\label{eq:m0}
\end{equation}
For our baseline IRX - $\beta$ relation, we make use of the consensus
low-redshift relation derived in Appendix B based on the following
three studies (Overzier et al.\ 2011; Takeuchi et al.\ 2012; Casey et
al.\ 2014).  The relation we derive is the following:
\begin{equation}
IRX_{z=0} = 1.7 (10^{0.4(1.86(\beta+1.85))} - 1)
\label{eq:z0}
\end{equation}
The infrared excess implied by the above relation are
$\approx$0.5$\times$ that of the Meurer et al.\ (1999) relation.  An
equivalent expression for a Reddy (similar to Calzetti et al.\ 2020)
and SMC-like dust law are the following:
\begin{equation}
IRX_{Reddy} = 1.7 (10^{0.4(1.84(\beta+1.85))} - 1)
\label{eq:reddy}
\end{equation}
and
\begin{equation}
IRX_{SMC} = 1.7 (10^{0.4(1.1(\beta+1.85))} - 1)
\label{eq:smc}
\end{equation}

Based on the above relations and observed $UV$ fluxes, we can compute
the equivalent flux at an observed wavelength of 1.26$\,$mm adopting a
modified blackbody form with a dust emissivity power-law spectral
index of $\beta_d = 1.6$ and dust temperature given by
Eq.~\ref{eq:tvsz}.  To account for the impact of the CMB at
$z\sim1.5$--10 on the expected flux densities we would measure, we
multiply the predicted flux (before consideration of CMB effects) by
$C_{\nu}$
\begin{equation}
C_{\nu} = \left[ 1 - \frac{B_{\nu} (T_{CMB} (z))}{B_{\nu} (T_d(z))} \right]
\end{equation}
following prescriptions given in da Cunha et al.\ (2013).

Using the above procedure, we calculated the expected flux for our
entire sample of 1362 $z=1.5$--10 galaxies identified over the 4.2
arcmin$^2$ ASPECS footprint alternatively making use of the consensus
IRX-stellar mass relation from Bouwens et al.\ (2016), our consensus
low-redshift IRX-$\beta$ relation, and also a SMC-like IRX-$\beta$
relation (Eqs.~\ref{eq:m0}-\ref{eq:smc}).  15, 28, and 8 sources,
respectively, are predicted to show $>$4$\sigma$ detections in the
ASPECS observations in the 1.2-mm continuum.  Assuming a fixed dust
temperature of 35 K, the predicted numbers would be 27, 42, and 11,
respectively.  Figure~\ref{fig:probe} shows the predicted IR
luminosities vs. redshift using either the aforementioned IRX-stellar
mass relation (\textit{left}) or the IRX-$\beta$ relation
(\textit{right}) for our fiducial dust temperature model.  The solid
red and dotted lines show the $4\sigma$ IR luminosity limit we probe
with the ASPECS data set adopting the fiducial dust temperature model
given in Eq.~\ref{eq:tvsz} (\textit{solid red line}) and assuming the
dust temperature remains fixed at 35 K for all of cosmic time
(\textit{dotted red line}).

\begin{figure}
\epsscale{1.15} \plotone{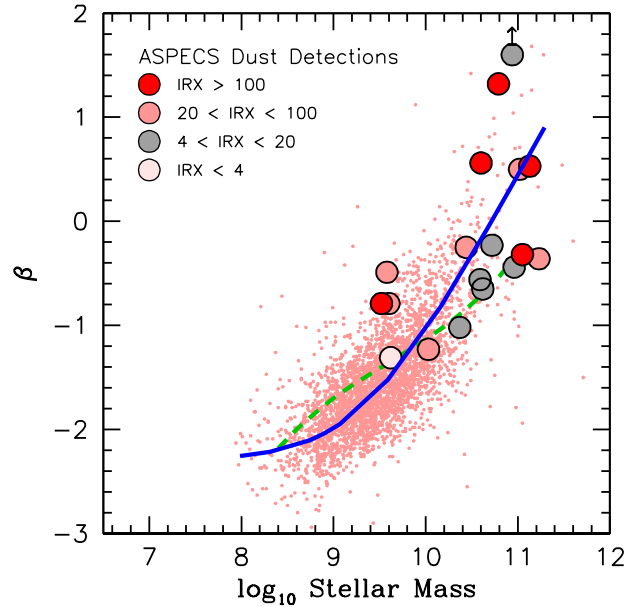}
\caption{$UV$-continuum slopes and stellar masses of detected galaxies
  in our ASPECS samples (\textit{solid circles}) shown relative to the
  slopes and stellar masses of $z\sim1.3$--2.5 galaxies from CANDELS
  shown for comparison.  The color of the solid circles indicates the
  IRX value derived for the corresponding galaxy.  The estimated
  stellar masses for sources from CANDELS are based on the new
  prospector catalogs (Leja et al.\ 2019).  A $+$0.12-dex correction
  has been applied to our FAST-inferred stellar mass estimates to make
  them consistent with \textsc{Prospector}-inferred estimates
  (Appendix A).  A black arrow has been included next to the circle
  representing the ASPECS source (XDFB-2380246263) which has a
  $UV$-continuum slope redder than our plotted boundaries.  The
  $UV$-continuum slope measurements for the CANDELS sources are based
  on fits to the measured rest-$UV$ fluxes (using the $B_{435}V_{606}$
  and $B_{435}V_{606}i_{775}$ bands for sources at $z=1.3$--1.9 and
  $z=1.9$--2.5, respectively) from the Skelton et al.\ (2014) 3D-HST
  catalogs.  The blue line shows the $\beta$ vs. stellar mass
  correlation we derive using the observed IRX-$\beta$ and IRX-stellar
  mass relations (\S3.4).  The stellar mass vs. $\beta$ relation
  derived by McLure et al.\ (2018) from a selection of $z=2$--3
  galaxies is given by the dashed green line.\label{fig:massbeta2}}
\end{figure}

\subsection{Continuum Detections of Individual Sources at 1.2$\,$mm}

Examination of the 1362 $z=1.5$--10 galaxies over our sensitive ASPECS
mosaic shows that 18 of these galaxies are detected at $>$4.0$\sigma$
in the 1.2$\,$mm-continuum images.  We use the flux densities and
uncertainties that Gonz{\'a}lez-L{\'o}pez et al.\ (2020) derive for
each source from the 1.2$\,$mm-continuum images.
Gonz{\'a}lez-L{\'o}pez et al.\ (2020) make use of flux density
measurements made from the tapered images, allowing for a more
complete account of the total dust-continuum flux density in sources,
many of which are spatially extended.  The coordinates and source
properties of the continuum detected sources are provided in
Table~\ref{tab:detect}.  1.2$\,$mm-continuum images of the
4$\sigma$-detected sources are presented in Figure~\ref{fig:stamps}
and shown with respect to the {\it HST} and {\it Spitzer}/IRAC images.

\begin{figure*}
\epsscale{1.15} 
\plotone{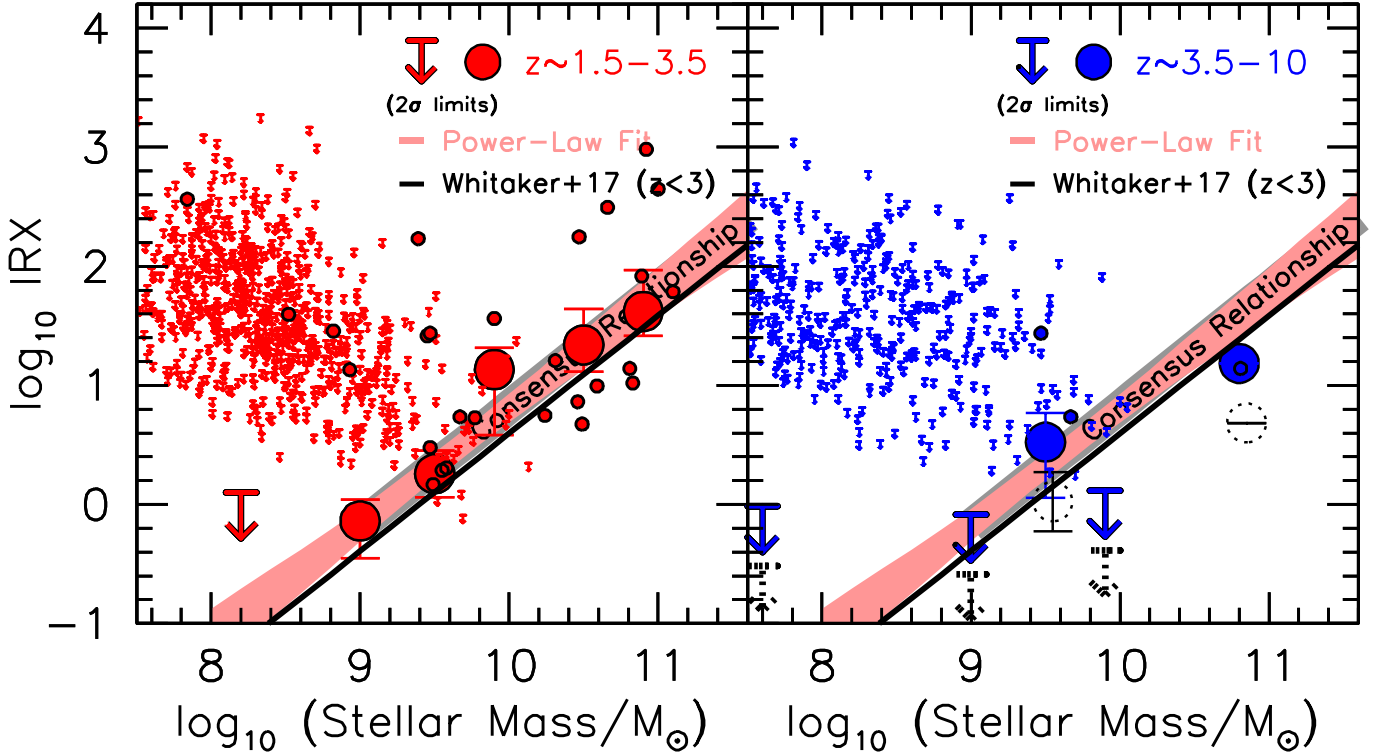}
\caption{Constraints on the infrared excess of $z=1.5$--3.5
  (\textit{left panel}) and $z=3.5$--10 (\textit{right panel})
  galaxies (\textit{large red and blue circles and downward arrows,
    respectively}) obtained by stacking the ALMA 1.2$\,$mm
  observations available for many individual sources over our 4.2
  arcmin$^2$ ASPECS footprint.  The small filled circles and downward
  arrows are for sources with a positive $3\sigma$ measurement of IRX
  and $3\sigma$ upper limit on IRX, respectively.  Upper limits and
  errorbars are $2\sigma$ and $1\sigma$, respectively for the stacked
  points.  The thick-shaded grey line shows the consensus dependence
  of IRX on galaxy stellar mass that had previously been derived for
  $z\sim2$--3 galaxies from the literature (Reddy et al.\ 2010;
  Whitaker et al.\ 2014; {\'A}lvarez-M{\'a}rquez et al.\ 2016) in
  Bouwens et al.\ (2016).  The light-red-shaded region included in the
  left panel shows the best-fit power-law relation we derive based on
  our ASPECS IRX mesurements at $z=1.5$--3.5; it is also included in
  the right panel to facilitate comparisons with the $z=3.5$--10
  results.  The black line shows the IRX vs. stellar mass relation
  found by Whitaker et al.\ (2017) to hold from $z\sim0$ to $z\sim3$.
  The fiducial results presented here from ASPECS are derived assuming
  that the dust temperature evolves as in Eq.~\ref{eq:tvsz}, but the
  dotted black circle and upper limits in the right panel show the
  impact of assuming no evolution in the dust temperature to $z>3$
  (i.e., fixing $T_d$ at 35 K).  Our ALMA stack results suggest that
  only galaxies with stellar masses in excess of $10^{9.0}$
  $M_{\odot}$ tend to output $>$50\% of their energy at far-infrared
  wavelengths.\label{fig:irxsm}}
\end{figure*}

\begin{figure*}
\epsscale{1.15}
\plotone{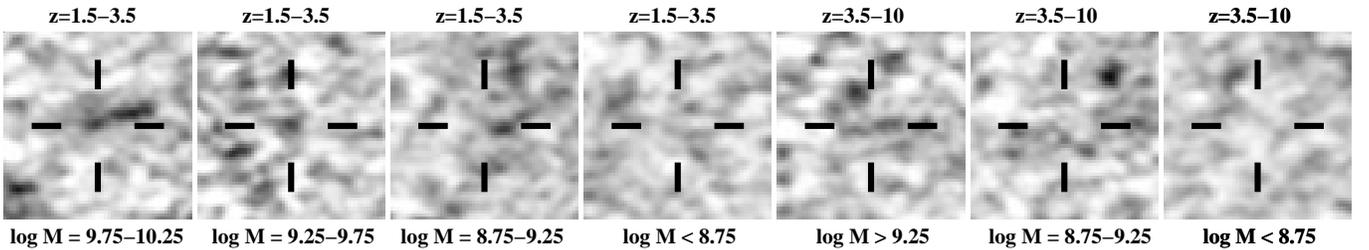}
\caption{Stacked 1.2$\,$mm-continuum images (12''$\times$12'') for all
  candidate $z=1.5$--3.5 galaxies falling in five different ranges of
  stellar mass ($>10^{10.25}$ $M_{\odot}$, 10$^{9.75}$ to $10^{10.25}$
  $M_{\odot}$, 10$^{9.25}$ to $10^{9.75}$ $M_{\odot}$, $10^{8.75}$ to
  $10^{9.25}$ $M_{\odot}$, and $<10^{8.75}$ $M_{\odot}$) and three
  different ranges of stellar mass at $z=3.5$--10 ($>10^{9.25}$
  $M_{\odot}$, $10^{8.75}$ to $10^{9.25}$ $M_{\odot}$, and
  $<10^{8.75}$ $M_{\odot}$).  In the stacks, sources are weighted
  according to the inverse square of the noise.  Note that the 18
  individually-detected sources from this analysis are not included in
  the presented stack results.\label{fig:massstack}}
\end{figure*}

The IR luminosities we estimated based on our far-IR SEDs and fiducial
dust temperature evolution (Eq.~\ref{eq:tvsz}) are presented in
Table~\ref{tab:detect} and range from 1.4 $\times$10$^{11}$
$L_{\odot}$ to 2.6 $\times$ 10$^{12}$ $L_{\odot}$.  Aravena et
al.\ (2020), in a separate analysis of these same sources using SED
fits from \textsc{MAGPHYS}, find the range to be 1.1 $\times$10$^{11}$
$L_{\odot}$ to 3.4 $\times$ 10$^{12}$ $L_{\odot}$.  Our derived IR
luminosities are just 0.01 dex higher in the mean than those employed
by Aravena et al.\ (2020), demonstrating that the modified blackbody
form we utilize here produce IR luminosities very similar to SED
analyses that include a mid-IR power-law.

The total number of $>$4$\sigma$ detections in the $z=1.5$--10 galaxies
found over the ASPECS footprint is 18.  In \S3.1, we had predicted
that 15, 28, and 8 sources would be found from this selection using the
consensus IRX-stellar mass relationship, the consensus low-redshift
IRX-$\beta$ relationship, and a SMC-like IRX-$\beta$ relationship.  If
in our use of the IRX-$\beta$ relationship, we only consider those
sources with stellar masses greater than $10^{9.5}$ $M_{\odot}$, the
predicted number of $4\sigma$ detections decreases to 16, almost
identical to the observed number.  As discussed in Bouwens et
al.\ (2016: \S3.1.1) and McLure et al.\ (2018), the impact of scatter
on the breadth of the $UV$-continuum slope $\beta$ distribution is to
increase the fraction of sources with redder $UV$-continuum slopes
$\beta$, increasing the predicted number of sources expected to be
detected in the dust continuum.

As in most previous work (Pannella et al.\ 2009; Bouwens et al.\ 2016;
Dunlop et al.\ 2017), detected sources from our selection tend to be
the star-forming galaxies with the highest stellar masses.  In
Figure~\ref{fig:massz} we present the stellar masses and redshifts
inferred for the 1362 $z=1.5$--10 galaxies over our ASPECS field,
indicating which sources are detected in ASPECS.  All 11
$z\sim1.5$--3.5 sources with high stellar masses ($>$10$^{10.0}$
$M_{\odot}$) and sensitive ALMA observations from ASPECS
($<$20$\mu$Jy$\,$beam$^{-1}$) are detected in our combined data set.
If we repeat this exercise on sources in our $z=1.5$--10 samples, 11
of 13 are detected, implying a 85$_{-18}^{+7}$\% detection fraction at
$>$10$^{10}$ $M_{\odot}$.

In Figure~\ref{fig:detsm}, we present the fraction of sources detected
at $>$$4\sigma$ as a function of stellar mass.  In computing this
fraction, we only consider those sources (939 out of 1362) over the
ASPECS field where the $1.2\,$mm-continuum sensitivities are the
highest, i.e., with $1\sigma$ RMS noise $<$20$\mu$Jy$\,$beam$^{-1}$.
As in previous work (e.g., Bouwens et al.\ 2016; Dunlop et al.\ 2017),
it is clear that stellar mass is a useful predictor of the
dust-continuum flux from star-forming galaxies.

Figure~\ref{fig:massbeta2} shows the continuum detections in our
sample relative to the stellar mass--$\beta$ trend found for galaxies
in CANDELS (see \S\ref{sec:massbeta}).  All $4\sigma$ detected sources
from ASPECS have $UV$-continuum slope $\beta$ of $-$1.3 or redder and
a stellar mass of $\gtrsim$$10^{9.4}$ $M_{\odot}$.  Detected sources
with the largest infrared excesses (\textit{red circles}) are
distributed towards the reddest $UV$ slopes and highest stellar
masses, as expected, but with a significant amount of scatter.

\begin{figure}
\epsscale{1.15}
\plotone{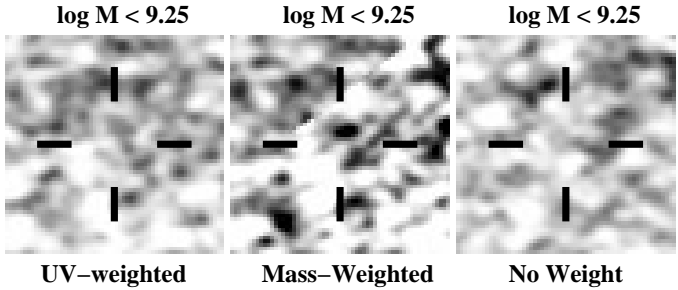}
\caption{1.2$\,$mm-continuum stack (12''$\times$12'') of 1253 candidate
  $z=1.5$--10 galaxies found with the ASPECS footprint with stellar
  masses less than $10^{9.25}$ $M_{\odot}$ (192 of these have stellar
  masses in the range $10^{8.75}$ $M_{\odot}$ to $10^{9.25}$
  $M_{\odot}$).  Left, center, and right panels show our stack results
  weighting the sources by their $UV$ flux, weighting sources by their
  stellar mass, and weighting sources equally, respectively.  Our deep
  stack results imply that the mean continuum flux for candidate
  $z=1.5$--10 galaxies with stellar masses less than 10$^{9.25}$
  $M_{\odot}$ is $-$0.1$\pm$0.4$\mu$Jy$\,$beam$^{-1}$.  This implies an average
  obscured SFRs for these sources of 0.0$\pm$0.1
  $M_{\odot}$$\,$yr$^{-1}$.\label{fig:deepstack}}
\end{figure}

\begin{figure*}
\epsscale{1.15}
\plotone{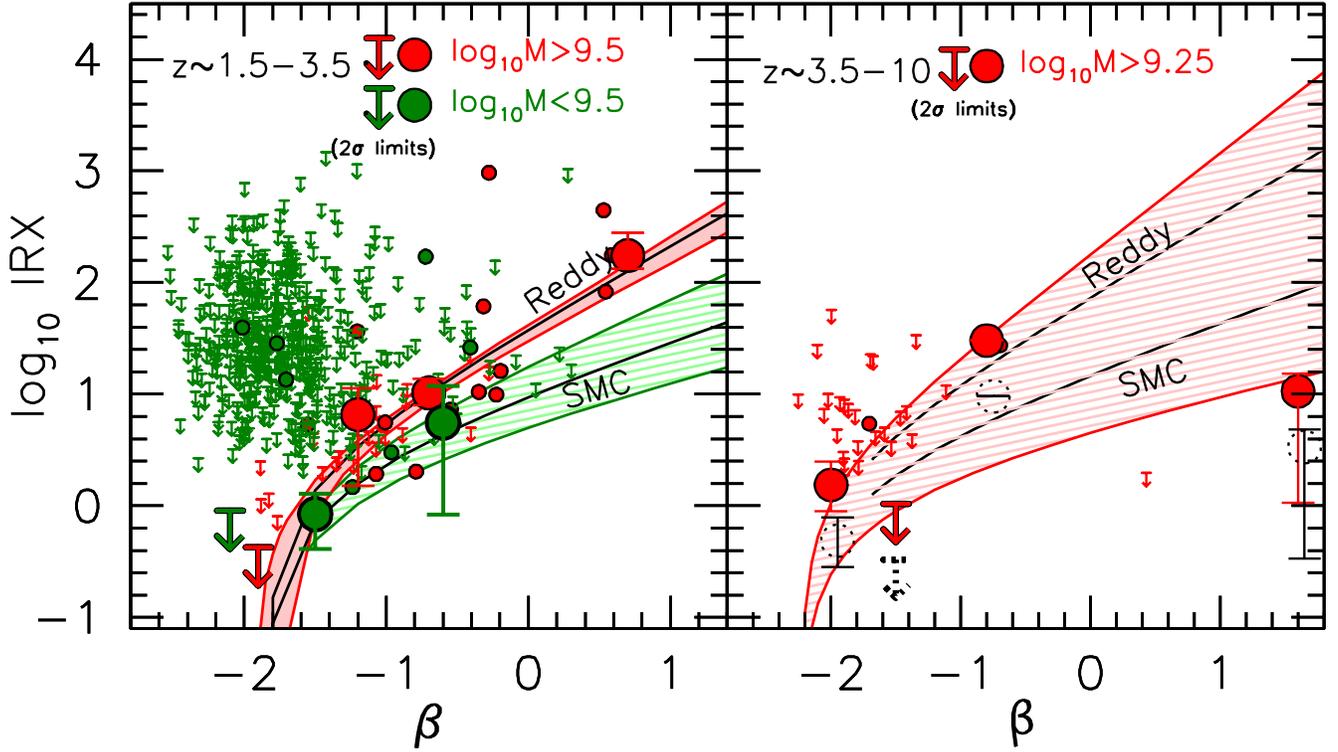}
\caption{(\textit{left panel}) Stacked constraints on the infrared
  excess in $z=1.5$--3.5 galaxies versus the $UV$-continuum slope
  $\beta$.  These results are shown for higher- and lower-mass
  subsamples ($>10^{9.5}$ $M_{\odot}$ and $<10^{9.5}$ $M_{\odot}$) of
  $z=1.5$--3.5 galaxies (\textit{red and green solid circles and
    downward arrows, respectively}) and are obtained by stacking the
  ALMA 1.2$\,$mm observations of individual sources over the ASPECS
  region.  Upper limits and errorbars on the stack results are
  $2\sigma$ and $1\sigma$, respectively.  The smaller solid circles
  and downward arrows indicate $>$3$\sigma$ measurements and 3$\sigma$
  upper limits for individual sources.  The black lines show the
  nominal IRX-$\beta$ relation for the Reddy (slightly steeper than
  Calzetti) and SMC dust laws (Eqs.~\ref{eq:reddy} and \ref{eq:smc}).
  The shaded red and light green regions indicate the 68\% confidence
  intervals on the IRX-$\beta$ relationship for sources with stellar
  masses of $>10^{9.5}$ $M_{\odot}$ and $<10^{9.5}$ $M_{\odot}$,
  respectively.  Our results are consistent with the IR emission from
  high-mass ($>10^{9.5}$ $M_{\odot}$) $z\sim1.5$--3.5 galaxies
  exhibiting a Calzetti-like IRX-$\beta$ relation.  The
    IRX-$\beta$ relation for lower-mass ($<10^{9.5}$ $M_{\odot}$)
    galaxies is more consistent with a SMC-like dust relation.
  (\textit{right panel}) Stacked constrants on the infrared excess in
  $z=3.5$--10 galaxies (for galaxies with $>$10$^{9.25}$ $M_{\odot}$
  in stellar mass) versus $\beta$.  The shaded red regions
    indicate the allowed range of IRX-$\beta$ relations alternatively
    fitting to the stacked detection at $\sim$$-$0.8 and $\sim$1.6.
    Our $z=3.5$--10 results are consistent with both a Reddy/Calzetti
    and SMC relation, but with much larger uncertainties.  While the
  fiducial results presented here from ASPECS assume an evolving dust
  temperature (Eq.~\ref{eq:tvsz}), the dotted black open circle and
  upper limits show the results if the dust temperature is assumed to
  have a similar temperature at $z>3$, i.e., $\sim$35 K, as is the
  case at $z<3$.\label{fig:irxbeta}}
\end{figure*}

\subsection{Stacked constraints on the Infrared Excess}

Fainter, lower mass sources in our selections are not sufficiently
bright in the dust continuum to be individually detected.  It is
therefore useful to stack the continuum observations from ASPECS to
derive constraints on their dust continuum properties.  We consider
various subdivisions of our samples in terms of the physical
properties.

For sources included in the stack, the ALMA continuum maps of the
relevant sources are mapped onto the same position and stacked in the
image plane, weighting each in proportion to the expected 1.2$\,$mm
continuum signal divided by the noise squared (per beam).  We derive a
flux density from the stack based on a convolution of the image stack
(3.3$''$$\times$3.3$''$ aperture) with the primary beam.  Individually
undetected sources are assumed to be unresolved at the resolution of
our observations.

\begin{figure*}
\epsscale{1.1}
\plotone{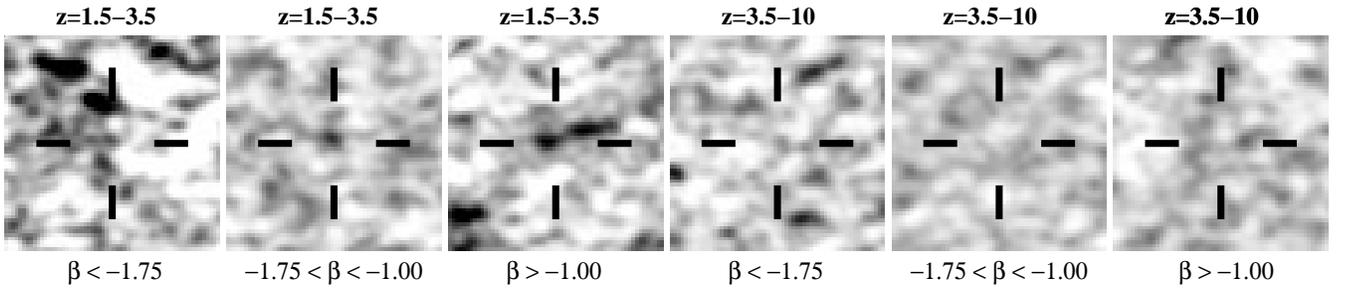}
\caption{Stacked 1.2$\,$mm-continuum images (12''$\times$12'') for
  $z=1.5$--3.5 and $z=3.5$--10 galaxies falling in different bins of
  $UV$-continuum slope $\beta$.  All sources that are individually
  detected at $\geq4\sigma$ are not included in the presented stack
  results.  Only the most massive ($>10^{9.5}$ $M_{\odot}$ and
  $>10^{9.25}$ $M_{\odot}$) sources are included in our $z= 1.5$--3.5
  and $z= 3.5$--10 stacks, respectively.  In the stacks, sources are
  weighted according to the inverse square of the
  noise.  \label{fig:betastack}}
\end{figure*}

\subsubsection{IRX vs. Stellar Mass\label{sec:irxmass}}

We first look at the average infrared excess of $z=1.5$--10 galaxies as
a function of stellar mass.  We consider six different bins of stellar
mass: $>$10$^{10.75}$ $M_{\odot}$, $10^{10.25}$ - $10^{10.75}$
$M_{\odot}$, $10^{9.75}$ - $10^{10.25}$ $M_{\odot}$,
10$^{9.25}$-10$^{9.75}$ $M_{\odot}$, 10$^{8.75}$-10$^{9.25}$
$M_{\odot}$, and $<$10$^{8.75}$ $M_{\odot}$.  For these stacks, we
weight sources according to the inverse square of the noise [in
  $\mu$Jy], i.e., $\sigma(f_{1.2mm})^{-2}$.

Our stack results are presented in Figure~\ref{fig:irxsm} for both our
$z=1.5$--3.5 and $z=3.5$--10 samples, including both the individually
detected and undetected sources.  Galaxies in our $10^{9.75}$ -
$10^{10.25}$ $M_{\odot}$ mass bin are detected at 10$\sigma$, while
sources in the $10^{9.25}$ - $10^{9.75}$ $M_{\odot}$ bin only show a
tentative $2\sigma$ detection.  Table~\ref{tab:irx} in the main text
and Table~\ref{tab:irxsm} from Appendix C presents these results in
tabular form.  Our stack results for star-forming galaxies which are
individually undetected ($<$4$\sigma$) are presented in
Figure~\ref{fig:massstack}.

Our $z=1.5$-3.5 stack results provide us with highest S/N results to
derive a dependence of the infrared excess on stellar mass.  In
quantifying the dependence, we made use of the power law relation
\begin{equation}
IRX_M = (M/M_s)^{\alpha}\label{eq:irxmass}
\end{equation}
where $M_s$ is the characteristic stellar mass for significant IR
emission ($L_{IR}$ = $L_{UV}$) and $\alpha$ gives the power by which
the infrared excess depends on mass.  We then fit our $z=1.5$-3.5
stacked IRX measurements to this relation and arrived at a best-fit
value for $M_s$ and $\alpha$ of $10^{9.15_{-0.16}^{+0.18}}$
$M_{\odot}$ and 0.97$_{-0.17}^{+0.17}$, respectively.  The best-fit
relation is shown in both the left and right panels of
Figure~\ref{fig:irxsm} with the light-red-shaded region.  Broadly, our
$z\sim1.5$--3.5 results are consistent with the consensus relation
that we derived in our earlier analysis based on results in the
literature (Bouwens et al.\ 2016).

At $z\sim3.5$--10, our stack results for the infrared excess show a
clear detection in the highest stellar mass bin and a tentative
$2\sigma$ detections in the third highest stellar mass bin, i.e.,
10$^{9.25}$ - 10$^{9.75}$ $M_{\odot}$, while at lower masses, there is
still no detection in our stack results.  Our new stack results for
the infrared excesses at $z=3.5$--10 seem consistent with what we
derive at lower redshift.  Previously, Pannella et al.\ (2015) had
found no strong evidence for evolution in the IRX-stellar mass
relation to $z\sim3.5$, and Whitaker et al.\ (2017) found this same
lack of evolution to $z\sim3$.  From first principles, one expect some
evolution in this relationship due to the observed evolution in the
mass-metallicity relation (e.g., Erb et al.\ 2006a); however, it is
possible that a higher gas and ISM mass in $z\gtrsim2$ galaxies
compensate for the lower metal content to produce a relatively
unevolving IRX-stellar mass relation (Tan et al.\ 2014).

However, we emphasize that this conclusion is sensitive to the dust
temperature evolution we adopt.  If there is no significant evolution
in the dust temperatures with redshift, then the infrared excesses at
$z=3.5$--10 would be lower by $\sim$0.4 dex than what we infer
$z=1.5$--3.5, and we would therefore infer that the IRX-stellar mass
relation increases at early cosmic times.  In Appendix D, we
investigated the extent to which our IRX vs. stellar mass relation
showed a dependence on the stellar population code used to estimate
the mass for individual sources and recovered a steeper IRX-stellar
mass relation using \textsc{Prospector} masses.

For stacks of sources with stellar masses less than $10^{9.25}$
$M_{\odot}$, we do not find a detection in the IR continuum.  In an
effort to provide a dramatic illustration of this, we include in
Figure~\ref{fig:deepstack} three different stacks of all 1253
$z=1.5$--10 sources with stellar mass estimates $<$10$^{9.25}$
$M_{\odot}$ over our ASPECS footprint.  Our first stack weights
sources by their $UV$ flux, our second stack weights sources by their
estimated stellar mass, and our third stack weights sources equally
(\textit{left}, \textit{center}, and \textit{right panels},
respectively).  None of the stacks show a significant detection, and
in our unweighted stack, the mean continuum flux density is
$-$0.1$\pm$0.4$\pm$0.4$\mu$Jy$\,$beam$^{-1}$.  Even weighting sources
in the stack by the measured $UV$-continuum slope $\beta$ fails to
result in a significant detection.  This demonstrates, rather
dramatically, that faint, UV-selected galaxies show essentially no
dust continuum emission (see also Carvajal et al.\ 2020).  Converting
this flux density constraint to a star formation rate for a galaxy at
$z\sim4$, we derive a SFR of 0.0$\pm$0.1 $M_{\odot}$$\,$yr$^{-1}$.

\subsubsection{Infrared Excess versus $\beta$\label{sec:irxbeta}}

Stacked results of $z=1.5$--3.5 and $z=3.5$--10 sources over our ASPECS
footprint are presented as a function of $UV$-continuum slope $\beta$
in Figure~\ref{fig:irxbeta} with the large solid circles and $2\sigma$
upper limits.  Five different bins in $\beta$ are utilized to better
map out the trend with $UV$-continuum slope $\beta$.

Separate stack results are presented for sources with stellar masses
$>10^{9.5}$ $M_{\odot}$ (\textit{large red circles and downward
  arrows, respectively}) and $<10^{9.5}$ $M_{\odot}$ (\textit{large
  green circles and downward arrows, respectively}) to evaluate
whether higher-mass galaxies show a different IRX-$\beta$ relationship
from lower-mass galaxies.  This treatment also ensures that results in
the redder, high-mass bins are not impacted by the inclusion of bluer,
lower-mass sources (but where the measured $UV$-continuum slopes
$\beta$ are much redder than the actual slopes due to the impact of
noise).  Figure~\ref{fig:betastack} presents our stack results for
star-forming galaxies which are individually undetected
($<$4$\sigma$).  Our IRX-$\beta$ stack results are presented in
Table~\ref{tab:irx} in the main text and Table~\ref{tab:irxbeta} in
Appendix C.

For our highest-mass $z\sim1.5$--3.5 samples, our stack results lie
closest to the Reddy (Calzetti-like) IRX-$\beta$ relations.  As in our
earlier analysis of the ASPECS pilot data, we formalize this analysis
by finding those parameters which best match the stacked IRX results
vs. $\beta$ and then computing 68\% confidence intervals on the
derived parameters.  Here we derive constraints on both
$dA_{UV}/d\beta$ and $\beta$ as 
\begin{equation}
IRX_{\beta} = 1.7 \times 10^{0.4(dA_{UV}/d\beta)(\beta-\beta_{int})}-1,
\label{eq:irxbeta}
\end{equation}
instead of just deriving constraints on $dA_{UV}/d\beta$ as in our
previous analysis.

Our maximum-likelihood derived values for $dA_{UV}/d\beta$ and
$\beta_{int}$ are 1.81$_{-0.14}^{+0.18}$ and $-1.86_{-0.10}^{+0.14}$
and presented in Table~\ref{tab:irxbetafit}.  The $dA_{UV}/d\beta$ we
derive is similar to the Calzetti or Reddy value, i.e., 1.97 or 1.84.
Meanwhile, the $\beta_{int}=-1.86$ we derive is not only redder than
the $\beta_{int}=-2.23$ implicit in the Meurer et al.\ (1999)
formulation, but also redder than what might be expected for dust-free
galaxies with a constant star formation rate for 100-500 Myr (e.g., as
in Reddy et al.\ 2018).  Both the $dA_{UV}/d\beta$ and $\beta_{int}$
we derive are consistent with the consensus low-redshift values for
these quantities (e.g., Eq~\ref{eq:z0}).  If we instead take
$\beta_{int} = -2.23$ as has been conventional (following Meurer et
al.\ 1999), the $dA_{UV}/d\beta$ we recover is 1.48$_{-0.11}^{+0.09}$.
In our pilot study, our best-fit determination for $dA_{UV}/d\beta$ is
1.26$_{-0.36}^{+0.27}$ when taking $\beta_{int}$ equal to $-2.23$.
For a $\beta_{int}=-2.30$, we recover $dA_{UV}/d\beta$ equal to
1.42$_{-0.11}^{+0.09}$.

\begin{deluxetable}{cccc}
\tablewidth{0cm}
\tablecolumns{4}
\tabletypesize{\footnotesize}
\tablecaption{Inferred IRX vs. Galaxy Stellar Mass and $\beta$ from ASPECS (assuming the dust temperature evolution specified in Eq.~\ref{eq:tvsz})\tablenotemark{$\dagger$}\label{tab:irx}}
\tablehead{\colhead{Stellar} & \colhead{} & \colhead{\# of} & \colhead{}\\
\colhead{Mass ($M_{\odot}$)} & \colhead{$\beta$} & \colhead{sources} & \colhead{IRX\tablenotemark{a}}}
\startdata
\multicolumn{4}{c}{$z=1.5$--3.5} \\
$>10^{10.75}$  & All &  5  &  51.34$_{-21.51}^{+65.82}$$\pm$1.29 \\
$10^{10.25}$ - $10^{10.75}$  & All &  6  &  26.99$_{-12.53}^{+27.18}$$\pm$0.64 \\
$10^{9.75}$ - $10^{10.25}$  & All &  11  &  16.73$_{-11.72}^{+9.37}$$\pm$0.51 \\
$10^{9.25}$ - $10^{9.75}$  & All &  33  &  2.23$_{-0.89}^{+1.17}$$\pm$0.23 \\
$10^{8.75}$ - $10^{9.25}$  & All &  123  &  0.90$_{-0.45}^{+0.43}$$\pm$0.38 \\
$< 10^{8.75}$  & All &  467  &  0.72$_{-0.80}^{+0.77}$$\pm$0.66 \\
\\
\multicolumn{4}{c}{$z=3.5$--10} \\
$M>10^{10.25}$  & All &  1  &  19.08$_{-0.00}^{+0.00}$$\pm$1.02 \\
$10^{9.75}$ - $10^{10.25}$  & All &  6  &  $-$0.22$_{-0.87}^{+0.76}$$\pm$1.11 \\
$10^{9.25}$ - $10^{9.75}$  & All &  31  &  4.12$_{-2.38}^{+3.23}$$\pm$0.49 \\
$10^{8.75}$ - $10^{9.25}$  & All &  69  &  0.41$_{-0.51}^{+0.50}$$\pm$0.61 \\
$< 10^{8.75}$  & All &  594  &  $-$0.72$_{-0.66}^{+0.59}$$\pm$0.59 \\\\
\multicolumn{4}{c}{$z=1.5$--10} \\
$< 10^{9.25}$  & All &  1253  &  0.50$_{-0.35}^{+0.34}$$\pm$0.31 \\
\\
\multicolumn{4}{c}{$z=1.5$--3.5} \\
$>$$10^{9.5}$ & $-4.0$$<$$\beta$$<$$-1.75$  &  4  &  0.02$_{-0.16}^{+0.12}$$\pm$0.21 \\
 & $-1.75$$<$$\beta$$<$$-1.00$  &  16  &  6.54$_{-4.97}^{+4.88}$$\pm$0.28 \\
 & $-1.00$$<$$\beta$$<$$-0.20$  &  14  &  10.27$_{-2.21}^{+3.74}$$\pm$0.30 \\
 & $-0.20$$<$$\beta$  &  4  &  174.57$_{-41.65}^{+104.96}$$\pm$3.32 \\
\\
$<$$10^{9.5}$ & $-4.0$$<$$\beta$$<$$-1.75$  &  369  &  0.83$_{-0.52}^{+0.54}$$\pm$0.43 \\
 & $-1.75$$<$$\beta$$<$$-1.00$  &  204  &  0.84$_{-0.44}^{+0.39}$$\pm$0.36 \\
 & $-1.00$$<$$\beta$  &  34  &  5.57$_{-4.73}^{+6.07}$$\pm$1.13 \\
\\
\multicolumn{4}{c}{$z=3.5$--10} \\
 & $-4.0$$<$$\beta$$<$$-1.75$  &  537  &  $-$0.24$_{-0.48}^{+0.39}$$\pm$0.37 \\
 & $-1.75$$<$$\beta$$<$$-1.00$  &  125  &  0.65$_{-0.54}^{+0.62}$$\pm$0.56 \\
 & $-1.00$$<$$\beta$  &  32  &  7.67$_{-4.98}^{+4.42}$$\pm$0.96
\enddata
\tablenotetext{$\dagger$}{See Tables~\ref{tab:irxsm}-\ref{tab:irxmuv} from Appendix C for a more detailed presentation of the stack results summarized here.}
\tablenotetext{a}{Both the bootstrap and formal uncertainties are quoted on the result (presented first and second, respectively).}
\end{deluxetable}

For lower-mass ($<$10$^{9.5}$ $M_{\odot}$) $z\sim1.5$--3.5 galaxies
found over ASPECS, significant ALMA continuum flux is found in two of
the three $\beta$ bins we consider.  Fixing $\beta_{int}$ to be the
same as for the higher-mass galaxies, we find a best-fit value for
$dA_{UV}/d\beta$ of 1.12$_{-0.30}^{+0.31}$.  This is most consistent
with an SMC-like dust curve, but is nevertheless consistent with our
constraints on the $dA_{UV}/d\beta$ value in the higher mass
$>$10$^{9.5}$ $M_{\odot}$ bin.

\begin{figure}
\epsscale{1.16} \plotone{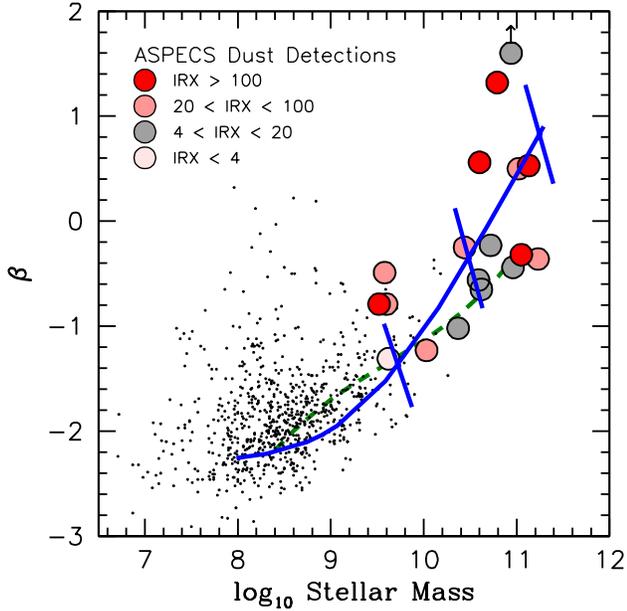}
\caption{$UV$-continuum slopes $\beta$ and stellar masses $M_{*}$ for
  $z\sim1.5$--3.5 galaxies from ASPECS.  The large solid circles show
  the sources from ASPECS that are detected and are presented as in
  Figure~\ref{fig:massbeta2}, while sources that are undetected are
  indicated with the small black circles.  The blue solid lines indicate
  those regions in parameter space where our bivariate relation for
  the infrared excess suggests values of 4, 20, and 100.  The dashed
  green line is as in Figure~\ref{fig:massbeta2}.\label{fig:massbeta}}
\end{figure}

\begin{deluxetable}{cccc}
\tablecolumns{4}
\tabletypesize{\footnotesize}
\tablecaption{Present Constraints on the IRX-$\beta$ relationship\label{tab:irxbetafit}}
\tablehead{\colhead{Sample} & \colhead{Mass Range} & \colhead{$dA_{UV}/d\beta$} & \colhead{$\beta_{int}$}}
\startdata
\multicolumn{4}{c}{Current Determinations}\\
$z\sim1.5$--3.5 & $>10^{9.5}$ $M_{\odot}$ & 1.81$_{-0.14}^{+0.18}$ & $-1.86_{-0.10}^{+0.14}$\\
$z\sim1.5$--3.5 & $<10^{9.5}$ $M_{\odot}$ & 1.12$_{-0.30}^{+0.31}$ & $-1.86$ (fixed)\\
$z\sim1.5$--3.5 & $>10^{9.5}$ $M_{\odot}$ & 1.48$_{-0.11}^{+0.09}$ & $-2.23$ (fixed)\\
$z\sim1.5$--3.5 & $>10^{9.5}$ $M_{\odot}$ & 1.42$_{-0.11}^{+0.09}$ & $-2.30$ (fixed) \\
\multicolumn{4}{c}{Canonical IRX-$\beta$ Relations}\\
\multicolumn{2}{c}{Consensus: $z\sim0$\tablenotemark{a}} & 1.86 & $-$1.87\\
\multicolumn{2}{c}{Reddy et al.\ 2015: $z\sim2$} & 1.84 & $-$2.43\\\\
\multicolumn{2}{c}{Overzier et al.\ 2011: $z\sim0$} & 1.96 & $-$1.96\\
\multicolumn{2}{c}{Takeuchi et al.\ 2012: $z\sim0$} & 1.58 & $-$1.94\\
\multicolumn{2}{c}{Casey et al.\ 2014: $z\sim0$} & 2.04 & $-$1.64\\
\multicolumn{2}{c}{Meurer et al.\ 1999: $z\sim0$} & 1.99 & $-$2.23\\\\
\multicolumn{4}{c}{Dust Laws}\\
\multicolumn{2}{c}{Calzetti} & 1.97 & ---\\
\multicolumn{2}{c}{SMC} & $\sim$1.10 & ---
\enddata
\tablenotetext{a}{Taking the median of the IRX-$\beta$ relations derived by Overzier et al.\ (2011), Takeuchi et al.\ (2012), and Casey et al.\ (2014).  See Appendix B.}
\end{deluxetable}

We now look at the constraints we can set on the IRX-$\beta$
relationship at $z\sim3.5$--10.  We focus on sources with the highest
stellar masses, i.e., $>$10$^{9.25}$ $M_{\odot}$ to minimize the
impact of intrisically blue, lower-mass sources scattering to redder
colors (see \S3.1.1 from Bouwens et al.\ 2016).  Our $z\sim3.5$--10
stack results for sources shows prominent detections in the reddest
two $\beta$ bins, one at $-0.8$ and 1.6.  Those two detections imply
very different IRX-$\beta$ relationships.  Fixing the value of
$\beta_{int}$ to be $-2.23$ and fitting to the two bluest $\beta$ bins
plus the $\beta\sim-0.8$ bin, we derive a $dA_{UV}/d\beta$ value of
2.27.  By contrast, if we fit to the two bluest $\beta$ bins plus the
$\beta\sim1.6$ bin, we derive a $dA_{UV}/d\beta$ value of 0.63.  Given
how different the two relations are and the fact that there are only
two significant detections at $z>3.5$ we can use from ASPECS, perhaps
it is best for us simply to quote our $z=3.5$--10 results as the range
spanned by these two relations.  As this range includes both
Reddy/Calzetti-like and SMC-like dust relations, the ASPECS data
provide us with very little information on how the IRX-$\beta$
relation evolves.

\subsubsection{Summary of Stack Results}

Our convenient summary of our main stack results as a function of
stellar mass, redshift, and $\beta$ is provided in
Table~\ref{tab:irx}.  For a more detailed breakdown of these stack
results and comparison with expectations, we refer the interested
reader to Appendix C.

\subsection{Infrared Excess as a bivariate function of stellar mass and $\beta$}

\subsubsection{Correlation with Stellar Mass and $UV$-continuum Slope $\beta$\label{sec:massbeta}}

Having looked at the correlation of the infrared excess with the
stellar mass and $UV$-continuum slope $\beta$, it is interesting to
try to link these relations based on the empirical correlation of
these two quantities with each other based on the large samples that
now exist based on various legacy data sets.  Given the significant
correlation between the dust content and metallicity of galaxies and
their stellar mass (e.g., Reddy et al.\ 2010; Pannella et al.\ 2015),
one would expect a strong correlation between the $UV$-continuum slope
of galaxies and their stellar mass, as in fact is observed (e.g.,
McLure et al.\ 2018; Carvajal et al.\ 2020).

For this exercise, we take all the $z=1.3$--2.5 sources identified
over the five CANDELS fields by 3DHST team (Skelton et al.\ 2014) and
compare their $UV$-continuum slopes $\beta$ with their stellar masses
derived by Prospector (Leja et al.\ 2017, 2019).  The results are
presented in Figure~\ref{fig:massbeta2}, and it is clear that for
sources with stellar masses to $10^{8.8}$ $M_{\odot}$ the
$UV$-continuum slopes $\beta$ of galaxies generally lie in the range
$-2.5$ to $-1.8$.  For sources with stellar masses $>$10$^{9}$
$M_{\odot}$, the $UV$-continuum slopes $\beta$ show a strong
correlation with stellar mass to $10^{11}$ $M_{\odot}$.

Using the correlations we derive between the infrared excess and the
stellar mass (\S3.3.1),
\begin{equation}
\log_{10} IRX = \alpha \log_{10} (M / M_s)
\end{equation}
and between the infrared excess and the $UV$-continuum slope $\beta$ (\S3.3.2)
\begin{equation}
\log_{10} IRX = \log_{10} (10^{0.4 (\frac{dA_{FUV}}{d\beta} (\beta -
  \beta_{int}))}-1) + 0.23.
\end{equation}
This results in
\begin{equation}
\beta = \beta_{int} + \frac{2.5}{\frac{dA_{FUV}}{d\beta}} \log_{10} (\frac{1}{1.7}(M/M_s)^{\alpha} + 1)
\label{eq:bm}
\end{equation}

Fixing $\beta_{int} = -2.3$ and taking the best-fit value we find for
$\frac{dA_{FUV}}{d\beta}$ (i.e., 1.42), we look for the optimal values
of $M_s$ and $\alpha$ to capture the observed relationship between
stellar mass and $UV$-continuum slope $\beta$ shown in
Figure~\ref{fig:massbeta2}.  In deriving this relationship, we
segregate sources into those above and below the $\beta$ vs. $M$
relation, determine the number of such sources in six distinct regions
along the relation, compute the square of the difference in the number
of sources on each side for each of the six regions, and then minimize
the square of the differences.  The best-fit values of $M_s$ and
$\alpha$ are 10$^{9.07}$ $M_{\odot}$ and 0.92, respectively.  This
best-fit relation is included in Figure~\ref{fig:massbeta2} with the
blue line.  For comparison, Figure~\ref{fig:massbeta2} also shows the
$\beta$ vs. stellar mass relationship derived by McLure et
al.\ (2018).  Encouragingly enough, the best-fit value for $M_s$ and
$\alpha$ are consistent (at $1\sigma$) with the values we derive from
our IRX-stellar mass analysis, i.e., $10^{9.15_{-0.16}^{+0.18}}$
$M_{\odot}$ and 0.97$_{-0.17}^{+0.17}$, respectively, demonstrating
that the IRX-$\beta$ and IRX-stellar mass relations we derive are
essentially equivalent.

\begin{figure*}
\epsscale{1.1}
\plotone{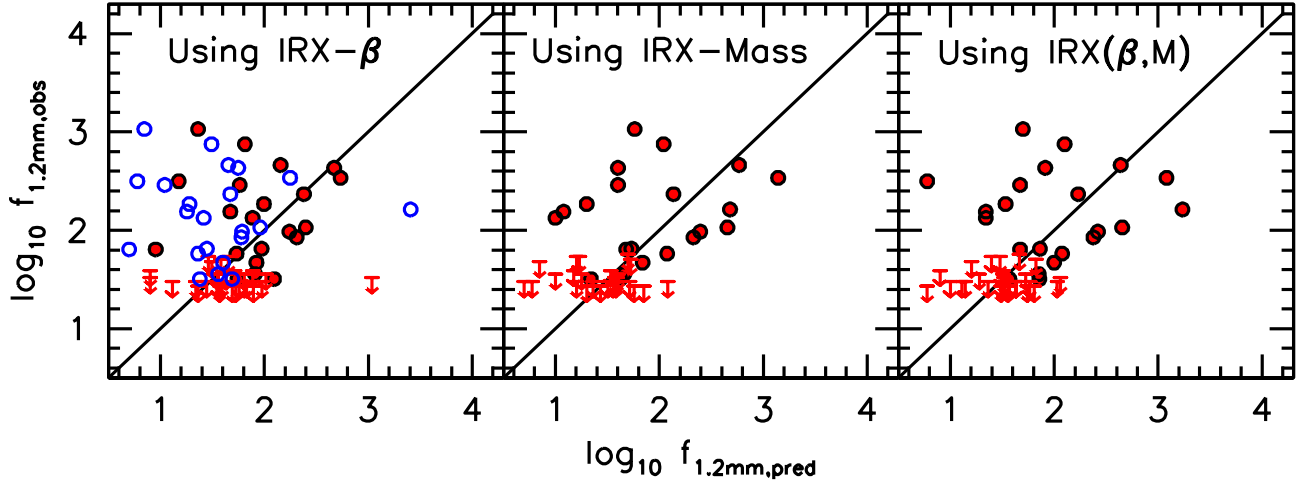}
\caption{Comparison of the predicted and measured flux densities of
  $z\gtrsim1.5$ galaxies at 1.2 mm within the ASPECS footprint.  The
  predicted flux densities (\textit{shown with the red solid circles})
  are based on the $UV$ magnitudes observed and the IRX-$\beta$,
  IRX-stellar mass, and $IRX(\beta,M)$ relations we derive here
  (Eqs.~\ref{eq:irxbeta}, \ref{eq:irxmass}, \ref{eq:irxbm}:
  \S\ref{sec:irxbeta},\ref{sec:irxmass}, \ref{sec:irxbetamass}).  The
  open blue circles in the left panel compare the predicted and
  measured flux densities based on a IRX-SMC relationship
  (Eq.\ref{eq:smc}).  The red downward pointing arrows correspond to
  3$\sigma$ upper limits.\label{fig:predict}}
\end{figure*}

\subsubsection{Infrared Excess of a Function of Stellar Mass and $UV$-continuum Slope $\beta$\label{sec:irxbetamass}}

Having quantified the approximate relationship between the stellar
mass and $UV$-continuum slope $\beta$ of galaxies at $z\sim1.5$-2.5,
we now move on to try to express the infrared excess as a bivariate
function of the $UV$-continuum slope $\beta$ and the stellar mass $M$.

One reason for pursuing such a parameterization would be to take
advantage of the greater information content present in both the
measured $UV$-continuum slope $\beta$ and the inferred stellar mass of
a galaxy.  While the two parameters are clearly correlated (e.g.,
\S\ref{sec:massbeta}), the two parameters do provide us with
independent information on sources and therefore theoretically should
be able to improve our estimates of the infrared excess.

We use the following functional form:
\begin{equation}
IRX(\beta,M) = 1.7 (10^{0.4(dA_{UV}/d\beta)(\beta+2.3)}-1)(M/M(\beta))^{\alpha}
\label{eq:irxbm}
\end{equation}
where $M(\beta)$ is as follows and gives the expected stellar mass for
a given $UV$-continuum slope (as derived in the previous subsection):
\begin{equation}
M(\beta) = (10^{9.07} M_{\odot})(1.7 \times 10^{0.4(1.42)(\beta+2.3)}-1)^{1/0.92}
\end{equation}
The expression we adopt for $IRX(\beta,M)$ is the standard form for
the IRX-$\beta$ relation, but then allows for a dependence on whether
a source is more or less massive than one would expect for a given
$UV$-continuum slope $\beta$.

Sources from ASPECS were divided in stellar mass and $\beta$ in the
same way as the previous sections, stacked using the same weighting
scheme as described in \S3.3, and then an average infrared excess
derived for each stellar mass-$\beta$ bin.  The derived infrared
excesses vs. $\beta$ and stellar mass were then fit using the
expression given in Eq.~\ref{eq:irxbm}.  The best-fit values we
recovered for $dA_{UV}/d\beta$ and $\alpha$ were 1.48$\pm$0.10 and
0.67$\pm$0.06.  Encouragingly enough, the best-fit value for
$dA_{UV}/d\beta$ is very similar to what we found expressing the
infrared excess as a function of the $UV$-continuum slope $\beta$
alone.  We do find a minor additional dependence on whether the
inferred stellar mass is greater or less than given by the general
correlation between stellar mass and $\beta$, but the dependence is
not particularly strong.  The blue lines in Figure~\ref{fig:massbeta}
presents the suggested regions in $\beta$/$M_{*}$ parameter space with
infrared excesses of 4, 20, and 100, shown relative to the detected
and undetected sources from ASPECS.

{\'A}lvarez-M{\'a}rquez et al.\ (2019) had previously attempted to
quantify the infrared excess as a function of both the UV-continuum
slope $\beta$ and stellar mass, as $\log_{10} (IRX) = (0.51 \pm
0.06)\beta_{UV} + (0.37 \pm 0.08) \log(M_{*} [M_{odot}]) - 1.89\pm
0.40$.  While the functional form {\'A}lvarez-M{\'a}rquez et
al.\ (2019) utilize is different from what we consider, it is
interesting to try to compute the logarithmic dependence of IRX on
$\beta_{UV}$ and $\log_{10}$ $M_{*}$ to investigate how similar the
results are.  For simplicity, we compute the dependence at a
$\beta=0.5$ and $\log_{10} M_{*}$ of 10$^{10.5}$ $M_{\odot}$.  For the
$IRX(\beta,M)$ function we derive, we compute a $d\log_{10}
(IRX)/d\beta$ of 0.18 and a $d\log_{10} (IRX)/d\log_{10} M_{*}$ of 0.67
vs. 0.51$\pm$0.06 and 0.37$\pm$0.08 found by {\'A}lvarez-M{\'a}rquez
et al.\ (2019).  These relations are in reasonably good agreement,
which is encouraging given the differences in approach (the
{\'A}lvarez-M{\'a}rquez et al.\ 2019 are based on deep Herschel
stacks).

\begin{deluxetable*}{cccccccccc}
\tablewidth{0cm}
\tablecolumns{10}
\tabletypesize{\footnotesize}
\tablecaption{Comparisons between the predicted and measured $1.2\mu$m flux densities for $z\gtrsim 1.5$ $UV$-selected galaxies showing 4$\sigma$ detections\tablenotemark{i}\label{tab:predict}}
\tablehead{
\colhead{} & \multicolumn{8}{c}{Predicted $f_{1.2mm}$ [$\mu$Jy]} & \colhead{Measured} \\
           & \colhead{} & \colhead{} & \colhead{} & \colhead{} & \colhead{} & \colhead{} & \colhead{} & \colhead{$(IRX_{\beta}~$} & \colhead{$f_{1.2mm}$}\\
\colhead{ID} & \colhead{$IRX_{M99}$\tablenotemark{a}} & \colhead{$IRX_{z=0}$\tablenotemark{b}} & \colhead{$IRX_{SMC}$\tablenotemark{c}} & \colhead{$IRX_{M,0}$\tablenotemark{d}} & \colhead{$IRX_{\beta}$\tablenotemark{e}} & \colhead{$IRX_M$\tablenotemark{f}} & \colhead{$IRX(\beta$,M)\tablenotemark{g}} & \colhead{$~IRX_M)^{1/2}$\tablenotemark{h}} & \colhead{[$\mu$Jy]}}\\
\startdata
     XDFU-2435246390 &    60 &    24 &     7 &    63 &    23 &    58 &    50 &    36 & 1071$\pm$46\\
     XDFU-2385446340 &   184 &    66 &    31 &   111 &    65 &   110 &   126 &    85 & 752$\pm$10\\
     XDFU-2397246112 &   380 &   151 &    45 &   642 &   143 &   587 &   436 &   290 & 461$\pm$14\\
     XDFU-2369747272 &  1572 &   534 &    56 &    43 &   469 &    40 &    82 &   137 & 432$\pm$9\\
     XDFU-2400547554 &  1434 &   571 &   177 &  1492 &   541 &  1391 &  1206 &   868 & 342$\pm$18\\
     XDFU-2410746315 &    41 &    16 &     6 &     3 &    15 &     3 &     6 &     7 & 316$\pm$11\\
     XDFU-2433446471 &   173 &    64 &    11 &    44 &    58 &    40 &    47 &    48 & 289$\pm$21\\
     XDFU-2350746475 &   710 &   264 &    47 &   148 &   240 &   137 &   170 &   182 & 233$\pm$11\\
     XDFU-2416846554 &   294 &   109 &    19 &    21 &    99 &    20 &    34 &    44 & 185$\pm$10\\
     XDFB-2380246263 & 262940 & 73791 &  2555 &   514 & 60164 &   480 &  1716 &  5373 & 163$\pm$10\\
     XDFB-2355547038 &   124 &    49 &    18 &    11 &    47 &    12 &    22 &    23 & 155$\pm$9\\
     XDFU-2387248103 &   203 &    81 &    26 &    10 &    77 &    10 &    22 &    28 & 134$\pm$24\\
     XDFU-2373546453 &   655 &   261 &    91 &   474 &   250 &   452 &   451 &   336 & 107$\pm$10\\
            XDFU4596 &   459 &   183 &    61 &   259 &   174 &   248 &   263 &   208 & 97$\pm$9\\
     XDFU-2361746276 &   552 &   218 &    60 &   225 &   205 &   213 &   238 &   209 & 85$\pm$12\\
            XDFU9838 &   253 &   100 &    28 &    56 &    94 &    54 &    73 &    71 & 65$\pm$15\\
     XDFU-2359847256 &   145 &    56 &    23 &   123 &    54 &   119 &   119 &    80 & 58$\pm$10\\
     XDFU-2370746171 &   244 &    85 &    40 &    67 &    84 &    69 &   100 &    76 & 47$\pm$11\\
\tableline
     \multicolumn{10}{c}{Performance\tablenotemark{i}}\\
     \multicolumn{10}{c}{$(f_{obs}-f_{pred})/ef_{obs}$}\\
     25\%/75\% Quartiles & [$-$23.2,$-$1.7] & [$-$8.0,0.8] & [$-$2.0,4.0] & [$-$4.8,1.1] & [$-$7.2,0.8] & [$-$4.8,1.0] & [$-$5.5,0.9] & [$-$3.8,1.3] \\
     \multicolumn{10}{c}{$(f_{obs}-f_{pred})/(f_{pred} ^2+ef_{obs} ^2)^{0.5}$}\\
    25\%/75\% Quartiles & [$-$1.1,$-$0.4] & [$-$1.1,0.6] & [$-$1.1,1.4] & [$-$1.1,0.7] & [$-$1.1,0.7] & [$-$1.1,0.7] & [$-$1.1,0.3] & [$-$1.1,0.6] \\
     \multicolumn{10}{c}{$\log_{10} (f_{obs}/f_{pred})$\tablenotemark{j}}\\
Mean / Std. Dev. & $-$0.42$\pm$0.81 & 0.01$\pm$0.80 & 0.54$\pm$0.70 & 0.29$\pm$0.69 & 0.03$\pm$0.79 & 0.30$\pm$0.68 & 0.14$\pm$0.65 & 0.17$\pm$0.67 \\
Median & $-$0.59 & $-$0.19 & 0.37 & 0.12 & $-$0.16 & 0.12 & $-$0.05 & $-$0.02 
\enddata
\tablenotetext{a}{From Eq.~\ref{eq:irxm99} (Appendix B), which is the Meurer et al.\ (1999) IRX-$\beta$ relationship.}
\tablenotetext{b}{From Eq.~\ref{eq:z0}, which is the consensus
low-redshift IRX-$\beta$ relation derived here in Appendix B from
literature results.}
\tablenotetext{c}{Eq.~\ref{eq:smc}, which gives an SMC-like IRX-$\beta$ relation}
\tablenotetext{d}{From Eq.~\ref{eq:m0}, which is the consensus
  IRX-stellar mass relation presented in our previous study Bouwens et
  al.\ (2016)}
\tablenotetext{e}{From Eq.~\ref{eq:irxbeta}, which is the IRX-$\beta$ relation we derived for
$>$10$^{9.5}$ $M_{\odot}$, $z\sim1.5$--3.5 galaxies (\S\ref{sec:irxbeta}).}
\tablenotetext{f}{From Eq.~\ref{eq:irxmass}, which is the IRX-stellar mass relation we derived for $z\sim1.5$--3.5 galaxies (\S\ref{sec:irxmass}).}
\tablenotetext{g}{From Eq.~\ref{eq:irxbm}, which is the IRX($\beta$, M) relation we derived (\S\ref{sec:irxbetamass}).}
\tablenotetext{h}{Geometric mean of our derived $z=1.5$--3.5 IRX-$\beta$ relation $IRX_{\beta}$ and our
IRX-stellar mass relationship $IRX_{M}$.} 
\tablenotetext{i}{See \S\ref{sec:performance} for a discussion}
\tablenotetext{j}{Only for those 25 sources where $f_{obs}/ef_{obs}>2$}
\end{deluxetable*} 

Given the strong correlation between both parameters, where $\Delta
\beta \sim 1.5 \Delta M_{*}$ (see \S\ref{sec:massbeta}), it is also
interesting to reformulate the {\'A}lvarez-M{\'a}rquez et al.\ (2019)
IRX relation to be just a single function of $\beta$.  We find
$d\log_{10} (IRX)/d\beta \sim 0.63$.  If we make the same change to
our bivariate $IRX(\beta,M)$ relation, we find $d\log_{10}
(IRX)/d\beta \sim 0.68$.  As with the previous comparison, the two
dependencies are similar, which is encouraging given differences in
the two approaches.

\subsection{Predictive Power of Different Estimators for IRX\label{sec:performance}}

Before concluding this section, it is useful to summarize the
predicted 1.2mm flux densities expected for different $z\gtrsim1.5$
galaxies over the ASPECS footprint and compare those predictions with
the observations.  A compilation of the results are presented in
Table~\ref{tab:predict} and include the predicted flux densities using
(1) the Meurer et al.\ (1999) IRX-$\beta$ relation
(Eq.~\ref{eq:irxm99}: Appendix B), (2) the consensus low-redshift
IRX-$\beta$ relation (Eq.~\ref{eq:z0}) derived here in Appendix B from
literature results, (3) an SMC-like IRX-$\beta$ relation
(Eq.~\ref{eq:smc}), (4) the consensus IRX-stellar mass relation
(Eq.~\ref{eq:m0}) presented in our previous study Bouwens et
al.\ (2016), (5) our derived IRX-$\beta$ relation for $>$10$^{9.5}$
$M_{\odot}$, $z\sim1.5$--3.5 galaxies (Eq.~\ref{eq:irxbeta}:
\S\ref{sec:irxbeta}), (6) our derived IRX-stellar mass relation for
$z\sim1.5$--3.5 galaxies (Eq.~\ref{eq:irxmass}: \S\ref{sec:irxmass}),
and (7) our derived IRX($\beta$, M) relation (Eq.~\ref{eq:irxbm}:
\S\ref{sec:irxbetamass}).  As one final predictor, we include a
comparison against the flux density predicted taking the geometric
mean of our derived $z=1.5$--3.5 IRX-$\beta$ relation and our
IRX-stellar mass relationship, i.e., $(IRX_{\beta} IRX_M)^{1/2}$, and
using Eqs.~\ref{eq:irxbeta} and \ref{eq:irxmass} while taking
$dA_{UV}/d\beta$, $\beta_{int}$, $M_s$, and $\alpha$ to be
1.81$_{-0.14}^{+0.18}$, $-1.86_{-0.10}^{+0.14}$,
$10^{9.15_{-0.16}^{+0.18}}$ $M_{\odot}$, and 0.97$_{-0.17}^{+0.17}$,
respectively.  This should provide for an alternate way of using both
the $UV$-continuum slopes $\beta$ and stellar masses in estimating the
infrared excess.

The observed fluxes are also explicitly compared against these many
estimators in Figure~\ref{fig:predict}.  A quantification of the mean,
median, and $1\sigma$ scatter in the logarithmic ratio of the
predicted and measured 1.2mm flux densities is presented in
Table~\ref{tab:predict}, and it is clear there is substantial scatter
between the observed and predicted flux densities.  The scatter ranges
from 0.65 to 0.81 dex, with the smallest dispersion found for the
$IRX(\beta,M)$ and $(IRX_{\beta} IRX_M)^{1/2}$ estimators, with only
slight increases in the dispersion for the other relations.  The
$IRX(\beta,M)$ and $(IRX_{\beta} IRX_M)^{1/2}$ estimators also provide
the best predictions of the observed flux densities in the median.

As a separate means of evaluating the estimators, we compare the
predicted 1.2mm flux densities from these estimators with the measured
flux densities using both the detected sources in
Table~\ref{tab:predict} and sources expected to be detected at
$>$2$\sigma$ averaging the IRX-$\beta$ and IRX-stellar mass relations
derived here (Eqs~\ref{eq:irxbeta} and \ref{eq:irxmass}), i.e., 70
sources in total.  For each of these sources, we computed the
difference between the measured and predicted flux for each source,
i.e., $f_{obs}$ and $f_{pred}$, divided the result by the measurement
error $ef_{obs}$, and then determined the average as well as the upper
and lower quartiles.  For almost every estimator, the difference
between the upper and lower quartiles is larger than the measurement
error by $\gtrsim$5$\times$.

For each of the estimators, we also computed the differences between
the measured and predicted flux densities for the same sources as the
previous exercise, divided the result by the root mean square of the
predicted flux densities and flux measurement uncertainties, and
finally computed the upper and lower quartiles.  This should give an
approximate relative uncertainty on the flux density predictions.  All
of our estimators perform comparably well, with only modest
differences between them.

In summary, as with previous work (e.g., Meurer et al.\ 1999; Reddy et
al.\ 2006), estimators of the infrared excess tend to be accurate in
predicting the obscured star formation rates or IR luminosities for
the \textit{average} source and tend to show at least $\sim$0.65 dex
scatter for individual sources.  Of those we consider, the different
estimators for the infrared excess all perform comparably, with
marginally better performance for the estimators that consider both
mass $M$ and $\beta$, i.e., $IRX(\beta,M)$ and $(IRX_{\beta}
IRX_M)^{1/2}$, while the $IRX_{M99}$ estimator performed the least
well.

\section{Discussion}

\subsection{Previous Reported Continuum Detections}

It is interesting to compare the present set of ALMA continuum
detections to those that were previously reported over the HUDF by
Aravena et al.\ (2016), Bouwens et al.\ (2016), and Dunlop et
al.\ (2017).  The reported detections and tentative detections by
Aravena et al.\ (2016) and Bouwens et al.\ (2016) made use of the 1
arcmin$^2$ pilot for ASPECS, while the Dunlop et al.\ (2017) results
were based on the 1.3mm ALMA continuum observations they obtained over
a 4.5 arcmin$^2$ region within the HUDF/XDF.

\begin{figure}
\epsscale{1.17} 
\plotone{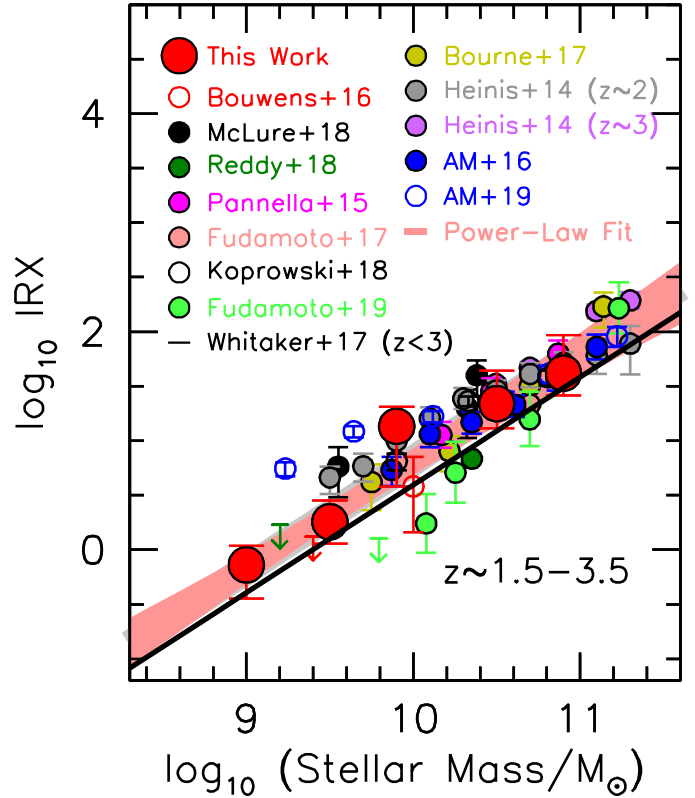}
\caption{Comparison of the present determinations of the IRX - stellar
  mass relation at $z\sim1.5$--3.5 with many previous determinations
  in the literature, including from the pilot study to ASPECS (Bouwens
  et al.\ 2016: \textit{open red circles}), McLure et al.\ (2018:
  \textit{solid black circles}), Reddy et al.\ (2018: \textit{solid
    green circles}), Pannella et al.\ (2015: \textit{solid magenta
    circles}), Fudamoto et al.\ (2017: \textit{solid light red
    circles}), Koprowski et al.\ (2018: \textit{open black circles}),
  Fudamoto et al.\ (2020a: \textit{solid light green circles}), Bourne
  et al.\ (2017: \textit{solid yellow circles}),
  {\'A}lvarez-M{\'a}rquez et al.\ (2016: \textit{solid blue circles}),
  {\'A}lvarez-M{\'a}rquez et al.\ (2019: \textit{open blue circles}),
  and Heinis et al.\ (2014) at both $z\sim2$ (\textit{solid gray
    circles}) and $z\sim3$ (\textit{solid violet circles}).  The solid
  black line gives the IRX vs. stellar mass trend Whitaker et
  al.\ (2017) derive for their results over the full range $z\sim0$ to
  $z\sim3$, while the shaded gray region gives the consensus
  IRX-stellar mass relation we derived for select literature results
  in our pilot study.  The light red shaded line is a fit to our IRX
  stack results vs. stellar mass.  Our new results are in agreement
  with previous work over the entire mass range well probed by this
  study ($10^{9}$ to $10^{11}$ $M_{\odot}$).\label{fig:irxsmlit}}
\end{figure}

\begin{figure}
\epsscale{1.17} 
\plotone{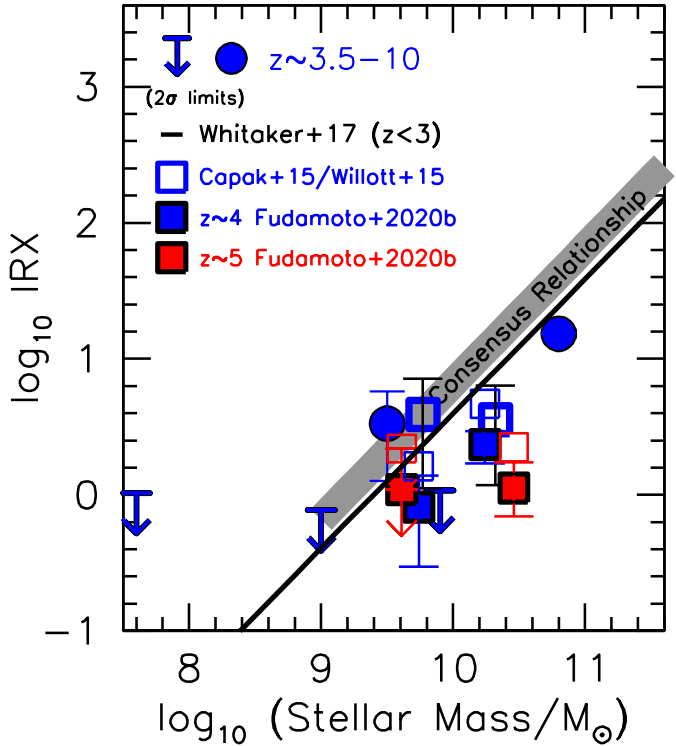}
\caption{Comparison of our IRX - stellar mass stack results with that
  inferred from the Capak et al.\ (2015) and Willott et al.\ (2015)
  observations assuming the fiducial dust temperature evolution given
  in Eq.~\ref{eq:tvsz}.  Also presented are the new results from
  ALPINE by Fudamoto et al.\ (2020b), both as quoted in that study and
  adopting the fiducial evolution in dust temperature adopted here
  (\textit{open blue and red squares} showing the results at
  $z\sim4.5$ and $z\sim5.5$, respectively).\label{fig:irxsm5}}
\end{figure}

Using the 1 arcmin$^2$ pilot observations for ASPECS, Aravena et
al.\ (2016) and Bouwens et al.\ (2016) detected 5 $z>1.5$ galaxies and
reported tentative detections for 3 more $z>1.5$ galaxies.  Our new
observations confirm all of our previously claimed detections at
$>$4$\sigma$, making it clear that those detections were real.  In
addition, one of the tentatively detected sources from our pilot
program, i.e., XDFU-2370746171, shows a $>$4$\sigma$ detection
(40$\pm$11$\mu$Jy$\,$beam$^{-1}$) in the new data, confirming that the
reported tentative detection (34$\pm$14$\mu$Jy$\,$beam$^{-1}$) from
our pilot was real.

The measured flux densities for the two other tentative detections
from our pilot, i.e., XDFU-2365446123 and XDFU-2384246384, are
$-$27$\pm$17 $\mu$Jy$\,$beam$^{-1}$ and 8$\pm$10 $\mu$Jy/beam vs. our
measurements of 38$\pm$16$\mu$Jy/beam and 36$\pm$14$\mu$Jy/beam,
respectively, in the pilot for these sources.  Combining the
measurements, the flux is 7$\pm$12$\mu$Jy$\,$beam$^{-1}$ for
XDFU-2365446123 and 17$\pm$8$\mu$Jy for XDFU-2384246384.  While the
new observations do not support the reality of either source,
XDFU-2384246384 still shows a tentative 2.1$\sigma$ detection in the
continuum in the combined data set and thus may be real.

In the Dunlop et al.\ (2017) search, 16 dust-continuum
($>$3.5$\sigma$) detections are identified, 11 of which have an
estimated redshift in excess of 1.5 and lie within the ASPECS
footprint.  8 of these 11 sources are clearly confirmed with our
ASPECS ALMA observations.  For the 3 reported continuum detections
from the Dunlop et al.\ (2017) which are not unambiguously confirmed
by our ASPECS observations, we measure $-9\pm 21\mu$Jy (UDF9), $-45\pm
31\mu$Jy (UDF12), and $-3\pm 9 \mu$Jy (UDF15).

\subsection{Comparison with Previous Determinations of the Infrared Excess}

It is interesting to compare the IRX-stellar mass and IRX-$\beta$
relations we derive with the many previous determinations in the
literature.  We focus on determinations at $z\sim1.5$--3.5 since this
is where our results are the most significant and where most of
previous results have been obtained.  In Figure~\ref{fig:irxsmlit}, we
compare the IRX-stellar mass relationship we find at $z\sim1.5$--3.5
with what we obtained in our pilot study (Bouwens et al.\ 2016) and
many other determinations in the literature (McLure et al.\ 2018;
Reddy et al.\ 2018; Pannella et al.\ 2015; Fudamoto et al.\ 2017,
2020a; Bourne et al.\ 2017; {\'A}lvarez-M{\'a}rquez et al.\ 2016,
2019; Heinis et al.\ 2014; Koprowski et al.\ 2018).

Overall, our new IRX-stellar mass results appear to be in agreement
with previous results as presented e.g. by Heinis et al.\ (2014),
Pannella et al.\ (2015), Bourne et al.\ (2017), and McLure et
al.\ (2018), or even as given by the consensus relation derived in our
pilot study (\textit{shown with the grey line}).  Our best-fit
IRX-stellar mass correlation is $\sim$0.2-0.3 dex higher at 10$^{10}$
$M_{\odot}$ than found in our earlier study (Bouwens et al.\ 2016) but
consistent within the quoted uncertainties.  Thanks to the larger
number of dust-continuum detected sources in the current ASPECS study
vs. our pilot study (18 vs. 3 $4\sigma$ detections), we are able to
significantly improve our quantification of the IRX-stellar mass
relation relative to our previous study.

The slope recovered for our new IRX-stellar mass relation, i.e.,
0.97$_{-0.17}^{+0.17}$, is very close to one.  We had previous adopted
a value of unity in Bouwens et al.\ (2016) for the consensus relation
(Eq.~\ref{eq:m0}) based on the IRX-stellar mass results of Reddy et
al.\ (2010), Whitaker et al.\ (2014), and {\'A}lvarez-M{\'a}rquez et
al.\ (2016).  The IRX - stellar mass relation derived by McLure et
al.\ (2018) using the shallower ALMA observations over the HUDF
(Dunlop et al.\ 2017) also find a slope (0.85$\pm$0.05), very close to
what we find here.  At one other extreme, Fudamoto et al.\ (2020a)
recover a much steeper slope (1.64$\pm$0.10) for the IRX-stellar mass
relation, similar to what we derive using \textsc{Prospector} for our
stellar mass estimates (Appendix D).  Meanwhile, earlier results
obtained from an analysis of Herschel data by Pannella et al.\ (2015)
find a much shallower IRX-stellar mass relation, with a slope of
$\sim$0.64, clearly shallower than what we find here (see also results
by {\'A}lvarez-M{\'a}rquez et al.\ 2019).  Given the current strong
constraints on the obscured SFR at low masses ($<$10$^{9.25}$
$M_{\odot}$) and the challenge that source confusion presents for the
lowest mass sources with {\it Herschel}, it seems likely that the
slope of the infrared excess is approximately unity or steeper, as
essentially all analyses relying on ALMA data have found.

\begin{figure}
\epsscale{1.17} 
\plotone{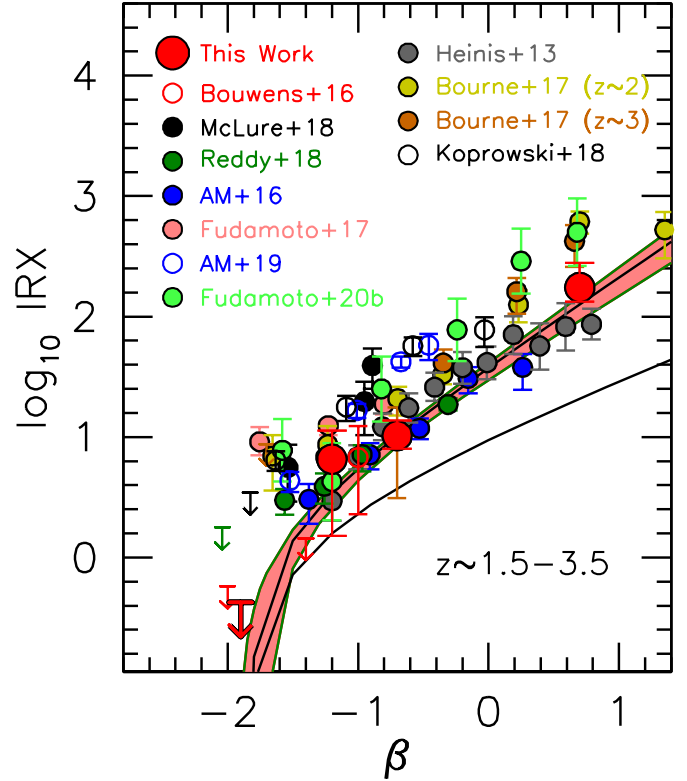}
\caption{Comparison of the present determinations of the IRX - $\beta$
  relation at $z\sim1.5$--3.5 with a wide variety of previous
  determinations, including the pilot study to ASPECS (Bouwens et
  al.\ 2016: \textit{open red circles}), McLure et al.\ (2018:
  \textit{solid black circles}), Reddy et al.\ (2018: \textit{solid
    green circles}), {\'A}lvarez-M{\'a}rquez et al.\ (2016:
  \textit{solid blue circles}), Fudamoto et al.\ (2017: \textit{solid
    light red circles}), {\'A}lvarez-M{\'a}rquez et al.\ (2019:
  \textit{open blue circles}), Fudamoto et al.\ (2020a: \textit{solid
    light green circles}), Koprowski et al.\ (2018: \textit{open black
    circles}), Heinis et al.\ (2013: \textit{solid grey circles}), and
  the Bourne et al.\ (2017) results at $z\sim2$ (\textit{solid yellow
    circles}) and $z\sim3$ (\textit{solid brown circles}).  The black
  lines show the Reddy (Calzetti-like) IRX-$\beta$ relationship
  (Eq.~\ref{eq:reddy}) and an SMC-like IRX-$\beta$ relation
  (Eq.~\ref{eq:smc}).\label{fig:irxbetalit}}
\end{figure}

\begin{deluxetable}{ccc}
\tabletypesize{\footnotesize}
\tablecaption{Estimated dust corrections to apply to the $UV$ luminosity density results integrated to various limiting luminosities\label{tab:dustcorr}}
\tablehead{
\colhead{} & \multicolumn{2}{c}{$\textrm{log}_{10}$ Dust Correction} \\
\colhead{Sample} & \colhead{($>$0.05 $L_{z=3}^{*}$)\tablenotemark{a}} & \colhead{($>$0.03 $L_{z=3}^{*}$)\tablenotemark{a}}}\\
\startdata
$z\sim3$ & 0.37\tablenotemark{*} & 0.34\tablenotemark{*}\\
$z\sim4$ & 0.33 & 0.31\\
$z\sim5$ & 0.30 & 0.27\\
$z\sim6$ & 0.20 & 0.17\\
$z\sim7$ & 0.09 & 0.07\\
$z\sim8$ & 0.07 & 0.06
\enddata
\tablenotetext{*}{For uniquely the $z\sim3$ sample, we make use of the
  finding by e.g. Reddy \& Steidel (2004) and Reddy et
  al.\ (2010) that the average infrared excess for galaxies brighter
  than 25.5 mag at $z\sim3$ is a factor of $\sim$5.}
\tablenotetext{a}{The specified limits 0.05 $L_{z=3}^{*}$ and 0.03 $L_{z=3}^{*}$ correspond to faint-end limits of $-17.7$ and $-17.0$, respectively, which is the limiting luminosity to which $z\sim7$ and $z\sim10$ galaxies can be found in current probes (Schenker et al.\ 2013; McLure et al.\ 2013; Ellis et al.\ 2013; Oesch et al.\ 2013; Bouwens et al.\ 2015).}
\end{deluxetable}

\begin{figure*}
\epsscale{1.1}
\plotone{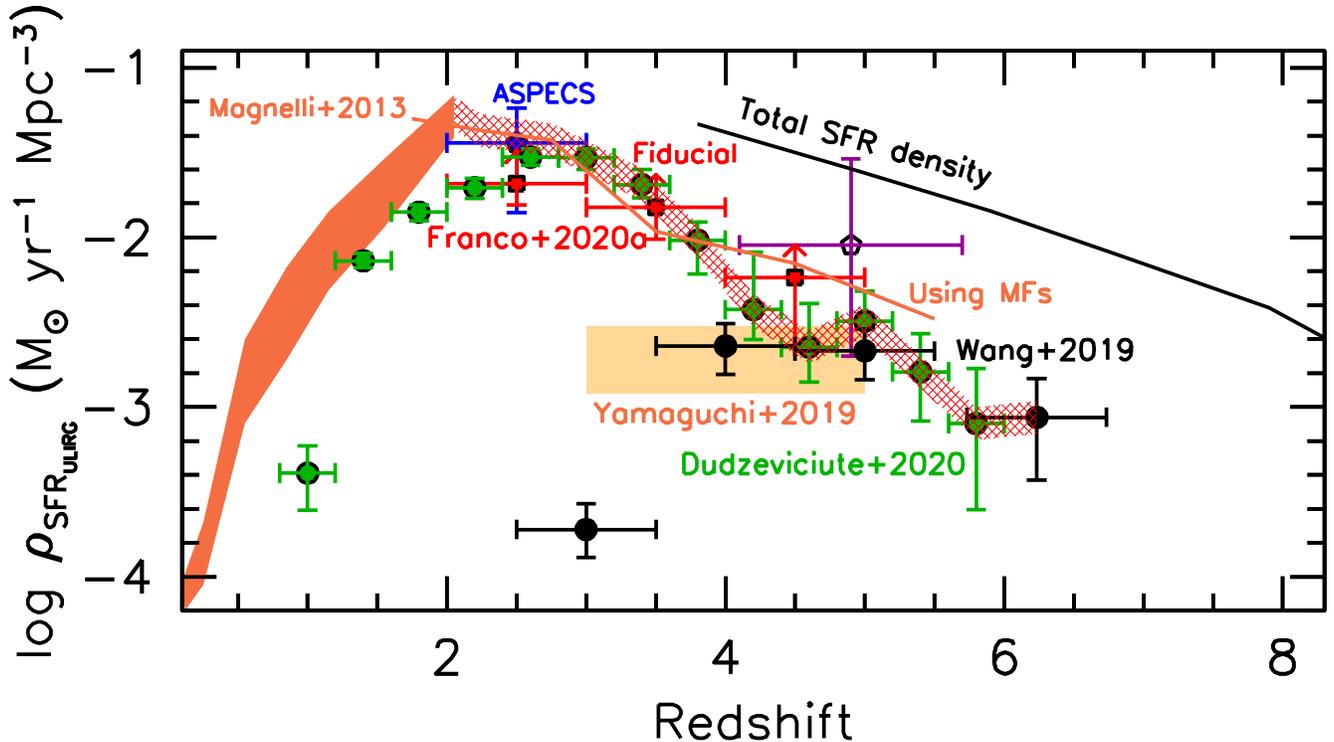}
\caption{Estimated SFR densities at $z=2$--8 from galaxies with $IR$
  luminosities greater than $10^{12}$ $L_{\odot}$ (corresponding to
  SFRs $>$100 $M_{\odot}$$\,$yr$^{-1}$) which is difficult to probe
  with $UV$-based searches (\S4.4).  Shown are the published
  determinations based on the Magnelli et al.\ (2013: \textit{dark
    orange shaded region}), Yamaguchi et al.\ (2019: \textit{light
    orange shaded region}), Williams et al.\ (2019: \textit{open
    purple pentagon}), and Wang et al.\ (2019: \textit{solid black
    circles}) probes.  The solid green circles indicate the SFR
  densities from Dudzevi{\v{c}}i{\={u}}t{\.{e}} et al.\ (2020), who
  extrapolated from a 870$\mu$m flux limit of 3.6 mJy to 1 mJy
  (equivalent to an $L_{IR}$ of $\approx$10$^{12}$ $L_{\odot}$).  The
  blue pentagon shows the SFR density of ULIRGs we compute from the
  ASPECS area (Gonz{\'a}lez-L{\'o}pez et al.\ 2020).  The estimates we
  show from Franco et al.\ (2020a: \textit{solid red squares}) are
  computed on the basis of the redshifts and fluxes from their sample
  and the cosmic volume included in a 69 arcmin$^2$ search area,
  assuming that $\sim$100\% of the far-IR flux is powered by star
  formation.  For reference, we also show the total SFR density we
  estimate for all galaxies at $z\geq4$ (brightward of $-$17 AB mag).
  In addition, we include an approximate prediction for the
  contribution of such galaxies to the cosmic SFR density
  (\textit{solid red line}) using the wide-area mass functions of
  Ilbert et al.\ (2013) and Davidzon et al.\ (2017) and the
  star-forming main sequence by Speagle et al.\ (2014).  Encouragingly
  enough, current observational constraints are consistent with the
  predicted contribution of such sources of cosmic SFR density at
  $z<4$ and moderately higher ($\sim$0.2 dex) at $z>4$.  The hatched
  red region shows the fiducial estimate of the obscured SFR density
  from ULIRGs we adopt here and relies on the Magnelli et al.\ (2013)
  determination at $z\sim2$, the mass function derived estimate at
  $z\sim2.75$, and the AS2UDS measurements
  (Dudzevi{\v{c}}i{\={u}}t{\.{e}} et al.\ 2020) at
  $z>3$.\label{fig:sfz_obsc}}
\end{figure*}

The IRX-stellar mass results we obtain at $z\sim3.5$--10 can be
compared with results obtained using a small sample of bright
$z\sim5$-6 galaxies from Capak et al.\ (2015) and Willott et
al.\ (2015) and assuming the dust temperature evolution given in
Eq.~\ref{eq:tvsz}.  Also included in this comparison are the new
ALPINE results from Fudamoto et al.\ (2020b), both as quoted in the
original study (\textit{solid colored points}) and adopting the
fiducial dust temperature evolution adopted here (Eq~\ref{eq:tvsz}).
This comparison is presented in Figure~\ref{fig:irxsm5}.  Our own
results appear to be most consistent with the consensus IRX-M$_{*}$
relationship we had derived in our pilot study (Bouwens et al.\ 2016)
and as now derived here as $z\sim1.5$--3.5.  While this suggests that
the IRX-stellar mass relation may extend to $z\sim5$--6 with little or
no evolution, the ASPECS field only contains a few bright, massive
sources to probe this well.  Additionally, this inference depends
critically on the dust temperature being relatively high, i.e.,
$\sim$50 K, at $z\sim4$--6.  If the temperature is instead $\sim$41 K
as Fudamoto et al.\ (2020b) adopt in their analysis, clearly the
IRX-stellar mass relation at $z>3.5$ is lower than what is found at
$z\sim1.5$--3.5.

In Figure~\ref{fig:irxbetalit}, we compare the IRX-$\beta$
relationship we derive for higher-mass, $z\sim1.5$--3.5 galaxies with
the results obtained in our pilot study (Bouwens et al.\ 2016) as well
as a wide variety of different determinations in the literature
(McLure et al.\ 2018; Reddy et al.\ 2018; {\'A}lvarez-M{\'a}rquez et
al.\ 2016, 2019; Fudamoto et al.\ 2017, 2020a; Heinis et al.\ 2013;
Bourne et al.\ 2017; Koprowski et al.\ 2018).  Similar to what we
found for the IRX-stellar mass relation, the larger number of
dust-continuum detections found here (vs. from the smaller-area ASPECS
pilot) results in our recovering a steeper IRX-$\beta$ relation than
in our pilot, i.e., 1.48$_{-0.11}^{+0.09}$ vs. $1.26_{-0.36}^{+0.26}$
when fixing $\beta_{int}=-2.23$.  The only apparently significant
difference occurs for our determination at $-$1.3 where the limit from
our pilot program was 1.31$_{-0.94}^{+0.67}$$\pm$0.72 (at
$\beta\sim-1.4$) and where our new measurement is
6.54$_{-4.97}^{+4.88}$$\pm$0.28 (at $\beta\sim-1.2$).  This difference
results both from the larger number of dust detected sources in the
$4\times$ larger area probed by ASPECS (vs. our PILOT) and from our
changing the $\beta$ binning scheme to exploit the larger number of
sources to improve our leverage for constraining the IRX-$\beta$
relation.

Relative to various determinations from the literature, the most
significant differences occur for the bluest values of $\beta$, i.e.,
$\beta\sim-1.8$, where our own determination of the infrared excess is
some 0.2-1.0 dex lower than the determinations of Reddy et
al.\ (2018), Fudamoto et al.\ (2017), Bourne et al.\ (2017), and
McLure et al.\ (2018).  It seems likely that the differences here are
due to the presence of blue, IR-luminous sources in many previous
selections.  While blue, IR-luminous galaxies are known to exist
(e.g., Reddy et al.\ 2006; Casey et al.\ 2014), especially at high IR
luminosities ($>$10$^{12}$ $L_{\odot}$) where there is less connection
between the $UV$ and $IR$ morphologies in galaxies, these sources are
not sufficiently common to be well sampled by the
$\sim$2.5$\times$10$^4$ comoving Mpc$^3$ volume probed by ASPECS at
$z=1.5$--3.5.

\begin{figure}
\epsscale{1.17} \plotone{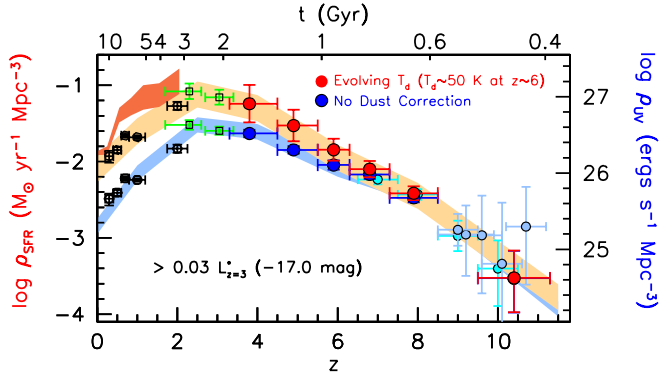}
\caption{Updated determinations of the derived SFR (\textit{left
    axis}) and $UV$ luminosity (\textit{right axis}) densities versus
  redshift (\S4.4).  The left axis gives the SFR densities we would
  infer from the measured luminosity densities, assuming the Madau et
  al.\ (1998) conversion factor relevant for star-forming galaxies
  with ages of $\gtrsim10^8$ yr (see also Kennicutt 1998).  The right
  axis gives the $UV$ luminosities we infer integrating the present
  and published LFs to a faint-end limit of $-17$ mag (0.03
  $L_{z=3}^{*}$) -- which is the approximate limit we can probe to
  $z\sim8$ in our deepest data set.  The upper and lower set of points
  (\textit{red and blue circles, respectively}) and shaded regions
  show the SFR and $UV$ luminosity densities corrected and uncorrected
  for the effects of dust extinction.  The dust correction we utilize
  relies on the bivariate $IRX(\beta,M_{*})$ relation derived here
  (Eq.~\ref{eq:irxbm}) for galaxies with solar masses $>$10$^{9}$
  $M_{\odot}$ and otherwise we take the correction to be zero.  The
  dust-corrected SFR density we quote includes the contribution of
  far-IR luminous ($>$10$^{12}$ $L_{\odot}$) galaxies, as indicated by
  the fiducial SFR density in Figure~\ref{fig:sfz_obsc}.  The dark red
  shaded region shows the implied SFR densities to $z<2$ from
  dust-obscured and IR luminous sources (Magnelli et al.\ 2013).  Also
  shown are the SFR densities at $z\sim 2$ and $z\sim 3$ from Reddy et
  al.\ (2009: \textit{green crosses}), at $z\sim0$--2 from
  Schiminovich et al.\ (2005: \textit{black hexagons}), at $z\sim7$--9
  from McLure et al.\ (2013) and Ellis et al.\ 2013: \textit{cyan
    solid circles}), and $z\sim9$--11 from CLASH (Bouwens et
  al.\ 2014b; Coe et al.\ 2013; Zheng et al.\ 2012: \textit{light blue
    circles}) and Oesch et al.\ (2013: \textit{light blue circles}).
  The $z\sim9$--11 constraints on the $UV$ luminosity density have
  been adjusted upwards to a limiting magnitude of $-17.0$ mag
  assuming a faint-end slope $\alpha$ of $-2.0$ (consistent with our
  constraints on $\alpha$ at both $z\sim7$ and at
  $z\sim8$).\label{fig:sfz}}
\end{figure}

\begin{figure*}
\epsscale{1.14}
\plotone{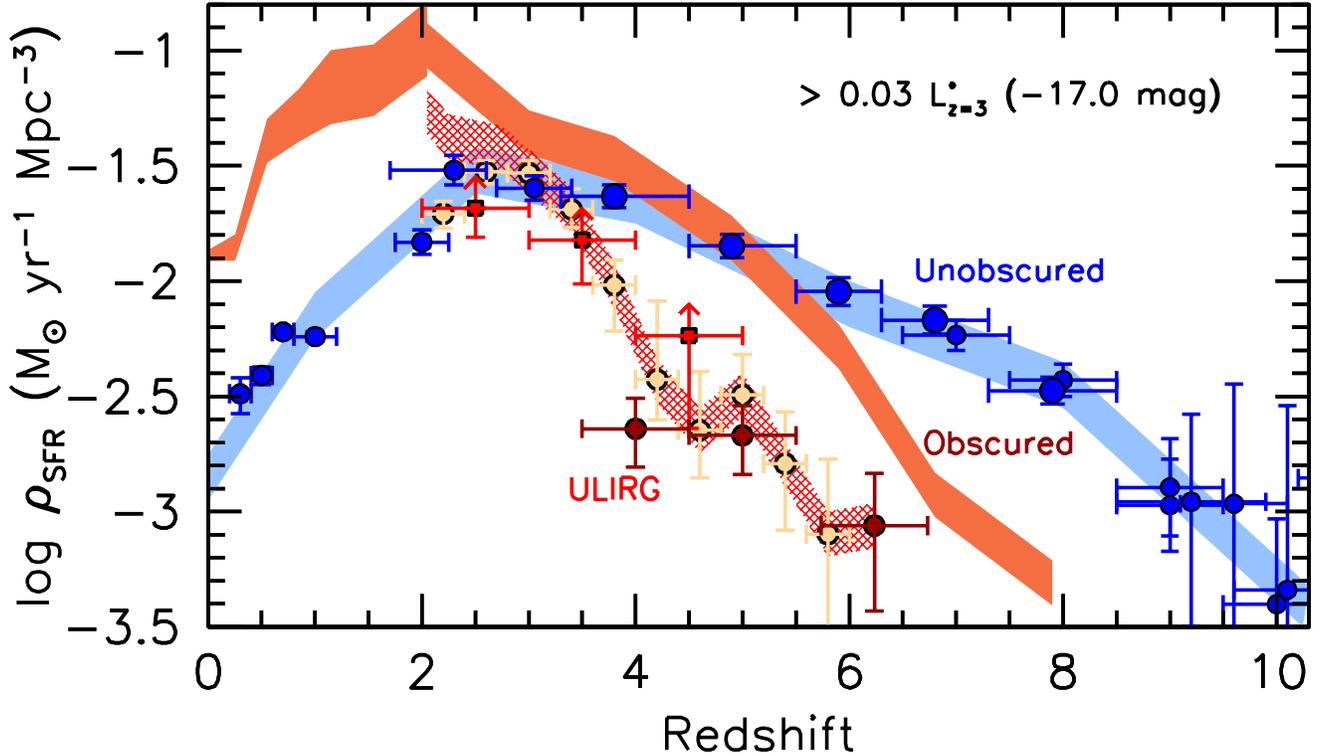}
\caption{Updated determinations of the SFR density vs. redshift shown
  in terms of the star formation which is unobscured (\textit{blue
    points and shaded region}) and obscured (\textit{red regions}: see
  \S4.4).  The contribution to the $z>2$ SFR density from obscured
  ULIRG-type galaxies, with $>$10$^{12}$ $L_{\odot}$ ($>$100
  $M_{\odot}$$\,$yr$^{-1}$) is shown with the red hatched region.  The
  solid red, light red, and brown circles shown at $z>2$ are from
  Franco et al.\ (2020a), Dudzevi{\v{c}}i{\={u}}t{\.{e}} et
  al.\ (2020), and Wang et al.\ (2019), respectively, and are as in
  Figure~\ref{fig:sfz_obsc} and \ref{fig:sfz}.  The SFR density of the
  universe is predominantly unobscured at $z>5$ and obscured at $z<5$.
  The approximate transition point between the two regimes is at
  $z\sim5$.\label{fig:sfzc}}
\end{figure*}

Otherwise, our IRX-$\beta$ results are broadly in agreement with the
results of Reddy et al.\ (2018), {\'A}lvarez-M{\'a}rquez et
al.\ (2016), and Heinis et al.\ (2013).  For redder values of $\beta$,
our IRX-$\beta$ results are lower than the results of McLure et
al.\ (2018), Fudamoto et al.\ (2017), Bourne et al.\ (2017), and
Fudamoto et al.\ (2020a) by $\sim$0.4 dex.  We expect that some
fraction of these differences, i.e., 0.3 dex, could result from
different calibrations to derive the IR luminosities and obscured SFRs
from the measured ALMA fluxes (e.g., Murphy et al.\ 2011 vs. Whitaker
et al.\ 2017).

\subsection{Dust Corrections for $z\gtrsim3$ Samples}

The purpose of this section is to take advantage of the results of our
analyses from the previous sections to derive dust corrections that we
can apply to the general star-forming galaxy population at $z\geq3.5$.  

We will focus on deriving these corrections as a function of the $UV$
luminosity of galaxies and derive a distribution of dust corrections
that make up each $UV$ luminosity bin.  To ensure a significant
sampling of each $UV$ luminosity bin, we leverage the large selections
of star-forming galaxies Bouwens et al.\ (2015) identified at
$z\sim4$, 5, 6, 7, 8, and 10 over the CANDELS GOODS-North and
GOODS-South.

Each of the sources over the CANDELS GOODS-North and GOODS-South
fields has sensitive {\it HST} optical/ACS and WFC3/IR photometry
available to derive $UV$-continuum slopes for each source in these
samples.  Another valuable aspect of sources in these fields is the
deep {\it Spitzer}/IRAC observations that exist from the 200-hour
GREATS program (Labb{\'e} 2014; Stefanon et al.\ 2020) to provide
rest-optical photometry for $z\sim4$--8 galaxies and thus to estimate
stellar masses.  {\it HST} and {\it Spitzer}/IRAC photometry is
performed on sources in these fields in a similar way to described in
\S2.2, and $UV$-continuum slopes $\beta$ and stellar masses are
estimated using the FAST stellar population fitting code as described
in \S2.5.

In deriving dust corrections for each bin in $UV$ luminosity, we make
use of the stellar masses and $UV$-continuum slopes $\beta$ derived
for our large CANDELS samples and utilize the new relation
Eq.~\ref{eq:irxbm} we derived in \S3.4.2 for the infrared excess IRX
expressed as a function of both $\beta$ and stellar mass $M_{*}$.  To
ensure that our extinction estimates are not overly impacted by noise
in the photometry scattering lower-mass sources to red $\beta$
measurements, we force the infrared excesses of sources with stellar
masses less than 10$^9$ $M_{\odot}$ to be zero, consistent with our
derived observational constraints.

For convenience, we present the dust corrections we have derived here
in Table~\ref{tab:dustcorr}.  If the dust temperatures of $z>3$
galaxies are in fact closer to 35 K than given by our fiducial dust
temperature model, the dust correction we compute would be
approximately half as large.  As in Bouwens et al.\ (2016), we assume
that the average dust correction for UV bright ($<$25.5 mag) galaxies
at $z\sim3$ is $\sim$5 following the findings of Reddy \& Steidel
(2004).

\begin{deluxetable*}{ccccccc}
\tablewidth{15.5cm} 
\tablecolumns{7} 
\tabletypesize{\footnotesize}
\tablecaption{Star Formation Rate Densities Inferred to $-17.0$ AB mag
  (0.03 $L_{z=3} ^{*}$)\label{tab:sfrdens}} \tablehead{
  \colhead{Lyman} & \colhead{} & \colhead{$\textrm{log}_{10}
    \mathcal{L}$} & \colhead{Dust} &
  \multicolumn{3}{c}{$\textrm{log}_{10}$ SFR density}
  \\ \colhead{Break} & \colhead{} & \colhead{(erg s$^{-1}$} &
  \colhead{Correction} & \multicolumn{3}{c}{($M_{\odot}$ Mpc$^{-3}$
    yr$^{-1}$)} \\ \colhead{Sample} & \colhead{$<z>$} &
  \colhead{Hz$^{-1}$ Mpc$^{-3}$)\tablenotemark{a}} &
  \colhead{(dex)\tablenotemark{b}} & \colhead{Uncorrected} &
  \colhead{Corrected} & \colhead{Incl. ULIRG\tablenotemark{b}}}
\startdata U & 3.0 & 26.55$\pm$0.06 & 0.44 & $-$1.60$\pm$0.03 &
$-$1.26$\pm$0.09 & $-$1.16$\pm$0.09\\ B & 3.8 & 26.52$\pm$0.06 & 0.39
& $-$1.63$\pm$0.06 & $-$1.32$\pm$0.06 & $-$1.24$\pm$0.06\\ V & 4.9 &
26.30$\pm$0.06 & 0.32 & $-$1.85$\pm$0.06 & $-$1.58$\pm$0.06 &
$-$1.53$\pm$0.06\\ i & 5.9 & 26.10$\pm$0.06 & 0.20 & $-$2.05$\pm$0.06
& $-$1.88$\pm$0.06 & $-$1.85$\pm$0.06\\ z & 6.8 & 25.98$\pm$0.06 &
0.07 & $-$2.17$\pm$0.06 & $-$2.10$\pm$0.06 & $-$2.10$\pm$0.06\\ Y &
7.9 & 25.67$\pm$0.06 & 0.06 & $-$2.48$\pm$0.06 & $-$2.42$\pm$0.06 &
$-$2.42$\pm$0.06\\ J & 10.4 & 24.62$_{-0.45}^{+0.36}$ & 0.00 &
$-3.28$$_{-0.45}^{+0.36}$ & $-3.28$$_{-0.45}^{+0.36}$ &
$-3.28$$_{-0.45}^{+0.36}$ \enddata \tablenotetext{a}{Integrated down
  to 0.03 $L_{z=3}^{*}$.  Based upon LF parameters in Table 6 of
  Bouwens et al.\ (2015).  The SFR density estimates
  assume $\gtrsim100$ Myr constant SFR and a Chabrier IMF (Madau \& Dickinson 2014).  Conversion to a Salpeter (1955) IMF would result in
  a factor of $\sim$1.6 (0.2 dex) increase in the SFR density
  estimates given here.}  \tablenotetext{b}{The contribution indicated
  here is our fiducial estimate from far-IR bright, ULIRG-like
  ($>10^{12}$ $L_{\odot}$) galaxies (see Fig.~\ref{fig:sfz_obsc}).
  The SFR density contribution for the far-IR bright population tends
  to be either missed completely due to these sources not being
  selected in Lyman-break galaxy probes (e.g., Simpson et al.\ 2014)
  or significantly underestimated due to the IR luminosities
  underestimated based on their $UV$ properties (e.g., Reddy \&
  Steidel 2009).}

\end{deluxetable*}

\subsection{Star Formation Rate Densities at $z\geq3$}

As in the analysis for our pilot study, we apply the dust corrections
we derive in the previous section to the $UV$ luminosity densities
integrating the $UV$ LF of Bouwens et al.\ (2015) to 0.05
$L_{z=3}^{*}$ ($-$17.7 mag) and to 0.03 $L_{z=3}^{*}$ ($-$17.0 mag).
As in previous work, the $UV$ luminosity densities are converted into
SFR densities using the Madau \& Dickinson (2014) conversion factor
$\kappa = 1.1x5 \times 10^{-28} M_{\odot} year^{-1} erg^{-1} s Hz$
(see also Madau et al. 1998 and Kennicutt 1998) modified to assume a
Chabrier (2003) IMF:
\begin{equation}
L_{UV} = \left( \frac{\textrm{SFR}}{M_{\odot} \textrm{yr}^{-1}} \right) 1.4 \times 10^{28} \textrm{erg}\, \textrm{s}^{-1}\, \textrm{Hz}^{-1}\label{eq:mad}
\end{equation}
This relationship assumes a constant SFR for 100 million years.  We
also apply these dust corrections to the Reddy \& Steidel (2009) and
McLure et al.\ (2013) LF results.  Our quantitative results for the
corrected and uncorrected SFR densities at $z\sim3$--10 are presented
in Table~\ref{tab:sfrdens}.  

In computing the SFR density, we must account not only for the impact
of dust extinction on the $UV$ luminosities themselves but also for
the more massive, far-infrared bright sources where standard dust
corrections are not effective or which are sufficiently faint in the
$UV$ to be entirely missed in standard LBG searches (e.g., Reddy et
al.\ 2006, 2008; Swinbank et al.\ 2014; Casey et al.\ 2018; Williams
et al.\ 2019; Dudzevi{\v{c}}i{\={u}}t{\.{e}} et al.\ 2020).  Such
sources are known to contribute a substantial fraction of the SFR
density at $z\sim0$--3 (Hughes et al.\ 1998; Blain et al.\ 1999; Lilly
et al.\ 1999; Chapman et al.\ 2005; Barger et al.\ 2012; Karim et
al.\ 2011; Magnelli et al.\ 2013; Madau \& Dickinson 2014; Swinbank et
al.\ 2014; Wang et al.\ 2019; Dudzevi{\v{c}}i{\={u}}t{\.{e}} et
al.\ 2020).  Perhaps the best way to account for these galaxies
(proposed earlier by Reddy et al.\ 2008) is to simply include them
based on dedicated searches for these sources in the IR.

We consider the results of Magnelli et al.\ (2013) at $z\sim0$--2
(which build on the results of Caputi et al.\ 2007 and Magnelli et
al.\ 2009, 2011), the Franco et al.\ (2020a) results at $z\sim2$--5
from a 69 arcmin$^2$ survey area, the Yamaguchi et al.\ (2019) results
at $z\sim3$--5 from the 26 arcmin$^2$ ASAGAO survey area, and the
Williams et al.\ (2019) serendipitious discovery of a probable dusty
SF source at $z\sim5$.  We compute the SFR density contribution from
ULIRG-type galaxies at $z\sim2.5$ from ASPECS volume by converting the
measured ALMA fluxes from detected sources in their survey area to
SFRs assuming 100\% of the energy comes from star formation, binning
the contributions by the derived redshifts for the sources, and then
dividing by the cosmic volume within a 4.2 arcmin$^2$ survey area,
finding 0.036$\pm$0.022 $M_{\odot}$$\,$yr$^{-1}$.  We use a similar
approach derive the SFR density contribution from ULIRGs at
$z\sim2$--5 from the 69-arcmin$^2$ Franco et al.\ (2020a) probe, but
given the limited depth of the Franco et al.\ (2020a) probe, we treat
this derived contribution as a lower limit.  The Franco et
al.\ (2020a) probe builds on the earlier Franco et al.\ (2018) study
using deeper search results presented in Franco et al.\ (2020b).

Additionally, we consider the integrated SFR density derived from
MAGPHYS fits to the $\sim$1 deg$^2$ AS2UDS sample by
Dudzevi{\v{c}}i{\={u}}t{\.{e}} et al.\ (2020), who corrected for
incompleteness using the number counts from Geach et al.\ (2017) and
extrapolated from the observed 870$\mu$m flux limit of 3.6 mJy for the
SCUBA-2 survey to 1 mJy using the slope of the number counts from
Hatsukade et al.\ (2018).  See \S5.4 and Figure 15 from
Dudzevi{\v{c}}i{\={u}}t{\.{e}} et al.\ (2020).  The uncertainties on
the values were calculated by resampling the SFR and redshift
probability distributions of each source.  The approximate survey
volume for the AS2UDS results is approximately 7$\times$10$^7$ Mpc$^3$
and so likely to be much more representative than smaller volume
studies.

Finally, we also estimate the SFR density from ULIRG-type sources by
assuming the star-forming main sequence results of Speagle et
al.\ (2014) apply to the wide-area $z\sim1.5$--3 mass functions of
Ilbert et al.\ (2013) and $z\sim3$--6 mass functions of Davidzon et
al.\ (2017).  Encouragingly enough, the estimated SFR contribution
provided by ULIRG-type galaxies using the observed mass functions
appears to plausibly consistent with that derived from constraints
available from direct searches for ULIRG-type sources at $z=2$--4 and
$\sim$0.2 dex higher at $z=4$--6.

\begin{figure}
\epsscale{1.15} \plotone{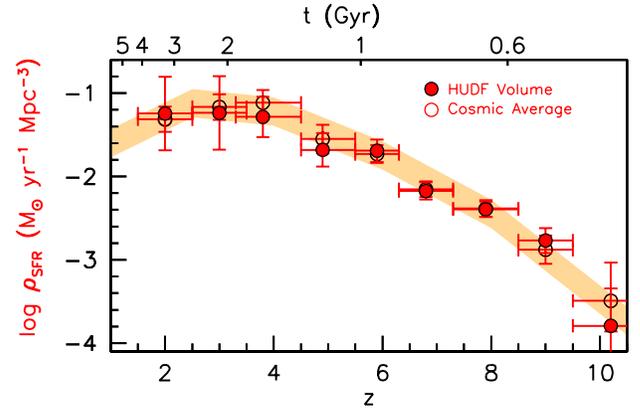}
\caption{Comparison of the inferred SFR density from the ASPECS volume
  (\textit{solid red circles}) with the present estimate based on
  much larger cosmic ($\sim$10$^6$ comoving Mpc$^{3}$) volumes
  (\textit{open circles}: see \S4.5).  The SFR density in the ASPECS
  volume is estimated by multiplying the cosmic SFR density by the
  relative normalization of the $UV$ LF over the ASPECS area to that
  derived over much wider areas and assuming that the same corrections
  for dust (and a missing ULIRG contribution) apply to
  both.\label{fig:sfz_rel}}
\end{figure}

A summary of the inferred SFR density for all the aforementioned ULIRG
probes is presented in Figure~\ref{fig:sfz_obsc}.  As our fiducial
estimate of the obscured SFR density from ULIRGs, we adopt the
Magnelli et al.\ (2013) constraints at $z\sim2$, our mass function
derived estimate at $z\sim2.75$, the AS2UDS estimates
(Dudzevi{\v{c}}i{\={u}}t{\.{e}} et al.\ 2020) at $z=3.4$--6, and the
Wang et al.\ (2019) at $z>6$.  We have indicated fiducial obscured SFR
densities in Figure~\ref{fig:sfz_obsc} with the hatched red area.
This fiducial model is most consistent with the dust poor model from
Casey et al.\ (2018).

We combine these SFR densities with those we derived by correcting the
$UV$ LFs at $z=3$--10 to present our best estimates for the SFR
density at $z=3$--10 in Table~\ref{tab:sfrdens} and
Figure~\ref{fig:sfz}, together with a few previous estimates
(Schiminovich et al.\ 2005; Reddy \& Steidel 2009; McLure et
al.\ 2013) in Figure~\ref{fig:sfz}.  It is interesting to compare the
contribution that unobscured and obscured star formation makes to the
total SFR density of the universe.  Figure~\ref{fig:sfzc} shows such a
breakdown of the SFR density.  The obscured SFR density shown at $z<2$
is from the Magnelli et al.\ (2009, 2011, 2013), while at $z\sim2$-3,
the obscured SFR density shown is the sum of the SFR density from
AS2UDS (Dudzevi{\v{c}}i{\={u}}t{\.{e}} et al.\ 2020) and from Reddy \&
Steidel (2009).  The contribution to the SFR density from ULIRGs is
presented for context.  From the presented breakdown, we can see that
star formation is mostly unobscured at $z>5$, mostly obscured at
$z<5$, and $z\sim5$ marks the approximate transition redshift between
the two regimes.  Previously, Bouwens et al.\ (2009), Bouwens et
al.\ (2016), and Dunlop et al.\ (2017) found that the approximate
transition point between the two regimes was $z\sim4$.

\subsection{Star Formation Rate Density in the ASPECS Volume}

Finally, before closing the discussion we provide in this paper on the
SFR density, it is interesting to try to estimate the SFR density
within the ASPECS HUDF/XDF volume itself.  Given the limited volume
probed by ASPECS and the impact of large scale structure, this is an
interesting issue to examine to help determine the extent to which
conclusions drawn from the HUDF volume are applicable to much larger
cosmic ($\sim$10$^6$ comoving Mpc$^3$) volumes of the universe (where
the impact of large-scale structure is less).

To estimate the approximate SFR density within the ASPECS volume, we
rederive the $UV$ LF at $z\sim2$, $z\sim3$, $z\sim4$, $z\sim5$,
$z\sim6$, $z\sim7$, $z\sim8$, and $z\sim10$ but only using sources in
the ASPECS/HUDF/XDF volume.  For simplicity, in deriving this LF, we
fix the faint-end slope $\alpha$ and characteristic luminosity $M^*$
to that derived from Bouwens et al.\ (2015) and Bouwens et al.\ (2020,
in prep) at these same redshifts and fit for the normalization
$\phi^*$.  The relative normalization we derive for the $UV$ LFs over
the ASPECS areas relative to the cosmic average, i.e., $\phi_{ASPECS}
^{*}/<\phi^*>$ is 1.17, 0.85, 0.68, 0.74, 1.10, 0.95, 0.98, 1.29, and
0.50 at $z\sim2$, $z\sim3$, $z\sim4$, $z\sim5$, $z\sim6$, $z\sim7$,
$z\sim8$, $z\sim9$, and $z\sim10$, respectively.  The rms logarithmic
scatter in these normalizations are 0.12 dex, i.e., fluctuations of
32\% (0.12 dex) in the volume density of galaxies in a given redshift
interval of the HUDF relative to the cosmic average.  32\% is fairly
similar to the expected variations one would expect for sources with a
volume density of $\sim$1$\times$10$^{-3}$ Mpc$^{-3}$ inside a 2
arcmin $\times$ 2 arcmin $\times$ $\Delta z \sim 1$ volume.
Figure~\ref{fig:sfz_rel} illustrates how the SFR density we infer from
the HUDF might compare with the cosmic average, if we assume that we
can apply the LF normalization factors just derived to the SFR density
as a whole.

To assess the impact of large scale structure on the present results
and other results from ASPECS, it is relevant to compare the observed
0.12-dex scatter with that expected from the relative small number of
dust detected and CO-detected sources over ASPECS.  In cases where the
number of sources per unit redshift is in the range 10-15, i.e.,
similar to the number of dust detected and CO detected sources in
ASPECS (e.g., Gonz{\'a}lez-L{\'o}pez et al.\ 2020; Boogaard et
al.\ 2019), the scatter expected from small number statistics will be
comparable to that seen in terms of large-scale structure.  This
suggests that any conclusions drawn from the HUDF ASPECS volume should
be applicable to much larger cosmic ($\sim$10$^6$ comoving Mpc$^{3}$)
volumes, with a relatively limited impact from large-scale structure.

\section{Summary}

Here we make use of sensitive observations we have obtained from the
ALMA large program ASPECS of far-IR continuum light for a large sample
of $z=1.5$--10 galaxies located over the Hubble Ultra Deep Field
(HUDF).  ASPECS probes with great sensitivity (9.3$\mu$Jy$\,$beam$^{-1}$:
$1\sigma$) the 1.2$\,$mm far-IR continuum of $z\geq2$ galaxies and
extends over a 4.2 arcmin$^2$ region using 90 hours of band-6
observations in total.

With these observations, we probe dust-enshrouded star formation to
7-28 $M_{\odot}$$\,$yr$^{-1}$ ($4\sigma$) from 1362 robust
$z=1.5$--10, $UV$-selected galaxies located over the ASPECS footprint.
These $z=1.5$--10 sources were either drawn from the literature
(Bouwens et al.\ 2015) or selected specifically for this study by
applying standard color selection or photometric redshift criteria to
the deep WFC3/UVIS observations over the HUDF from the UVUDF program
(Teplitz et al.\ 2013; Rafelski et al.\ 2015).

Eighteen of the $z>1.5$ galaxies within our ASPECS footprint are
detected at $>$4$\sigma$ in our 1.2-mm continuum observations.  12 of
the 18 $>$4$\sigma$ detections were previously identified as part of
the ASPECS pilot program (Aravena et al.\ 2016; Bouwens et al.\ 2016)
or the Dunlop et al.\ (2017) program.  Six of the reported continuum
detections are new discoveries from the ASPECS large program (see
Gonz{\'a}lez-L{\'o}pez et al.\ 2020; Aravena et al.\ 2020).

The observed number of continuum detections is in agreement with the
predictions obtained by applying a consensus low-redshift IRX-$\beta$
relationship derived here (Appendix B) to the highest-mass $z=1.5$--10
galaxies found over ASPECS suggests a likely sample of 28 continuum
detections, while only 16 continuum detections is predicted if only
sources with stellar masses in excess of $10^{9.5}$ $M_{\odot}$ are
considered.  This consensus IRX-$\beta$ relationship is constructed by
combining the IRX-$\beta$ relations derived in Overzier et
al.\ (2011), Takeuchi et al.\ (2012), and Casey et al.\ (2014).

In agreement with previous studies, we find that the fraction of
detected galaxies in our samples increases sharply with increasing
stellar mass, with the detection fraction rising from 0\% at
$10^{9.0}$ $M_{\odot}$ to 85$_{-18}^{+9}$\% at $>$10$^{10}$
$M_{\odot}$ for sources probed to a sensitivity of
$<$20$\mu$Jy$\,$beam$^{-1}$.  Interestingly, at low stellar masses,
i.e., $<$10$^{9.25}$ $M_{\odot}$, stacking all 1253 sources in our
catalogs over the ASPECS footprint, we recover an average 1.2mm flux
density of $-$0.1$\pm$0.4$\mu$Jy$\,$beam$^{-1}$, implying that the
obscured star formation rate of lower-mass galaxies is essentially
zero, i.e., 0.0$\pm$0.1 $M_{\odot}$$\,$yr$^{-1}$ (converting the flux
density constraint to SFR at $z\sim4$).

The infrared excess ($IRX = L_{IR}/L_{UV}$) of galaxies in our
$z=1.5$--3.5 sample shows a strong correlation with the estimated
stellar mass $M$, with a best-fit relation $IRX =
(M/10^{9.15_{-0.16}^{+0.18}} M_{\odot})^{0.97_{-0.17}^{+0.17}}$.  Both
the recovered normalization and slope of this relation is in agreement
with previous work.  The infrared excess of galaxies in our $z=3.5$--10
sample seems to show approximately the same relationship with stellar
mass.  Unfortunately, there are an insufficient number of high-mass
star-forming galaxies within the ASPECS volume to constrain the
relation.

However, we do note that, for our particular sample of galaxies, the
IRX versus stellar mass relation we derive does show some dependence
on which stellar population we use to estimate stellar masses.  If we
instead use Prospector (Leja et al.\ 2017) stellar population model to
estimate masses for sources in our sample instead of FAST (Kriek et
al.\ 2009), we derive a steeper IRX stellar mass relationship.

The IRX-$\beta$ relation we recover for higher-mass ($>10^{9.5}$
$M_{\odot}$) $z\sim1.5$--3.5 galaxies is most consistent with a
Calzetti-like IRX-$\beta$ relation (here represented with the Reddy et
al.\ 2015 dust curve).  The relation we derive is somewhat steeper
than we previously derived (Bouwens et al.\ 2016), but is nevertheless
consistent.  Our new IRX-$\beta$ relation is similar to that derived
by many previous teams (Reddy et al.\ 2018; {\'A}lvarez-M{\'a}rquez et
al.\ 2016; Heinis et al.\ 2013), but lower than some others (McLure et
al.\ 2018), especially at blue $\beta$'s (i.e., $\beta\sim-1.8$).  For
lower-mass $z\sim 1.5$--3.5 galaxies, the IRX-$\beta$ relation we
derive is most consistent with an SMC-like relation.

Using stellar-mass and $\beta$ measurements for $z\sim2$ galaxies over
CANDELS, we derive the following relation between $\beta$
and stellar mass:
\begin{equation}
M(\beta) = (10^{9.07} M_{\odot})(1.7 \times 10^{0.4(1.42)(\beta+2.3)}-1)
\end{equation}
We then use this correlation to show that our IRX-$\beta$ and
IRX-stellar mass relations are closely connected (see also McLure et
al.\ 2018; Carvajal et al.\ 2020).  We then use these constraints to
express the infrared excess as the following bivariate function of
$\beta$ and stellar mass:
\begin{displaymath}
IRX(\beta,M) = 1.7 (10^{0.4(dA_{UV}/d\beta)(\beta+2.3)}-1)(M/M(\beta))^{\alpha}
\end{displaymath}
The best-fit values we derive for $dA_{UV}/d\beta$ and $\alpha$ are
1.48$\pm$0.10 and 0.67$\pm$0.06, respectively, using our ASPECS
measurements.

We quantify the stacked constraints on the infrared excess in $z>3.5$
galaxies as a function of stellar mass and $\beta$ results and recover
results at $z>3.5$ consistent with what we find at $z=1.5$--3.5 if we
assume a significant evolution in dust temperature with redshift
(e.g., as found by Schreiber et al.\ 2018 or using our
Eq.~\ref{eq:tvsz}).  If the dust temperature of $z\sim3.5$--10
galaxies instead remains fixed at 35 K
(e.g. Dudzevi{\v{c}}i{\={u}}t{\.{e}} et al.\ 2020), we recover
infrared excesses at $z>3.5$ that are 0.4 dex lower than at
$z=1.5$--3.5.

Finally, we make use of our improved constraints on the dependence of
the infrared excess on $\beta$ and stellar mass to provide new
estimates of the dust corrections for the general star-forming galaxy
population at $z\geq4$.  We determine these dust corrections as a
function of $UV$ luminosity and use the measured $UV$ continuum
slopes, stellar masses, and $UV$ luminosities for large numbers of
$z\sim4$, 5, 6, 7, and 8 galaxies identified over the CANDELS
GOODS-South and GOODS-North fields to compute these corrections.

We then leverage these new dust corrections and the $UV$ LF
determinations from Bouwens et al.\ (2015) to provide updated
estimates of the SFR density at $z=4$--10.  We explicitly subdivide
these SFR density estimates into the obscured and unobscured
contributions and show that the SFR density transitions from being
primarily unobscured to obscured at $z\sim5$.  Previously, Bouwens et
al.\ (2009), Bouwens et al.\ (2016), and Dunlop et al.\ (2017) found
that the approximate transition point between the two regimes was
$z\sim4$.

In the future, we can look forward to further significant progress in
our understanding of obscured star formation at high redshift from
targeting large numbers of moderate to high mass galaxies at $z>3.5$
as is being done with the ALPINE program (Le Fevre et al.\ 2019;
Fudamoto et al.\ 2020b).  Improvements in our constraints on the dust
temperatures of $z>3$ galaxies from shorter and longer wavelength
observations will be valuable in computing more accurate IR
luminosities of individual sources.  Also important will be the
discovery of larger, statistical samples of IR-luminous, dusty star
forming galaxies (e.g., Dudzevi{\v{c}}i{\={u}}t{\.{e}} et al.\ 2020)
to achieve a more complete census of the total SFR density at $z>3$.
Finally, at the extreme low-luminosity end, further progress will be
made in searching for obscured star formation in individual low
luminosity sources through the ALMA Lensing Cluster Survey large
program (2018.1.00035.L, PI: Kohno).

\acknowledgements

We thank Matthieu B{\'e}thermin for helpful discussions.  This paper
benefited greatly from a helpful report from an anonymous referee.
This paper makes use of the ALMA data from the program
2016.1.00324.L. ALMA is a partnership of ESO (representing its member
states), NSF (USA) and NINS (Japan), together with NRC (Canada), NSC
and ASIAA (Taiwan), and KASI (Republic of Korea), in cooperation with
the Republic of Chile. The Joint ALMA Observatory is operated by ESO,
AUI/NRAO and NAOJ.  R.J.B., M.S., and T.N.  acknowledge support from
NWO TOP grant TOP1.16.057.  J.G-L. acknowledges partial support from
ALMA-CONICYT project 31160033.  I.R.S. acknowledges support from STFC
(ST/P000541/1).  F.W. and M.N. acknowledge support from ERC Advanced
Grant 740246 (Cosmic Gas).  U.D. acknowledges the support of STFC
studentship (ST/R504725/1).  D.R. acknowledges support from the
National Science Foundation under grant numbers AST-1614213 and
AST-1910107 and from the Alexander von Humboldt Foundation through a
Humboldt Research Fellowship for Experienced Researchers.  H.I.
acknowledges support from JSPS KAKENHI Grant Number JP19K23462.  Este
trabajo cont{\'o} con el apoyo de CONICYT + PCI + INSTITUTO Max Planck
de Astronomia MPG190030.

\appendix

\begin{deluxetable*}{ccccccccc}
\tablewidth{0cm}
\tablecolumns{9}
\tabletypesize{\footnotesize}
\tablecaption{Stacked Results: IRX versus Stellar Mass\label{tab:irxsm}}
\tablehead{
\colhead{} & \colhead{} & \colhead{$\log_{10}$} & \colhead{} & \colhead{Measured} & \colhead{Predicted $f_{1.2mm}$} & \colhead{} & \colhead{Measured}\\
\colhead{} & \colhead{\# of} & \colhead{M$_{wht}/$} & \colhead{} & \colhead{$f_{1.2mm}$ } & \colhead{flux [$\mu$Jy]} & \colhead{Measured} & \colhead{$f_{1.2mm}$/}\\
\colhead{Mass (M)} & \colhead{sources} & \colhead{M$_{\odot}$} & \colhead{$\beta_{wht}$} & \colhead{flux [$\mu$Jy]\tablenotemark{a,b}} & \colhead{Mass\tablenotemark{c,d}} & \colhead{IRX\tablenotemark{a,b,d}} & \colhead{$f_{UV}$\tablenotemark{a,b,e}}}
\startdata
\multicolumn{9}{c}{$z=1.5$--3.5} \\
$>10^{10.75} M_{\odot}$ & 5 & 10.9 & 0.3 & 337$_{-54}^{+103}$$\pm$8 & 603 & 41.70$_{-16.08}^{+56.17}$$\pm$1.01 & 465$_{-142}^{+873}$$\pm$13\\
$10^{10.25} M_{\odot}$ - $10^{10.75} M_{\odot}$ & 6 & 10.5 & 0.0 & 190$_{-50}^{+51}$$\pm$4 & 214 & 21.88$_{-9.18}^{+21.27}$$\pm$0.50 & 193$_{-31}^{+154}$$\pm$8\\
$10^{10.25} M_{\odot}$ - $10^{10.75} M_{\odot}$ (ind $<$4$\sigma$) & 0 & 0.0 & 0.0 & 0.0$_{-0.0}^{+0.0}$$\pm$0.0 & 0 & 0.00$_{-0.00}^{+0.00}$$\pm$0.00 & 0$_{-0}^{+0}$$\pm$0\\
$10^{9.75} M_{\odot}$ - $10^{10.25} M_{\odot}$ & 11 & 9.9 & $-$1.0 & 166$_{-128}^{+133}$$\pm$4 & 78 & 13.56$_{-9.73}^{+7.58}$$\pm$0.42 & 448$_{-381}^{+298}$$\pm$9\\
$10^{9.75} M_{\odot}$ - $10^{10.25} M_{\odot}$ (ind $<$4$\sigma$) & 9 & 9.9 & $-$0.9 & 21$_{-4}^{+5}$$\pm$6 & 51 & 2.47$_{-0.86}^{+0.85}$$\pm$0.65 & 31$_{-14}^{+31}$$\pm$11\\
$10^{9.25} M_{\odot}$ - $10^{9.75} M_{\odot}$ & 33 & 9.5 & $-$1.4 & 26$_{-12}^{+17}$$\pm$3 & 39 & 1.81$_{-0.62}^{+0.96}$$\pm$0.19 & 30$_{-10}^{+13}$$\pm$4\\
$10^{8.75} M_{\odot}$ - $10^{9.25} M_{\odot}$ & 123 & 9.0 & $-$1.6 & 2.4$_{-1.4}^{+1.4}$$\pm$1.2 & 4 & 0.73$_{-0.38}^{+0.39}$$\pm$0.31 & 22$_{-7}^{+7}$$\pm$6\\
$< 10^{8.75} M_{\odot}$ & 467 & 8.2 & $-$1.9 & 0.6$_{-0.7}^{+0.8}$$\pm$0.6 & 0 & 0.59$_{-0.62}^{+0.60}$$\pm$0.52 & 0$_{-9}^{+9}$$\pm$8\\\\
\multicolumn{9}{c}{$z=3.5$--10} \\
$M>10^{10.25} M_{\odot}$ & 1 & 10.8 & 2.9 & 180$_{-0}^{+0}$$\pm$10 & 428 & 19.08$_{-0.00}^{+0.00}$$\pm$1.02 & 708$_{-0}^{+0}$$\pm$38\\
$10^{9.75} M_{\odot}$ - $10^{10.25} M_{\odot}$ & 6 & 9.9 & $-$1.1 & $-$1$_{-5}^{+5}$$\pm$5 & 30 & $-$0.22$_{-0.87}^{+0.76}$$\pm$1.11 & $-$7$_{-30}^{+27}$$\pm$47\\
$10^{9.25} M_{\odot}$ - $10^{9.75} M_{\odot}$ & 31 & 9.5 & $-$1.6 & 10$_{-4}^{+6}$$\pm$2 & 11 & 4.12$_{-2.38}^{+3.23}$$\pm$0.49 & 85$_{-50}^{+74}$$\pm$15\\
$10^{8.75} M_{\odot}$ - $10^{9.25} M_{\odot}$ & 69 & 9.0 & $-$1.9 & 0.6$_{-1.6}^{+1.6}$$\pm$1.5 & 2 & 0.41$_{-0.51}^{+0.50}$$\pm$0.61 & 39$_{-14}^{+14}$$\pm$15\\
$< 10^{8.75} M_{\odot}$ & 594 & 7.6 & $-$2.2 & $-$0.6$_{-0.6}^{+0.5}$$\pm$0.6 & 0 & $-$0.72$_{-0.66}^{+0.59}$$\pm$0.59 & 23$_{-15}^{+14}$$\pm$14\\
$< 10^{9.75} M_{\odot}$ & 694 & 7.9 & $-$2.1 & 0.2$_{-0.6}^{+0.7}$$\pm$0.5 & 1 & 0.27$_{-0.58}^{+0.68}$$\pm$0.39 & 47$_{-18}^{+23}$$\pm$8\\\\
\multicolumn{9}{c}{$z=1.5$-10} \\
$< 10^{9.75} M_{\odot}$ & 1317 & 8.0 & $-$2.0 & 0.7$_{-0.5}^{+0.6}$$\pm$0.4 & 1 & 0.98$_{-0.35}^{+0.34}$$\pm$0.24 & 27$_{-6}^{+7}$$\pm$3\\
$< 10^{9.25} M_{\odot}$ & 1253 & 7.9 & $-$2.1 & $-$0.1$_{-0.4}^{+0.5}$$\pm$0.4 & 0 & 0.50$_{-0.35}^{+0.34}$$\pm$0.31 & 18$_{-5}^{+5}$$\pm$4\\
All & 1346 & 8.0 & $-$2.0 & 2.2$_{-0.8}^{+0.8}$$\pm$0.4 & 3 & 3.84$_{-0.90}^{+0.95}$$\pm$0.22 & 93$_{-28}^{+36}$$\pm$2
\enddata
\tablenotetext{a}{This column presents stack results.  Each source is
  weighted according to the inverse square of the noise.  The
  weightings are therefore independent of stellar mass and
  $UV$-continuum slope $\beta$.}
\tablenotetext{b}{Both the bootstrap and formal uncertainties are quoted on the result (presented first and second, respectively).}
\tablenotetext{c}{The 1.2$\,$mm continuum flux predicted from the consensus $z\sim2$--3 IRX-stellar mass relationship weighting individual
sources in exactly the same way as for the measured 1.2$\,$mm continuum flux.  This column should therefore be directly comparable with
the column directly to the left, i.e., giving the measured flux.}
\tablenotetext{d}{Assuming a standard modified blackbody SED with our evolving dust temperature model and accounting for the impact of the CMB on the
  measured flux (da Cunha et al.\ 2013).}
\tablenotetext{e}{Results do not depend on the assumed far-IR SED template.}
\end{deluxetable*}

\begin{deluxetable*}{cccccccccc}
\tablewidth{0cm}
\tablecolumns{10}
\tabletypesize{\footnotesize}
\tablecaption{IRX versus Apparent Magnitude in the Rest-frame $UV$ ($m_{UV,AB}$)\label{tab:irxmuv}}
\tablehead{
\colhead{} & \colhead{} & \colhead{$\log_{10}$} & \colhead{} & \colhead{Measured } & \multicolumn{3}{c}{Predicted} & \colhead{} & \colhead{}\\
\colhead{} & \colhead{\# of} & \colhead{$M_{med}$/} & \colhead{} & \colhead{$f_{1.2mm}$ } & \multicolumn{3}{c}{$f_{1.2mm}$ [$\mu$Jy]} & \colhead{} & \colhead{$f_{1.2mm}$/}\\
\colhead{$m_{UV}$} & \colhead{sources} & \colhead{$M_{\odot}$} & \colhead{$\beta_{med}$} & \colhead{[$\mu$Jy]\tablenotemark{a}} & \colhead{Calz\tablenotemark{a}} & \colhead{SMC\tablenotemark{a}} & \colhead{Mass\tablenotemark{a}} & \colhead{IRX\tablenotemark{a}} & \colhead{$f_{UV}$\tablenotemark{a}}}
\startdata
\multicolumn{9}{c}{$z=1.5$--3.5} \\
$<25$ & 35 & 9.5 & $-$1.4 & 92$_{-52}^{+58}$$\pm$2 & 171 & 24 & 108 & 5.73$_{-2.06}^{+2.20}$$\pm$0.16 & 109$_{-43}^{+54}$$\pm$3\\
$<25$ (ind $<$4$\sigma$) & 29 & 9.4 & $-$1.5 & 12$_{-5}^{+5}$$\pm$3 & 99 & 14 & 43 & 1.14$_{-0.37}^{+0.42}$$\pm$0.21 & 18$_{-6}^{+7}$$\pm$4\\
25-31 & 610 & 8.4 & $-$1.8 & 4.7$_{-1.4}^{+1.6}$$\pm$0.6 & 18 & 2 & 3 & 4.14$_{-1.34}^{+1.51}$$\pm$0.35 & 52$_{-15}^{+16}$$\pm$5\\
All & 645 & 8.5 & $-$1.7 & 8.4$_{-2.6}^{+3.1}$$\pm$0.5 & 24 & 3 & 7 & 4.67$_{-1.08}^{+1.32}$$\pm$0.24 & 95$_{-33}^{+39}$$\pm$3\\\\
\multicolumn{9}{c}{$z=3.5$--10} \\
$<26$ & 33 & 9.1 & $-$1.5 & 13$_{-8}^{+10}$$\pm$3 & 11442 & 124 & 38 & 1.87$_{-1.35}^{+1.40}$$\pm$0.29 & 73$_{-45}^{+46}$$\pm$10\\
26-31 & 668 & 7.9 & $-$2.1 & 0.1$_{-0.7}^{+0.7}$$\pm$0.5 & 4 & 0 & 1 & 0.20$_{-0.73}^{+0.84}$$\pm$0.50 & 80$_{-42}^{+50}$$\pm$13\\
All & 701 & 7.9 & $-$2.1 & 0.5$_{-0.7}^{+0.6}$$\pm$0.5 & 387 & 5 & 2 & 0.67$_{-0.65}^{+0.70}$$\pm$0.38 & 75$_{-31}^{+37}$$\pm$8
\enddata
\tablenotetext{a}{Calculated identically to the columns in Table~\ref{tab:irxsm}, but using the subdivisions of sources indicated in the rows of this table.}
\end{deluxetable*}

\section{A.  Comparison of our Fiducial Stellar Mass Estimates with those 
from the MAGPHYS and Prospector}

In this appendix, we compare the fiducial stellar masses we derive for
sources in our study using \textsc{FAST} (Kriek et al.\ 2009) with
those derived from the MAGPHYS software (da Cunha et al.\ 2008) which
is used in many of the other ASPECS analyses (e.g., Magnelli et
al.\ 2020).  The stellar masses we estimated with FAST are in
reasonable agrement with \textsc{MAGPHYS}, with the median and mean
stellar mass derived by \textsc{MAGPHYS} being 0.07 dex and 0.36 dex
higher, respectively, with a median absolute difference between the
two mass estimates of 0.38 dex.  This is consistent with their being
no major systematic biases in the results from the present study --
which rely on \textsc{FAST}-estimated masses -- relative to other
papers in the ASPECS series -- where the reliance is on
\textsc{MAGPHYS}-estimated masses.

We also compared our stellar mass estimates with those we derived from
the Prospector code (Leja et al.\ 2017) to $z<2.5$ where Leja et
al.\ (2019) publish stellar mass results based on the Skelton et
al.\ (2014) photometry.  Prospector has many advantages for deriving
robust mass estimates for sources given its flexibility in accounting
for a wide variety of different star formation histories, dust
extinction and reradiation, dust extinction curves, stellar
metallicities, and nebular emission.  The median and mean stellar mass
found with PROSPECTOR from the Leja et al.\ (2019) compilation is 0.12
dex and 0.19 higher, respectively, than what we find from FAST for
sources over the ASPECS HUDF area.  The root mean square difference is
0.28 dex.

\section{B.  Consensus $z\sim0$ IRX-$\beta$ Relationship}

Results from the $z\sim0$ universe provide us with an important
baseline for interpreting dust continuum results in the $z>1.5$
universe.  This is especially the case given the little evolution in
the relationship between the infrared excess and $UV$-continuum slope
$\beta$ from $z\sim2$ to $z\sim0$ (Reddy et al.\ 2006; McLure et
al.\ 2018) and also limited evolution in the infrared excess - stellar
mass relationship (e.g., Pannella et al.\ 2009; Whitaker et
al.\ 2017).

\begin{deluxetable*}{cccccccccc}
\tablewidth{0cm}
\tablecolumns{10}
\tabletypesize{\footnotesize}
\tablecaption{IRX versus $\beta$\label{tab:irxbeta}}
\tablehead{\colhead{} & \colhead{} & \colhead{$\log_{10}$} & \colhead{} & \colhead{Measured} & \multicolumn{2}{c}{Predicted} & \colhead{} & \colhead{} & \colhead{Measured}\\
\colhead{} & \colhead{\# of} & \colhead{M$_{wht}/$} & \colhead{} & \colhead{$f_{1.2mm}$ } & \multicolumn{2}{c}{$f_{1.2mm}$ [$\mu$Jy]} & \colhead{Measured} &  \colhead{Predicted} & \colhead{$f_{1.2mm}$/}\\
\colhead{$\beta$} & \colhead{sources} & \colhead{M$_{\odot}$} & \colhead{$\beta_{med}$} & \colhead{[$\mu$Jy]\tablenotemark{a,b}} & \colhead{Calz\tablenotemark{c,d}} & \colhead{SMC\tablenotemark{c,d}} & \colhead{IRX\tablenotemark{a,b,d}} & \colhead{IRX$_{SMC}$\tablenotemark{c}} & \colhead{$f_{UV}$\tablenotemark{a,b,e}}}
\startdata
\multicolumn{9}{c}{$z=1.5$--3.5 (All Masses)} \\
$-4.0 < \beta < -1.75$ & 373 & 8.3 & $-$2.1 & 0.8$_{-0.8}^{+0.8}$$\pm$0.7 & 3 & 0 & 0.67$_{-0.44}^{+0.44}$$\pm$0.39 & 0.01 & 5$_{-4}^{+4}$$\pm$4\\
$-1.75 < \beta < -1.00$ & 220 & 8.6 & $-$1.5 & 10.8$_{-7.1}^{+7.9}$$\pm$1.0 & 32 & 5 & 2.18$_{-1.14}^{+1.37}$$\pm$0.27 & 0.88 & 117$_{-76}^{+78}$$\pm$4\\
$-1.00 < \beta$ & 52 & 9.0 & $-$0.5 & 50$_{-15}^{+16}$$\pm$2 & 211 & 22 & 17.33$_{-5.24}^{+7.14}$$\pm$0.37 & 5.97 & 182$_{-49}^{+73}$$\pm$5\\
$-1.00 < \beta$ (ind $<$4$\sigma$) & 39 & 8.7 & $-$0.6 & 4$_{-3}^{+3}$$\pm$2 & 101 & 13 & 1.44$_{-0.99}^{+0.76}$$\pm$0.69 & 4.54 & 31$_{-14}^{+9}$$\pm$9\\\\
\multicolumn{9}{c}{$z=1.5$--3.5 ($>10^{9.5} M_{\odot}$)} \\
$-4.0 < \beta < -1.75$ & 4 & 9.6 & $-$1.9 & $-$0$_{-6}^{+6}$$\pm$6 & 48 & 0 & 0.02$_{-0.16}^{+0.12}$$\pm$0.21 & 0.00 & 3$_{-5}^{+3}$$\pm$6\\
$-1.75 < \beta < -1.00$ & 16 & 9.7 & $-$1.2 & 114$_{-88}^{+93}$$\pm$4 & 143 & 23 & 6.54$_{-4.97}^{+4.88}$$\pm$0.28 & 1.65 & 243$_{-201}^{+198}$$\pm$6\\
$-1.75 < \beta < -1.00$ (ind $<$4$\sigma$) & 14 & 9.6 & $-$1.2 & 13$_{-8}^{+8}$$\pm$4 & 132 & 21 & 0.89$_{-0.51}^{+0.49}$$\pm$0.35 & 1.58 & 17$_{-11}^{+15}$$\pm$7\\
$-1.00 < \beta < -0.20$ & 14 & 10.1 & $-$0.7 & 86$_{-25}^{+33}$$\pm$4 & 398 & 51 & 10.27$_{-2.21}^{+3.74}$$\pm$0.30 & 4.23 & 175$_{-47}^{+75}$$\pm$6\\
$-1.00 < \beta < -0.20$ (ind $<$4$\sigma$) & 7 & 9.8 & $-$0.8 & 19$_{-3}^{+2}$$\pm$5 & 155 & 22 & 3.00$_{-0.65}^{+1.12}$$\pm$0.67 & 3.38 & 44$_{-6}^{+44}$$\pm$10\\
$-0.20 < \beta$ & 4 & 10.7 & 0.7 & 289$_{-43}^{+57}$$\pm$6 & 799 & 42 & 174.57$_{-41.65}^{+104.96}$$\pm$3.32 & 22.35 & 4855$_{-1150}^{+1838}$$\pm$90\\\\
\multicolumn{9}{c}{$z=1.5$--3.5 ($<10^{9.5} M_{\odot}$)} \\
$-4.0 < \beta < -1.75$ & 369 & 8.3 & $-$2.1 & 0.8$_{-0.8}^{+0.8}$$\pm$0.7 & 2 & 0 & 0.83$_{-0.52}^{+0.54}$$\pm$0.43 & 0.01 & 10$_{-9}^{+9}$$\pm$7\\
$-1.75 < \beta < -1.00$ & 204 & 8.5 & $-$1.5 & 2.3$_{-1.2}^{+1.2}$$\pm$1.0 & 23 & 3 & 0.84$_{-0.44}^{+0.39}$$\pm$0.36 & 0.82 & 44$_{-15}^{+12}$$\pm$5\\
$-1.00 < \beta$ & 34 & 8.5 & $-$0.6 & 14$_{-11}^{+13}$$\pm$2 & 93 & 12 & 5.57$_{-4.73}^{+6.07}$$\pm$1.13 & 4.71 & 53$_{-56}^{+115}$$\pm$18\\\\
\multicolumn{9}{c}{$z=3.5$--10 (All Masses)} \\
$-4.0 < \beta < -1.75$ & 537 & 7.9 & $-$2.3 & $-$0.3$_{-0.6}^{+0.6}$$\pm$0.6 & 1 & 0 & $-$0.24$_{-0.48}^{+0.39}$$\pm$0.37 & 0.01 & 26$_{-11}^{+11}$$\pm$10\\
$-1.75 < \beta < -1.00$ & 125 & 8.1 & $-$1.5 & 2.0$_{-1.3}^{+1.2}$$\pm$1.2 & 12 & 2 & 0.65$_{-0.54}^{+0.62}$$\pm$0.56 & 0.77 & 38$_{-16}^{+16}$$\pm$16\\
$-1.00 < \beta$ & 32 & 8.4 & $-$0.5 & 7$_{-6}^{+8}$$\pm$2 & 7875 & 89 & 7.67$_{-4.98}^{+4.42}$$\pm$0.96 & 10.76 & 407$_{-265}^{+267}$$\pm$23\\\\
\multicolumn{9}{c}{$z=3.5$--10 ($>10^{9.25} M_{\odot}$, $m_{UV}<28.5$)} \\
$-4.0 < \beta < -1.75$ & 18 & 9.5 & $-$2.0 & 11$_{-4}^{+4}$$\pm$3 & 11 & 0 & 1.54$_{-0.65}^{+0.95}$$\pm$0.39 & 0.02 & 34$_{-19}^{+33}$$\pm$18\\
$-1.75 < \beta < -1.00$ & 8 & 9.7 & $-$1.5 & $-$4$_{-4}^{+4}$$\pm$4 & 38 & 6 & $-$0.62$_{-0.65}^{+0.52}$$\pm$0.74 & 0.83 & $-$30$_{-31}^{+25}$$\pm$34\\
$-1.00 < \beta < -0.2$ & 1 & 9.5 & $-$0.8 & 171$_{-0}^{+0}$$\pm$9 & 124 & 18 & 30.26$_{-0.00}^{+0.00}$$\pm$1.73 & 3.27 & 1362$_{-0}^{+0}$$\pm$74\\
$-0.20 < \beta$ & 2 & 10.2 & 1.6 & 98$_{-80}^{+82}$$\pm$7 & 136731 & 1449 & 10.65$_{-9.58}^{+4.66}$$\pm$0.55 & 114.37 & 356$_{-299}^{+352}$$\pm$26\\
\enddata
\tablenotetext{a}{This column presents stack results.  Each source is
  weighted according to the inverse square of the noise.  The weightings are therefore independent
  of stellar mass and $UV$-continuum slope $\beta$.}
\tablenotetext{b}{Both the bootstrap and formal uncertainties are quoted on the result (presented first and second, respectively).}
\tablenotetext{c}{The 1.2$\,$mm continuum flux predicted using the M99 or SMC IRX-$\beta$ relationship weighting individual
sources in exactly the same way as for the measured 1.2$\,$mm continuum flux, so these two quantities should be directly comparable.}
\tablenotetext{d}{Assuming a standard modified blackbody SED with our evolving dust temperature model and accounting for the impact of the CMB on the
  measured flux (da Cunha et al.\ 2013).}
\tablenotetext{e}{Results do not depend on the assumed far-IR SED template.}
\end{deluxetable*}

The conventional $z\sim0$ reference point has been the Meurer et
al.\ (1999) relation.  However, it is now clear based on a large
amount of work that the actual $z\sim0$ relation should shift both to
redder $\beta$'s and lower infrared excesses (Overzier et al.\ 2011;
Takeuchi et al.\ 2012; Casey et al.\ 2014).\footnote{This shift in the
  $z\sim0$ relation from Meurer et al.\ (1999) is a result of the fact
  that the effective aperture of the IUE observations was too small to
  probe the full $UV$ luminosities of sources in the Meurer et
  al.\ (1999) sample.}  Instead of debating the merits of three recent
determinations of the IRX-$\beta$ relationship at $z\sim0$ by Overzier
et al.\ (2011), Takeuchi et al.\ (2012), and Casey et al.\ (2014),
perhaps the easiest approach is just to find the mean of the
parameters derived in these studies, and use that as our relation.
The means we derive for the intrinsic (unreddened) $UV$-continuum
slope of stellar populations, i.e., $\beta_{int}$, and
$\frac{dA_{FUV}}{d\beta}$ (with no weighting) are $-1.85$ and 1.86,
respectively, such that $A_{FUV} = 1.86 (\beta + 1.85)$ for $\beta <
-1.85$.

Following Meurer et al.\ (1999), the expression for the infrared
excess ($L_{IR}/L_{UV}$) is
\begin{equation}
\log_{10} IRX = \log_{10} (10^{0.4 A_{FUV}} - 1) + \log_{10} \frac{BC_{FUV,*}}{BC_{dust}}
\end{equation}
In this treatment, $BC_{FUV,*}$ and $BC_{dust}$ are the bolometric
corrections from the $L_{UV}$ and $L_{IR}$ luminosities to the total
luminosities in the $UV$ and $IR$.  Taking $L_{UV}$ to be equal to
$\lambda f_{\lambda}$ evaluated at 1600\AA, typical estimates for
$BC_{FUV,*}$ have been in the range 1.66 to 1.71 (Meurer et
al.\ 1999), and we will take $BC_{FUV,*}$ to be equal to 1.7.  If we
also treat $L_{IR}$ as the total IR luminosity (8-1000$\mu$m), we see
that $BC_{dust}$ is approximately equal to 1.

With these inputs, the fiducial $z\sim0$ IRX-$\beta$ relation we
utilize in this study is the following:
\begin{equation}
IRX_{z=0} = 1.7 (10^{0.4(1.86(\beta+1.85))} - 1)
\end{equation}

Despite the significant amount of evidence pointing to a greyer
Calzetti-like extinction curve for high-mass galaxies (Reddy et
al.\ 2006; Daddi et al.\ 2007; Pannella et al.\ 2009), at least some
lower-mass mass galaxies appear to show a steeper SMC-like extinction
curve (Baker et al.\ 2001; Reddy et al.\ 2006, 2010; Siana et
al.\ 2008, 2009).  

Using the observational results of Lequeux et al.\ (1982), Prevot et
al.\ (1984), and Bouchet et al.\ (1985: see also Pei 1992; Pettini et
al.\ 1998; Gordon et al.\ 2003), we earlier obtained the following
representation of the SMC extinction relation in Bouwens et
al.\ (2016): $A_{FUV} = 1.1 (\beta + 2.23)$.  To make this extinction
relation more consistent with the one obtained from the Overzier et
al.\ (2011), Takeuchi et al.\ (2012), and Casey et al.\ (2014)
results, we adjust the $\beta$ intercept to be $-$1.85.  This results
in the following relation:
\begin{equation}
IRX_{SMC} = 1.7 (10^{0.4(1.1(\beta+1.85))} - 1)
\label{eq:irxsmc}
\end{equation}

One other IRX-$\beta$ relationship we compare with in the present
study is the canonical Meurer et al.\ (1999) IRX-$\beta$ relation:
\begin{equation}
IRX_{M99} = 1.7 (10^{0.4(1.99(\beta+2.23))} - 1)
\label{eq:irxm99}
\end{equation}

\section{C.  Comprehensive Presentation of Stack Results}

The purpose of this appendix is to provide a much more comprehensive
presentation of the stack results from ASPECS than is convenient for
the main text.  Tables~\ref{tab:irxsm}-\ref{tab:irxbeta} show our
results for $z\sim2$--10 samples split by stellar mass, $UV$-continuum
slope $\beta$, and apparent magnitude in the $UV$.  

\begin{figure}
\epsscale{0.6} 
\plotone{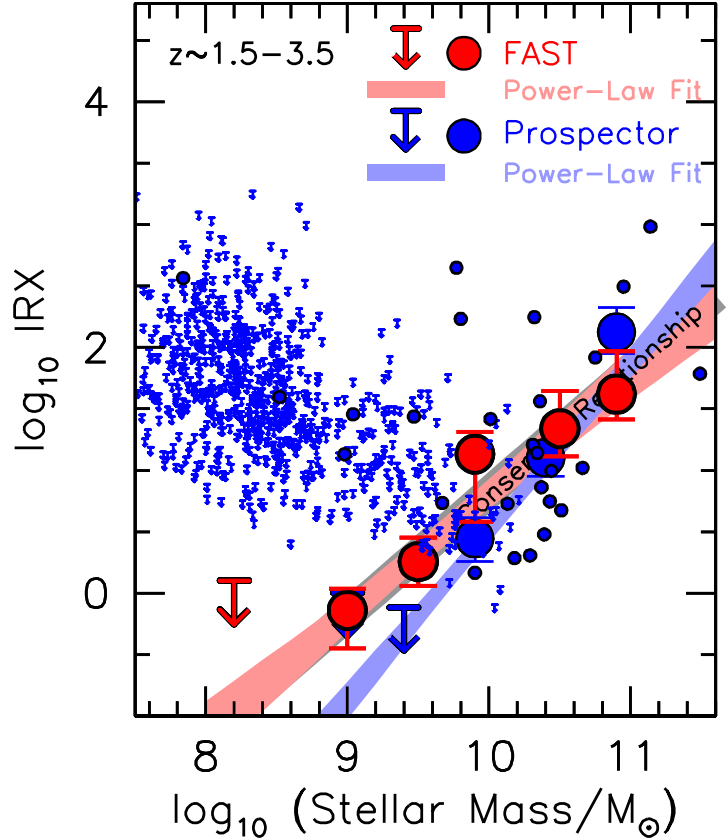}
\caption{Illustration on how the stacked infrared excess vs. stellar
  mass relationship of $z=1.5$--3.5 galaxies depends on whether the
  FAST or \textsc{Prospector} stellar population modeling software is
  used to derive stellar masses for sources over the HUDF
  (\textit{large red and blue circles and downward arrows,
    respectively}).  The shaded red and blue regions are the best-fit
  power-law relations to IRX-stellar mass relation adopting the
  \textsc{FAST} and \textsc{Prospector} stellar masses,
  respectively.\label{fig:irxsmp}}
\end{figure}

\section{D.  Sensitivity of IRX vs. Stellar Mass Relation to Stellar Population Model}

While exploring the relationship between IRX and stellar mass, we
experimented with the use of different codes to estimate the stellar
mass for individual sources over our HUDF ASPECS field.  As found
e.g. in Appendix A, stellar population codes like MAGPHYS (da Cunha et
al.\ 2008) and \textsc{Prospector} (Leja et al.\ 2017) find $\sim$0.12
dex higher stellar masses in general than the FAST (Kriek et
al.\ 2007) stellar population we use for our fiducial stellar mass
estimates.

Our determination of the IRX vs. stellar mass relation can potentially
depend on the stellar population modeling code we use to estimate the
stellar masses of specific sources.  To investigate the dependence on
the stellar mass estimates, we made use of the stellar mass estimates
that Leja et al.\ (2019) provide for sources in our sample to
$z\sim2.5$ from \textsc{Prospector} (for every case where a match can
be found) and rederive the stacked infrared excess vs. stellar mass
relation at $z=1.5$--3.5.  The best-fit values we find for $M_s$ and
$\alpha$ as applies to Eq.~\ref{eq:irxmass} is
$10^{9.63_{-0.12}^{+0.12}}$ $M_{\odot}$ and 1.37$_{-0.15}^{+0.18}$,
respectively, and is shown with the blue solid circles and light-blue
power-law fit in Figure~\ref{fig:irxsmp}.  The derived relationship is
significantly steeper than our fiducial determination shown with the
red solid points and red-shaded region, with a much lower implied IRX
at stellar masses $<$10$^{9.5}$ $M_{\odot}$.

\end{document}